\documentclass[review]{elsarticle} 
\usepackage[colorlinks,citecolor=blue,linktoc=all,linkcolor=cyan]{hyperref}
\usepackage{longtable}
\usepackage{graphicx}
\usepackage{multirow}
\usepackage{bm}
\usepackage{tensor}
\usepackage{tabularx}
\usepackage{textcomp}
\usepackage{color}
\usepackage{geometry}
\usepackage{tensor}
\usepackage{url}
\usepackage{amssymb}
\usepackage{amsmath}
\newcolumntype{Y}{>{\centering\arraybackslash}X}
\usepackage{threeparttable}
\usepackage{epstopdf}
\usepackage{gensymb}
\usepackage{booktabs}
\usepackage{color}
\usepackage{url}
\usepackage{siunitx}
\usepackage{subfigure}
\usepackage{longtable}
\usepackage{array}
\usepackage{booktabs}
\usepackage{verbatim}
\usepackage{float}
\usepackage{enumerate}
\usepackage{arydshln} 
\usepackage{ulem}
\usepackage{lineno}
\usepackage[T1]{fontenc}
\usepackage{dsfont}               
\usepackage{mathrsfs}             
\usepackage{slashed}              
\usepackage{amsbsy}
\usepackage{amsfonts}

\numberwithin{equation}{section}
\numberwithin{table}{section}
\numberwithin{figure}{section}
\journal{Progress in Particle and Nuclear Physics}

\usepackage{titlesec}
\usepackage{sectsty}
\titleformat{\section}{\normalfont\Large\bfseries}{\thesection}{1em}{}
\titleformat{\subsection}{\normalfont\large\bfseries}{\thesubsection}{1em}{}
\titleformat{\subsubsection}{\normalfont\normalsize\bfseries}{\thesubsubsection}{1em}{}

\begin{document}
\begin{frontmatter}

\title{Laser Spectroscopy for the Study of Exotic Nuclei}

\author[PKU]{X.F.~Yang\corref{mycorrespondingauthor}}
\cortext[mycorrespondingauthor]{Corresponding authors}
\ead{xiaofei.yang@pku.edu.cn}
\author[PKU]{S.J.~Wang}
\author[MIT]{S.G.~Wilkins\corref{mycorrespondingauthor}}
\ead{wilkinss@mit.edu}
\author[MIT]{R.F.~Garcia Ruiz\corref{mycorrespondingauthor}}
\ead{rgarciar@mit.edu}
\address[PKU]{School of Physics and State Key Laboratory of Nuclear Physics and Technology, Peking University, Beijing 100871, China}
\address[MIT]{Massachusetts Institute of Technology, Cambridge, MA 02139, USA}

\begin{abstract} 
Investigation into the properties and structure of unstable nuclei far from stability is a key avenue of research in modern nuclear physics. These efforts are motivated by the continual observation of unexpected structure phenomena in nuclei with unusual proton-to-neutron ratios. In recent decades, laser spectroscopy techniques have made significant contributions in our understanding of exotic nuclei in different mass regions encompassing almost the entire nuclear chart. This is achieved through determining multiple fundamental properties of nuclear ground and isomeric states, such as nuclear spins, magnetic dipole and electric quadrupole moments and charge radii, via the measurement of hyperfine structures and isotope shifts in the atomic or ionic spectra of the nuclei of interest. These properties offer prominent tests of recently developed state-of-the-art nuclear theory and help to stimulate new developments in improving the many-body methods and nucleon-nucleon interactions at the core of these models. With the aim of exploring more exotic short-lived nuclei located ever closer to the proton and neutron driplines, laser spectroscopy techniques, with their continuous technological developments towards higher resolution and higher sensitivity, are extensively employed at current- and next-generation radioactive ion beam facilities worldwide. Ongoing efforts in parallel promise to improve the availability of these even more exotic species at next-generation facilities. Very recently, an innovative application of laser spectroscopy on molecules containing short-lived nuclei has been demonstrated offering additional opportunities for several fields of research, e.g. fundamental symmetry studies and astrophysics. In this review, the basic nuclear properties measurable with laser spectroscopy will be introduced. How these observables are associated with nuclear structure and nucleon-nucleon interactions will be discussed. Following this, a general overview of different laser spectroscopy methods will be given with particular emphasis on technical advancements reported in recent years. The main focus of this article is to review the numerous highlights that have resulted from studying exotic nuclei in different mass regions with laser spectroscopy techniques since the last edition in this series. Finally, the challenges facing the field in addition to future opportunities will be discussed.
\end{abstract}


\end{frontmatter}

\newpage

\thispagestyle{empty}
\tableofcontents


\section{Introduction\label{sec:intro}}

Nuclear electromagnetic properties of ground and long-lived isomeric states are fundamental properties of atomic nuclei that offer complementary insights into their multi-faceted nature. Measurements of these observables in nuclei are essential to elucidate details about their structure and the inter-nucleon interactions that govern them. Several new radioactive ion beam (RIB) facilities are commencing operation in the coming years with their development motivated largely by the many examples of surprising structure phenomena that have been observed in exotic nuclei.
Investigating the properties and structure of unstable nuclei far from stability at RIB facilities remains a large and key avenue of nuclear physics research today.
Unexpected physical phenomena are continuously observed in different regions of the nuclear chart. These observations continue to challenge our global understanding of nuclear structure and provide a testing ground for state-of-the-art nuclear theory. In parallel to advances in experimental techniques, impressive progress has been made in nuclear theory which allows our knowledge of the nuclear force to be much more deeply connected with the fundamental forces of nature.

\begin{figure*}[t!]
\begin{center}
\includegraphics[width=0.99\textwidth]{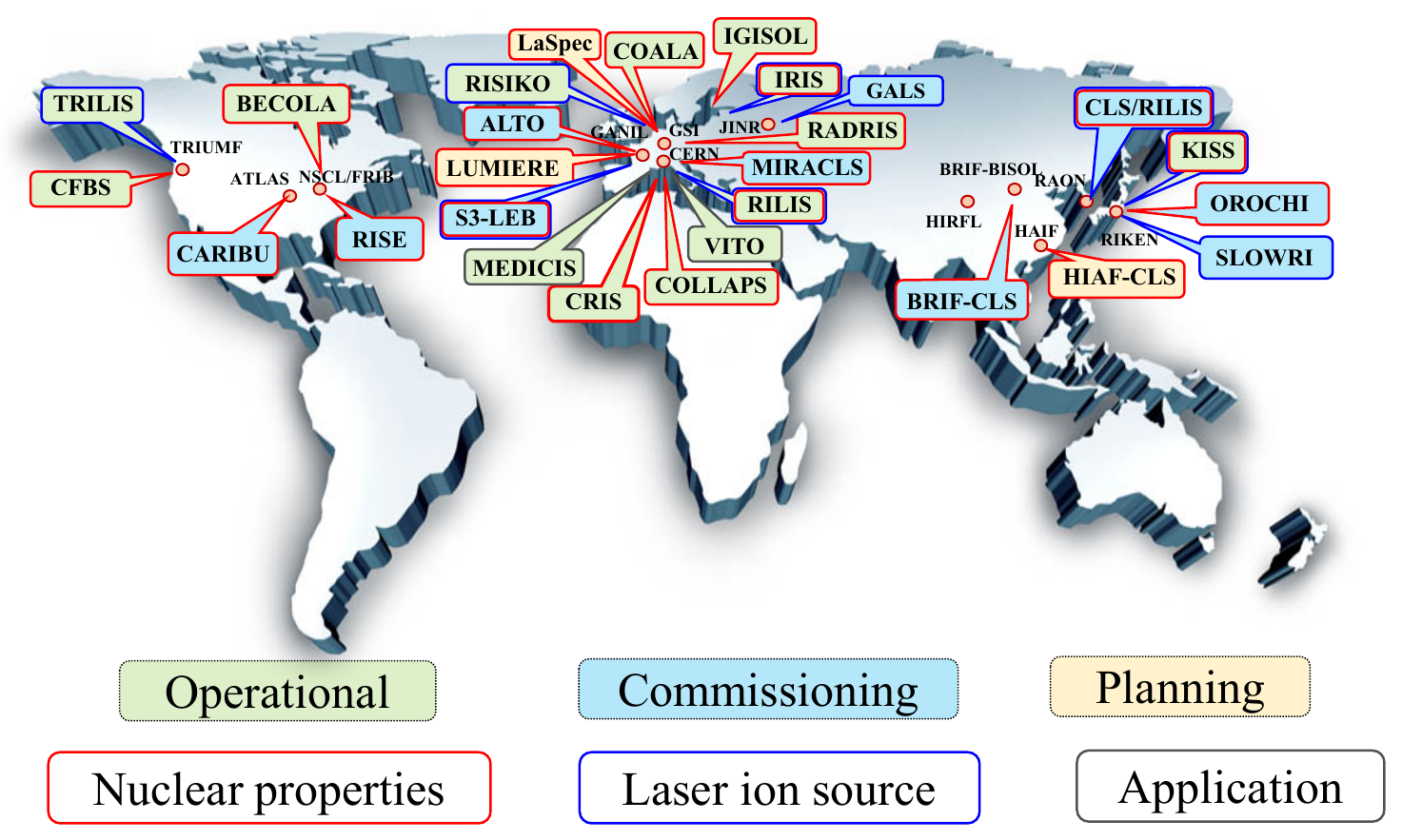}
\caption{\label{fig:fig1.1}\footnotesize{Distribution of laser spectroscopy setups at the different radioactive ion beam facilities around the world.}}
\end{center}
\end{figure*}

Laser spectroscopy techniques are a class of powerful tools that can, in some cases, simultaneously access multiple fundamental properties of atomic nuclei (spins, electromagnetic moments, charge radii) in a nuclear model-independent way. Measurement of these quantities is realized by probing the hyperfine structure and isotope shift of atomic/ionic energy levels. The application of laser spectroscopy methods in nuclear physics research has a long history with its origins involving the study of stable and long-lived radioactive isotopes found in nature. The integration of laser spectroscopy setups into RIB facilities, first initiated in the 1970s, opened up a new era of so-called \lq on-line' experiments, allowing systematic studies of nuclear properties over long isotopic chains. The first review of this endeavour was published in 1979~\cite{PR1979}. A remarkable example from these formative years of on-line laser spectroscopy is the observation of the large odd-even staggering effect and isomer shifts of neutron-deficient mercury isotopes~\cite{Hg-radii1972,Hg-moment1976,Hg-moment1977,Hg-radii1979}. This behavior provided early, compelling and intuitive experimental evidence of the phenomenon known as shape coexistence~\cite{RMP2011}, which remains a key theme in nuclear structure research today~\cite{SC-2022}.

On-line laser spectroscopy experiments at RIB facilities also stimulated the development of collinear laser spectroscopy, which was validated as an approach at TRIGA, Mainz~\cite{CLS1,CLS2}, achieving both a high resolution and sensitivity for the time. Soon after, the collinear laser spectroscopy method was applied at on-line facilities and used extensively in the study of nuclear properties of unstable isotopes~\cite{COLLAPS1,COLLAPS2,HI1985}. 
Concurrently, the highly sensitive resonance ionization spectroscopy technique was developed and also used to measure radioactive species~\cite{RIS1983}. On-line laser spectroscopy methods then entered a prolific period where they provided experimental data on a rich variety of nuclei and nuclear phenomena. Earlier reviews by Otten~\cite{Otten1989}, and then Billowes and Campbell~\cite{JPG1995} summarize much of this work. 

Later, the introduction of gas-filled radiofrequency quadrupole ion traps to produce high-quality, bunched ion beams in the 2000s dramatically enhanced the general sensitivity of standard collinear laser spectroscopy~\cite{Hf-radii2002,Zr-radii2002-1,RFQ} and enabled the potential of the newly implemented collinear resonance ionization spectroscopy to be realized~\cite{CRIS01,Kieran-phd,CRIS-NIM2013}. In parallel, developments of in-source laser spectroscopy and laser ion sources have played a major role in both the production and study of exotic nuclei~\cite{RIS2012}. These technical developments have seen the application of laser spectroscopy experiments across a wide range of fields in fundamental and applied science. Past reviews of this were given by Cheal and Flanagan~\cite{JPG2010}, and by Campbell, Moore and Pearson~\cite{PPNP2016}. 

Efforts to investigate exotic nuclei at the extremes of stability have motivated major developments in RIB production during the last decade. Upgrades to several existing facilities are underway including RIBF at RIKEN, Japan~\cite{RIKEN}, ISAC at TRIUMF, Canada~\cite{ISAC-TRIUMF2014}, ISOLDE at CERN, Switzerland~\cite{ISOLDE2017}, IGISOL at Jyv\"askyl\"a, Finland~\cite{IGISOL2014}. Entirely new-generation RIB facilities are also becoming operational or under construction, including FRIB at MSU, USA~\cite{FRIB}, FAIR at Darmstadt, Germany~\cite{FAIR}, SPIRAL2 at GANIL, France~\cite{SPIRAL2-GANIL}, RISP at RAON, South Korean~\cite{RISP-RAON}, HIAF at IMP, China~\cite{HIAF}, across the globe.
New concepts for next-generation RIB facilities have also been proposed and planned, such as EURISOL in Europe~\cite{EURISOL} and BISOL in China~\cite{BISOL}. 

RIB facilities are defined by their operating principle employing either the Isotope Separation On-Line (ISOL) technique~\cite{ISOLDE2017, ISAC-TRIUMF2014}, the in-flight Projectile Fragmentation (PF) technique~\cite{FRIB,RIKEN}, or an unprecedented combination of both~\cite{EURISOL,BISOL}. Due to their recognized potential to make meaningful contributions across multiple fields of science in addition to the study of exotic nuclei~\cite{JPG2010,PPNP2016,PPNP2021}, on-line laser spectroscopy setups are either already in operation or planned at nearly every current and upcoming RIB facility around the world and are shown in Fig.~\ref{fig:fig1.1}. 

Furthering our knowledge towards nuclei at the limits of existence poses significant experimental challenges for laser spectroscopy techniques, as these isotopes have shorter lifetimes, are produced in vanishing quantities and are often accompanied by large amounts of isobaric contamination (especially when produced at thick-target ISOL facilities). These challenges constantly drive and stimulate innovation to allow the limits of spectral resolution and sensitivity to be continuously improved upon.
Examples of well-established laser spectroscopy setups are the COLlinear LAser SPectroscopy  (COLLAPS) at ISOLDE-CERN~\cite{COLLAPS1}; collinear laser spectroscopy at IGISOL-JYFL~\cite{Y-radii2018,IGISOL-CLS}; the Collinear Fast Beam laser Spectroscopy (CFBS) at ISAC-TRIUMF~\cite{CFBS}; the Collinear Resonance Ionization Spectroscopy (CRIS) setup at ISOLDE-CERN, which combines the high resolution offered by fast-beam collinear geometry with the high sensitivity of multi-step resonance ionization process~\cite{CRIS-NIM2020}; the BEam COoler and LAser spectroscopy (BECOLA) setup at the PF-type RIB facility at NSCL-MSU~\cite{BECOLA}; the Resonance Ionization Laser Ion Source (RILIS) at ISOLDE-CERN~\cite{RILIS2017} and TRILIS at ISAC-TRIUMF~\cite{TRILIS} which could reach extremely high sensitivities; in-gas-cell laser ionization spectroscopy at KISS, RIKEN~\cite{KISS-2}; in-gas-jet laser ionization and spectroscopy (IGLIS) at KU Leuven~\cite{Ac-moment2017} enabling a high spectral resolution while maintaining high efficiency; and the RAdiation Detected Resonance Ionization Spectroscopy (RADRIS) dedicated for the study of heavy actinides at GSI~\cite{No-atom2016,No-radii2018}. 

\begin{figure*}[t!]
\begin{center}
\includegraphics[width=0.95\textwidth]{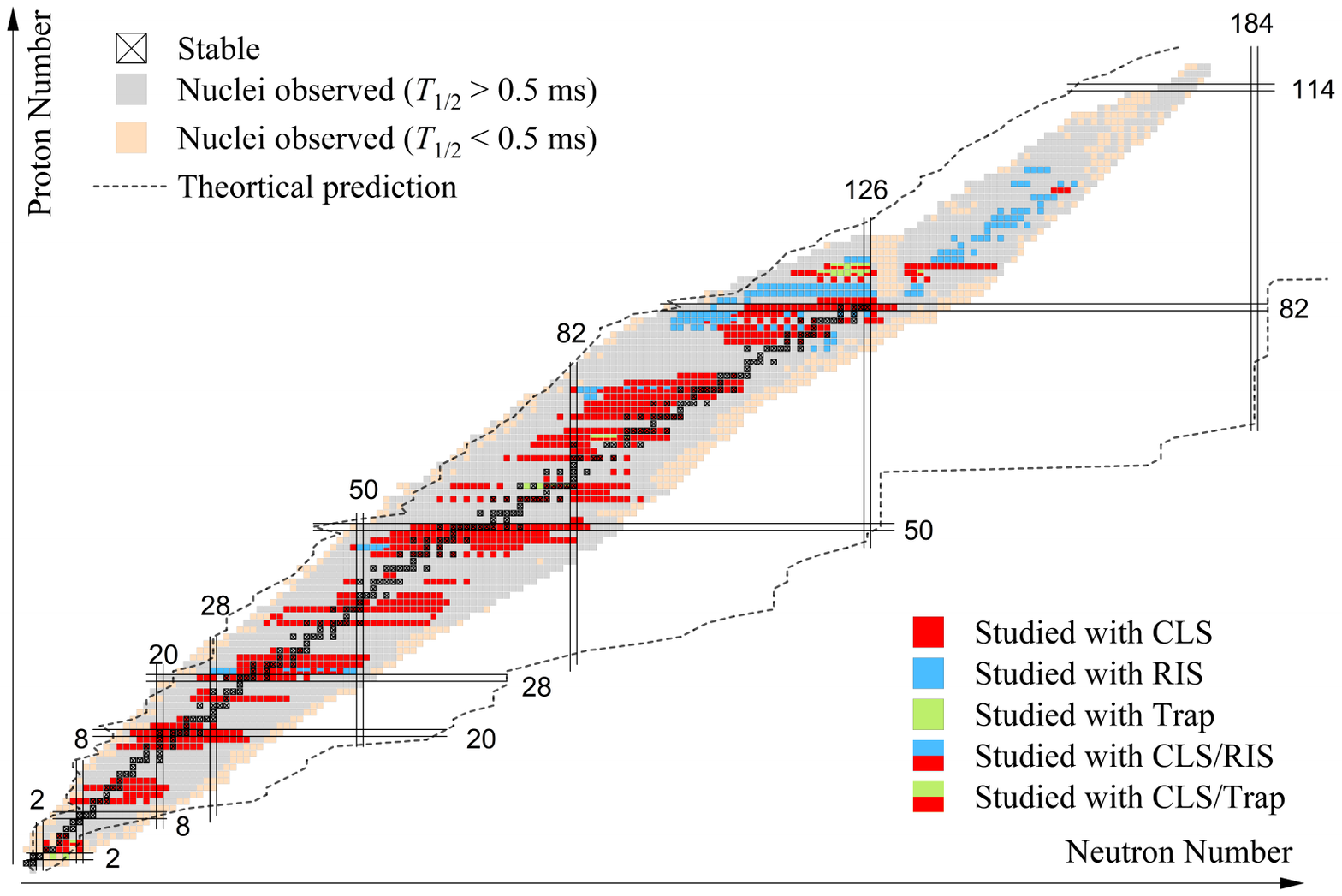}
\caption{\label{fig:fig1.2}\footnotesize{The chart of the nuclides. Stable and long-lived isotopes that exist in large quantities on Earth are indicated in black. The dashed line indicates the region within which bound nuclei are predicted to exist by nuclear theory~\cite{XIA2018}. Unstable isotopes produced at RIB facilities are shown in grey (for nuclei with $T_{1/2}>0.5$~ms) and light yellow (for nuclei with $T_{1/2}<0.5$~ms). Around a thousand ground- and long-lived isomeric states of unstable nuclei have been studied by laser spectroscopy experiments so far, which are indicated with red, blue and green squares, depending on the technique employed.}}
\end{center}
\end{figure*}

More than one thousand isotopes and isomers out of the roughly three thousand species producible at RIB facilities have been investigated so far by laser spectroscopy techniques, as is summarized in Fig.~\ref{fig:fig1.2} and in Table~\ref{tab:table1}. These measurements have long played an indispensable role in the study of nuclear structure and inter-nucleon interactions, and have led to a series of breakthroughs. Some prominent examples that deserve highlighting are the following: the nuclear halo structure in light-mass nuclei confirmed through their nuclear charge radii~\cite{He-radii2004,He-radii2007,Li-radii2006,Be-radii2009,Be-radii2012}, nuclear shell evolution and changes of magicity investigated through the nuclear spins, moments and charge radii of medium-mass nuclei~\cite{K-moment2013, Ca-radii2016,Cu-moment2009-2}; nuclear deformation and shape coexistence evidenced with the nuclear moments and radii of medium- and heavy-mass nuclei~\cite{Y-radii2007, Zn-radii2016,Hg-radii2018}. Additionally, nuclear electromagnetic properties of exotic nuclei constitute stringent tests for state-of-the-art nuclear and atomic theory~\cite{Ca-moment2015,Ca-radii2016,Ca-radii2019, Cu-radii2020,K-radii2021,Ni-radii2022,Ni-radii2022-2}. In the last few years, laser spectroscopy studies of heavy nuclei, for example those of actinide elements, have made significant progress~\cite{No-atom2016,Ac-moment2017,No-radii2018}. Such studies are not only of importance for understanding nuclear structure in the region, but also allow the role of relativistic effects, electron correlations, and quantum electrodynamics on atomic structure to be investigated, as has been reviewed recently by Block, Laatiaoui and Raeder~\cite{PPNP2021}.

Recent laser spectroscopy measurements of radium monofluoride (RaF) at ISOLDE-CERN have motivated a growing interest in the study of short-lived radioactive molecules for fundamental physics research~\cite{Gar20,Udre21,Isa21}. The structure of certain molecules can be used to drastically enhance their sensitivity to parity- and time reversal-violating nuclear properties by more than three orders of magnitude with respect to atomic systems~\cite{Chu19,cairncross:2017,Cairncross19}. Moreover, molecules containing octupole-deformed nuclei, such as radium, can provide a further enhancement of more than two orders of magnitude on top of this~\cite{flambaum2019,flambaum2020electric,Yu21}. 

The present review aims to give an overview of recent progress in the field since the last edition in 2016~\cite{PPNP2016}), in addition to future perspectives opened up by technical developments enabling a higher resolution and/or sensitivity. A focus is given on the physics impact of recent laser spectroscopy measurements of exotic nuclei across the nuclear chart. The manuscript is structured as follows. First, we present the concepts of the atomic hyperfine structure (HFS) and isotope shift and their relationship to nuclear properties in Sec.~\ref{sec:atoms}. The importance of these properties in contributing to our understanding of nuclear structure and inter-nucleon interactions is discussed in Sec.~\ref{sec:properties}. Different aspects and recent developments of laser spectroscopy techniques are presented in Sec.~\ref{sec:hfsmethod}, with Sec.~\ref{sec:cls} dedicated to collinear laser spectroscopy, and Sec.~\ref{sec:insource} to laser resonance ionization spectroscopy close to the RIB production site, Sec.~\ref{sec:trap} to laser spectroscopy of trapped ions/atoms, and Sec.~\ref{sec:RIB} to a brief introduction of RIB production at on-line facilities. A summary of recent physics results from the study of exotic nuclei in different mass regions of the nuclear chart is detailed in Sec.~\ref{physics}. An outlook on the future opportunities and challenges of laser spectroscopy measurements of more exotic nuclei, technical developments, as well as radioactive molecules will be given in Sec.~\ref{outlook}, followed by a summary of this review in Sec.~\ref{summary} 

\begin{center}
\scriptsize{
\begin{longtable}{p{0.3cm}<{\centering}p{0.3cm}<{\centering}p{6cm}<{\centering}p{1.8cm}<{\centering}p{1.8cm}<{\centering}p{2.4cm}<{\centering}}

\caption{\label{tab:table1}\footnotesize{Table of isotopes for which published laser spectroscopy measurements exist before October 2022. Full references of measurements covered by previous reviews are given. New measurements reported since the last review in this series~\cite{PPNP2016} are given in a separate column.}}\\ 
\hline
	& $Z$ &	$A$	&	Observables	&	New Refs.	&	Full Refs.\\
\hline
\endfirsthead
\hline
	& $Z$ &	$A$	&	Observables	&	New Refs.	&	Full Refs.	\\
\hline
\endhead
\hline
\multicolumn{6}{l}{Continued on next page}\\ 
\endfoot
\hline
\endlastfoot
\hline
He	&	2	&	6,8	&	$\delta \langle r^2\rangle$	&		&	\cite{He-radii2004,He-radii2007} \\
\hdashline[0.2pt/2pt]

Li	&	3	&	6-9,11	&	$I$, $\mu$, $Q_{\rm s}$,$\delta \langle r^2\rangle$ 	&		&	\cite{Li-radii2006,Li-moment1987,Li-moment1992,Li-moment1994,Li-radii2004,Li-moment2005,Li-moment2008,Li-radii2011-1,Li-radii2011-2,Li-moment2011-3,Li-moment2014}	\\
\hdashline[0.2pt/2pt]

\multirow{1}{*}{Be}	&	\multirow{1}{*}{4}	&	7-11(odd-$A$)	&	$\mu$,$\delta \langle r^2\rangle$	&	\cite{Be-radii2017}	&	\cite{Be-radii2009,Be-moment1999,Be-moment2008,Be-moment2014}	\\
	&		&	10,12	&	$\delta \langle r^2\rangle$	&	\cite{Be-radii2017}	&	\cite{Be-radii2009,Be-radii2012,Be-radii2017}	\\
\hdashline[0.2pt/2pt]

B	&	5	&	10,11	&	$\delta \langle r^2\rangle$	&	\cite{B-radii2019}	&	\cite{B-radii2019} 	\\
\hdashline[0.2pt/2pt]

Ne	&	6	&	17-26,28	&	$I$, $\mu$, $Q_{\rm s}$,$\delta \langle r^2\rangle$	&		&	\cite{Ne-radii2000,Ne-moment2005,Ne-radii2008,Ne-radii2011}	\\
\hdashline[0.2pt/2pt]

\multirow{1}{*}{Na}	&	\multirow{1}{*}{11}	&	20	&	$\mu$	&		&	\cite{Na-k-moment1975} 	\\
	&		&	21-31	&	$I$, $\mu$, $Q_{\rm s}$,$\delta \langle r^2\rangle$	&	\cite{Na-moment2020}	&	\cite{Na-moment1978,Na-moment1999,Na-moment2000,Na-moment2020} 	\\
\hdashline[0.2pt/2pt]

\multirow{1}{*}{Mg}	&	\multirow{1}{*}{12}	&	21,23,27,29,31	&	$I$, $\mu$, $\delta \langle r^2\rangle$	&	\cite{Mg-moment2017}	&	\cite{Mg-moment2005,Mg-moment2009,Mg-moment2008,Mg-moment2017,Mg-moment2012}	\\
	&		&	22,24-26,28-30,33	&	$I$, $\mu$	&		&	\cite{Mg-moment2012,Mg-moment2007}	\\
	&		&	32	&	$\delta \langle r^2\rangle$	&		&	\cite{Mg-moment2012}	\\
\hdashline[0.2pt/2pt]

\multirow{1}{*}{Al}	&	\multirow{1}{*}{13}	&	26	&	$\mu$	&		&	\cite{Al-moment1996,Al-quadrupole1997}	\\
	&		&	27-32	&	$\mu$, $Q_{\rm s}$,$\delta \langle r^2\rangle$ 	&\cite{Al-radii2021}	&	\cite{Al-radii2021}	\\
\hdashline[0.2pt/2pt]

Ar	&	18	&	32-39,40-44,46	&	$I$,$\mu$, $Q_{\rm s}$,$\delta \langle r^2\rangle$ 	&		&	\cite{Ar-radii1996,Ar-radii2008}	\\
\hdashline[0.2pt/2pt]

\multirow{1}{*}{K}	&	\multirow{1}{*}{19}	&	36-47	&	$I$, $\mu$, $Q_{\rm s}$,$\delta \langle r^2\rangle$	&	\cite{K-radii2015,K-radii2019}	&	\cite{Na-k-moment1975,K-radii1997,K-moment1969,K-radii1982,K-radii2014-1,K-moment2014,K-radii2015,K-radii2019}	\\
	&		&	49,51	&	$I$, $\mu$, $\delta \langle r^2\rangle$	&		&	\cite{K-moment2013}	\\
	&		&	$\rm 38^{m}$,52	&	$\delta \langle r^2\rangle$	&	\cite{K-radii2021}	&	\cite{K-radii2014-2,K-radii2021}	\\
\hdashline[0.2pt/2pt]

\multirow{1}{*}{Ca}	&	\multirow{1}{*}{20}	&	36,38,40,50,52	&	$\delta \langle r^2\rangle$	&	\cite{Ca-radii2016,Ca-radii2019}	&	\cite{Ca-radii1992-1,Ca-radii1992-2,Ca-radii2016,Ca-radii2019}	\\
	&		&	37,39,41-49,51	&	$I$, $\mu$, $Q_{\rm s}$,$\delta \langle r^2\rangle$	&	\cite{Ca-moment2015,Ca-radii2016,Ca-moment2019,Ca-radii2019}	&	\cite{Ca-radii1980,Ca-radii1982,Ca-moment2015,Ca-radii2016,Ca-radii1996,Ca-moment2019,Ca-radii2019}	\\
\hdashline[0.2pt/2pt]

\multirow{1}{*}{Sc}	&	\multirow{1}{*}{21}	&	42-46, $\rm 44^{m}$,$\rm 45^{m}$	&	$I$,$\mu$, $Q_{\rm s}$,$\delta \langle r^2\rangle$	& \cite{Sc-radii2021}		&	\cite{Sc-radii2011,Sc-radii2021}	\\
	&		&	41,47,49	&	$\mu$, $Q_{\rm s}$	& \cite{Sc-moment2022-40Sc,Sc-moments2022}		&	\cite{Sc-moment2022-40Sc,Sc-moments2022}	\\
\hdashline[0.2pt/2pt]

Ti	&	22	&	44,45	&	$\delta \langle r^2\rangle$	&		&	\cite{Ti-radii2002,Ti-radii2004}	\\
\hdashline[0.2pt/2pt]

Cr	&	24	&	50,52-54	&	$\delta \langle r^2\rangle$	&		&	\cite{radii-data}	\\
\hdashline[0.2pt/2pt]

\multirow{1}{*}{Mn}	&	\multirow{1}{*}{25}	&	51-57,59,61,63,$\rm 50^{m}$,$\rm 52^{m}$	&	$I$, $\mu$, $Q_{\rm s}$,$\delta \langle r^2\rangle$ 	&	\cite{Mn-moment2015,Mn-radii2016,Mn-moment2016}	&	\cite{Mn-moment2010,Mn-moment2015,Mn-radii2016,Mn-moment2016}	\\
	&		&	58-64,58-62$^{\rm m}$(even-$A$)	&	$I$, $\mu$, $\delta \langle r^2\rangle$ 	&	\cite{Mn-moment2-2015,Mn-radii2016}	&	\cite{Mn-moment2-2015,Mn-radii2016}	\\
\hdashline[0.2pt/2pt]

\multirow{1}{*}{Fe}	&	\multirow{1}{*}{26}	&	52	&	$\delta \langle r^2\rangle$	&	\cite{Fe-radii2016}	&	\cite{Fe-radii2016}	\\
	&		&	53	&	$\mu$, $Q_{\rm s}$,$\delta \langle r^2\rangle$	&	\cite{Fe-radii2016,Fe-moment2017}	&	\cite{Fe-radii2016,Fe-moment2017}	\\
\hdashline[0.2pt/2pt] 

Ni	&	28	&	54-56, 58-68,70	&	$\delta \langle r^2\rangle$	&	\cite{Ni-radii2020,Ni-radii2021,Ni-radii2022,Ni-radii2022-2}	&	\cite{Ni-radii2020,Ni-radii2021,Ni-radii2022,Ni-radii2022-2}	\\
\hdashline[0.2pt/2pt]

\multirow{1}{*}{Cu}	&	\multirow{1}{*}{29}	&	58-78,$\rm 68^{m}$,$\rm 70^{m1,m2}$	&	$I$, $\mu$, $Q_{\rm s}$,$\delta \langle r^2\rangle$ 	&	\cite{Cu-radii2016,Cu-moment2017,Cu-radii2020}	&	\cite{Cu-moment2009-1,Cu-moment2008,Cu-moment2009-2,Cu-moment2010-1,Cu-moment2011-1,Cu-moment2011-2,Cu-radii2016,Cu-moment2017,Cu-radii2020,Cu-moment2002}	\\
	&		&	57	&	$Q_{\rm s}$,$\delta \langle r^2\rangle$	&		&	\cite{Cu-moment2009-1,Cu-moment2010-2}	\\
\hdashline[0.2pt/2pt]

\multirow{1}{*}{Zn}	&	\multirow{1}{*}{30}	&	63-79,71-79$^{\rm m}$(odd-$A$)	&	$I$, $\mu$, $Q_{\rm s}$,$\delta \langle r^2\rangle$ 	&	\cite{Zn-radii2016,Zn-moment2017,Zn-spin2018}	&	\cite{Zn-radii2016,Zn-moment2017,Zn-spin2018}	\\
	&		&	62-80	&	$\delta \langle r^2\rangle$	&	\cite{Zn-radii2019}	&	\cite{Zn-radii2019}	\\
\hdashline[0.2pt/2pt]

Ga	&	31	&	63-82,$\rm 80^{m}$	&	$I$, $\mu$, $Q_{\rm s}$,$\delta \langle r^2\rangle$ 	&	\cite{Ga-radii2017}	&	\cite{Ga-moment2010-1,Ga-moment2010-2,Ga-moment2011,Ga-radii2012,Ga-moment2012,Ga-radii2017}	\\
\hdashline[0.2pt/2pt]

Ge	&	32	&	69,71,73	&	$\mu$, $Q_{\rm s}$	&	\cite{Ge-moment2020}	&	\cite{Ge-moment2020}	\\
\hdashline[0.2pt/2pt]

Kr	&	36	&	72,74-96,$\rm 79^{m}$,$\rm 81^{m}$,$\rm 83^{m}$,$\rm 85^{m}$	&	$I$, $\mu$, $Q_{\rm s}$,$\delta \langle r^2\rangle$ 	&		&	\cite{Kr-radii1990,Kr-moment1995,Kr-radii1996}	\\
\hdashline[0.2pt/2pt]

Rb	&	37	&	76-98,$\rm 78^{m}$,$\rm 81^{m}$,$\rm 82^{m}$,$\rm 84^{m}$,$\rm 86^{m}$,$\rm 90^{m}$,$\rm 98^{m1,m2}$	&	$I$, $\mu$, $Q_{\rm s}$,$\delta \langle r^2\rangle$ 	&		&	\cite{Rb-radii1981,Rb-radii2011,Rb-radii2015}	\\
\hdashline[0.2pt/2pt]

Sr	&	38	&	77-100,$\rm 83^{m}$,$\rm 85^{m}$,$\rm 87^{m}$	&	$I$, $\mu$, $Q_{\rm s}$,$\delta \langle r^2\rangle$ 	&		&	\cite{Sr-radii1985,Sr-radii1986,Sr-radii1987-1,Sr-radii1987-2,Sr-radii1988,Sr-moment1987-1,Sr-moment1987-2,Sr-radii1990,Sr-radii1991,Sr-radii1992,Kr-radii1996}	\\
\hdashline[0.2pt/2pt]

\multirow{1}{*}{Y}	&	\multirow{1}{*}{39}	&	86-90,92-102,$\rm 87^{m}$-$\rm 90^{m}$,$\rm 93^{m}$,$\rm 96^{m}$,$\rm 97^{m1,m2}$,$\rm 98^{m}$	&	$\mu$, $Q_{\rm s}$,$\delta \langle r^2\rangle$ 	& \cite{Y-radii2018} &	\cite{Y-radii2007,Y-radii2018}	\\
	&		&	$\rm 100^{m}$	&	$I$, $\mu$, $Q_{\rm s}$	&		&	\cite{Y-moment2010}	\\
\hdashline[0.2pt/2pt]

Zr	&	40	&	87-89,96-102,$\rm 87^{m}$,$\rm 89^{m}$	&	$I$, $\mu$, $Q_{\rm s}$,$\delta \langle r^2\rangle$ 	&		&	\cite{Ti-radii2002,Zr-radii2002-2,Zr-radii2002-1}	\\
\hdashline[0.2pt/2pt]

Nb	&	41	&	90-93,99,101,103,$\rm 90^{m}$,$\rm 91^{m}$	&	$\delta \langle r^2\rangle$	&		&	\cite{Nb-radii2009}	\\
\hdashline[0.2pt/2pt]
Mo	&	42	&	90,91,102-106,108	&	$\mu$, $\delta \langle r^2\rangle$ 	&		&	\cite{Mo-radii2009}	\\
\hdashline[0.2pt/2pt]

Tc	&	43	&	97-99	&	$I$, $\mu$, $Q_{\rm s}$,$\delta \langle r^2\rangle$ 	&	\cite{Tc-radii2020}	&	\cite{Tc-radii2020}	\\
\hdashline[0.2pt/2pt]

Ru	&	44	&	96,98-102,104	&	$\delta \langle r^2\rangle$	&		&	\cite{Ru-radii2014}	\\
\hdashline[0.2pt/2pt]

Pd	&	46	&	98-102, 104-106, 108, 110, 112, 114, 116, 118 &	$\delta \langle r^2\rangle$	&	\cite{Pd-radii2022}	&	\cite{Pd-radii2022}	\\
\hdashline[0.2pt/2pt]

\multirow{1}{*}{Ag}	&	\multirow{1}{*}{47}	&	101,103-105,107,109,$\rm 105^{m}$,$\rm 106^{m}$	&	$I$, $\mu$, $Q_{\rm s}$,$\delta \langle r^2\rangle$ 	&		&	\cite{Ag-radii1989}	\\
	&		&	97-100	&	$I$, $\mu$, $\delta \langle r^2\rangle$ 	&		&	\cite{Ag-radii2014}	\\
	&		&	96-104,107,109,114-121	&	 $\delta \langle r^2\rangle$ 	&	\cite{Ag-radii2021}	&	\cite{Ag-radii2021}	\\
\hdashline[0.2pt/2pt]

\multirow{1}{*}{Cd}	&	\multirow{1}{*}{48}	&	101-121,123-129 (odd-$A$),111-129$^{\rm m}$ (odd-$A$)	&	$I$, $\mu$, $Q_{\rm s}$,$\delta \langle r^2\rangle$ 	&	\cite{Cd-moment2015,Cd-radii2016,Cd-moment2018,Cd-radii2018}	&	\cite{Cd-radii1987,Cd-moment2013,Cd-moment2015,Cd-radii2016,Cd-moment2018,Cd-radii2018}	\\
	&		&	100,122-130(even-$A$),$\rm 112^{m}$-$\rm 128^{m}$(even-A)	&	$\delta \langle r^2\rangle$	&	\cite{Cd-radii2016,Cd-radii2018}	&	\cite{Cd-radii2016,Cd-radii2018}	\\
\hdashline[0.2pt/2pt]

\multirow{1}{*}{In}	&	\multirow{1}{*}{49}	&	104-109,111,112,116-127,$\rm 108^{m}$,$\rm 110^{m}$,$\rm 115^{m}$,$\rm 126^{m}$	&	$I$, $\mu$, $Q_{\rm s}$,$\delta \langle r^2\rangle$ 	&	\cite{In-moment2018,In-radii2020}	&	\cite{In-radii1985-1,In-radii1987,In-radii2020,In-radii1985-2,In-radii1986,In-moment2018,In-radii2020}	\\
	&		&	105-131,$\rm 113^{m}$-$\rm 131^{m}$ (odd-$A$)	&	$I$, $\mu$, $Q_{\rm s}$ 	&	\cite{In-moment2022}	&	\cite{In-moment2022}	\\
\hdashline[0.2pt/2pt]

\multirow{1}{*}{Sn}	&	\multirow{1}{*}{50}	&	108-132,$\rm 117^{m}$-$\rm 131^{m}$(odd-$A$),$\rm 130^{m}$	&	$\mu$, $Q_{\rm s}$,$\delta \langle r^2\rangle$ 	&\cite{Sn-radii2019,Sn-radii2020}		&	\cite{Sn-radii1986,Sn-radii1987,Sn-radii2002,Sn-radii2005,Sn-radii2019,Sn-radii2020}	\\
	&		&	108-134(even-$A$)	&	$\delta \langle r^2\rangle$	&	\cite{Sn-radii2019}	&	\cite{Sn-radii2019}	\\
	&		&	133	&	$\mu$, $Q_{\rm s}$	&	\cite{Sn-moment2020}	&	\cite{Sn-moment2020}	\\
\hdashline[0.2pt/2pt]

Sb	&	51	&	121,123,133,134	&	$\mu$, $Q_{\rm s}$,$\delta \langle r^2\rangle$ 	&	\cite{Sb-moment2021}	&	\cite{Sb-moment2021}	\\
\hdashline[0.2pt/2pt]

\multirow{1}{*}{Te}	&	\multirow{1}{*}{52}	&	133,135,$\rm 125^{m}$-$\rm 133^{m}$(odd-$A$)	&	$I$, $\mu$, $Q_{\rm s}$	&		&	\cite{Te-moment2006}	\\
	&		&	132,134,136	&	$\delta \langle r^2\rangle$	&		&	\cite{Te-radii2007}	\\
\hdashline[0.2pt/2pt]

Xe	&	54	&	116-146 (even-$A$),129,131,137-143 (odd-$A$)	&	$I$, $\mu$, $Q_{\rm s}$,$\delta \langle r^2\rangle$ 	&		&	\cite{Xe-radii1989}	\\
\hdashline[0.2pt/2pt]

\multirow{1}{*}{Cs}	&	\multirow{1}{*}{55}	&	118-146,$\rm 119^{m}$,$\rm 121^{m}$,$\rm 122^{m}$,$\rm 130^{m}$,$\rm 134^{m}$-$\rm 136^{m}$,$\rm 138^{m}$	&	$I$, $\mu$, $Q_{\rm s}$,$\delta \langle r^2\rangle$ 	&		&	\cite{CLS2,Cs-radii1978-2,CLS2-2,Cs-radii1981,Cs-radii1987,Cs-moment2005}	\\
\hdashline[0.2pt/2pt]

\multirow{1}{*}{Ba}	&	\multirow{1}{*}{56}	&	120-123,125-146,148,$\rm 127^{m}$-$\rm 137^{m}$(odd-A),$\rm 130^{m}$	&	$I$, $\mu$, $Q_{\rm s}$,$\delta \langle r^2\rangle$ 	&		&	\cite{COLLAPS2,Ba-radii1988,Ba-radii1992,Ba-radii1979,Ba-radii1987,Ba-radii2002,Otten1989}		\\
\hdashline[0.2pt/2pt]

La	&	57	&	135,137	&	$\mu$, $Q_{\rm s}$,$\delta \langle r^2\rangle$ 	&		&	\cite{La-radii2003}	\\
\hdashline[0.2pt/2pt]

Ce	&	58	&	146,148	&	$\delta \langle r^2\rangle$	&		&	\cite{Ce-radii2003}	\\
\hdashline[0.2pt/2pt]

Pr	&	59	&	135-137,141	&	$\mu$	& 	\cite{Pr-moment2019}	&	\cite{Pr-moment2019}	\\
\hdashline[0.2pt/2pt]

Nd	&	60	&	132,134-146,148,150	&	$I$, $\mu$, $Q_{\rm s}$,$\delta \langle r^2\rangle$ 	&		&	\cite{Nd-radii1992,Otten1989}	\\
\hdashline[0.2pt/2pt]

Pm	&	61	&	143-146	&	$\mu$, $Q_{\rm s}$,$\delta \langle r^2\rangle$ 	&	\cite{Pm-radii2020}	&	\cite{Pm-radii2020}	\\
\hdashline[0.2pt/2pt]

Sm	&	62	&	138-146,151,153,$\rm 141^{m}$	&	$I$, $\mu$, $Q_{\rm s}$,$\delta \langle r^2\rangle$ 	&		&	\cite{Nd-radii1992,Otten1989}	\\
\hdashline[0.2pt/2pt]

\multirow{1}{*}{Eu}	&	\multirow{1}{*}{63}	&	140-153,155-159,$\rm 142^{m}$,$\rm 150^{m}$,$\rm 152^{m}$	&	$\mu$, $Q_{\rm s}$,$\delta \langle r^2\rangle$ 	&		&	\cite{Eu-radii1983-1,Eu-radii1983-2,Eu-radii1984-1,Eu-radii1984-2,Eu-radii1985,Eu-radii1990,Nd-radii1992,Eu-moment1997}	\\
	&		&	154	&	$Q_{\rm s}$,$\delta \langle r^2\rangle$ 	&		&	\cite{Eu-radii1983-2}	\\
	&		&	138,139	&	$I$, $\mu$, $\delta \langle r^2\rangle$ 	&		&	\cite{Eu-radii1983-2,Nd-radii1992,Otten1989}	\\
\hdashline[0.2pt/2pt]

\multirow{1}{*}{Gd}	&	\multirow{1}{*}{64}	&	146-160,$\rm 143^{m}$	&	$\delta \langle r^2\rangle$	&		&	\cite{Gd-radii1988}	\\
	&		&	145,$\rm 145^{m}$	&	$\mu$, $Q_{\rm s}$,$\delta \langle r^2\rangle$ 	&		&	\cite{Gd-moment2005}	\\
\hdashline[0.2pt/2pt]

Tb	&	65	&	147-155,157,159	&	$\mu$, $Q_{\rm s}$,$\delta \langle r^2\rangle$ 	&		&	\cite{Tb-radii1990}	\\
\hdashline[0.2pt/2pt]

\multirow{1}{*}{Dy}	&	\multirow{1}{*}{66}	&	149-159 (odd-$A$)	&	$I$, $\mu$, $Q_{\rm s}$,$\delta \langle r^2\rangle$ 	&		&	\cite{Otten1989}	\\
	&		&	146-164 (even-$A$)	&	$\delta \langle r^2\rangle$	&		&	\cite{Otten1989}	\\
\hdashline[0.2pt/2pt]

Ho	&	67	&	151-163,$\rm 151^{m}$-$\rm 154^{m}$-$\rm 162^{m}$(even-A)	&	$I$, $\mu$, $Q_{\rm s}$,$\delta \langle r^2\rangle$ 	&		&	\cite{Ho-radii1989,Otten1989}	\\
\hdashline[0.2pt/2pt]

\multirow{1}{*}{Er}	&	\multirow{1}{*}{68}	&	153-167 (odd-$A$)	&	$I$, $\mu$, $Q_{\rm s}$	&	&	\cite{Otten1989}	\\
	&		&	150-164(even-$A$)	&	$\delta \langle r^2\rangle$	&		&	\cite{Otten1989}	\\
\hdashline[0.2pt/2pt]

Tm	&	69	&	153,154,156-172,$\rm 154^{m}$	&	$I$, $\mu$, $Q_{\rm s}$,$\delta \langle r^2\rangle$ 	&		&	\cite{Tm-radii1988,Tm-radii2000,Otten1989}	\\
\hdashline[0.2pt/2pt]

\multirow{1}{*}{Yb}	&	\multirow{1}{*}{70}	&	153-177,$\rm 176^{m}$,$\rm 177^{m}$	&	$I$, $\mu$, $Q_{\rm s}$,$\delta \langle r^2\rangle$ 	&  &	\cite{Otten1989,Yb-radii1983,Yb-radii1989,CRIS01,Yb-radii1992,Tm-radii2000,Yb-radii2002,Yb-radii2007,Yb-radii2012}	\\
	&		&	152	&	$\delta \langle r^2\rangle$ 	&		&	\cite{Yb-radii1989}	\\
\hdashline[0.2pt/2pt]

Lu	&	71	&	161-179,$\rm 166^{m1,m2}$,$\rm 167^{m}$-$\rm 169^{m}$,$\rm 171^{m}$,$\rm 172^{m}$,$\rm 174^{m}$,$\rm 176^{m}$,$\rm 177^{m}$,$\rm 178^{m}$	&	$\mu$, $Q_{\rm s}$,$\delta \langle r^2\rangle$ 	&		&	\cite{Yb-radii1989}	\\
\hdashline[0.2pt/2pt]

\multirow{1}{*}{Hf}	&	\multirow{1}{*}{72}	&	170,172-174	&	$\delta \langle r^2\rangle$	&		&	\cite{Hf-radii1999}	\\
	&		&	171,175,$\rm 171^{m}$,$\rm 178^{m1}$	&	$I$,$\mu$, $Q_{\rm s}$,$\delta \langle r^2\rangle$ 	&		&	\cite{Hf-radii2000,Hf-radii2002,Yb-radii2007}	\\
	&		&	$\rm 178^{m2}$	&	$\mu$, $Q_{\rm s}$	&		&	\cite{Hf-moment1994}	\\
\hdashline[0.2pt/2pt]

\multirow{1}{*}{Ta}	&	73	&	179	&	$I$, $\mu$, $Q_{\rm s}$	&		&	\cite{Ta-moment1996}	\\
	&		& $\rm 180^{m}$, 181	&	$I$, $\mu$&		&	\cite{Ta-moment2006-181Ta}	\\
\hdashline[0.2pt/2pt]

W	&	74	&	180,182-184,186	&	IS and HFS	&		&		\cite{W-A-IS-2013}\\
\hdashline[0.2pt/2pt]

Re	&	75	&	185,187	& IS and HFS		&		&		\cite{Re-A-IS-1982,Re-A-IS-1986}\\
\hdashline[0.2pt/2pt]

Os	&	76	&	194,196	&	$\delta \langle r^2\rangle$	& \cite{Os-radii2020} &	\cite{Os-radii2020}	\\
\hdashline[0.2pt/2pt]

\multirow{1}{*}{Ir}	&	\multirow{1}{*}{77}	&	182-189,191,193,$\rm 186^{m}$	&	$\mu$, $Q_{\rm s}$,$\delta \langle r^2\rangle$ 	&		&	\cite{Ir-radii2000}	\\
	&		&	196-198	&	$\mu$, $\delta \langle r^2\rangle$ 	&	\cite{Ir-radii2020}	&	\cite{Ir-radii2020}	\\
\hdashline[0.2pt/2pt]

Pt	&	78	&	183-196,198,199,$\rm 183^{m}$,$\rm 185^{m}$,$\rm 199^{m}$	&	$\mu$, $Q_{\rm s}$,$\delta \langle r^2\rangle$ 	& \cite{Pt-radii2017}	&	\cite{Pt-radii1988,Pt-radii1989,Pt-radii1992,Pt-radii1999,Pt-radii2000,Pt-radii2017}	\\
\hdashline[0.2pt/2pt]

\multirow{1}{*}{Au}	&	\multirow{1}{*}{79}	&	183-197,$\rm 184^{m}$	&	$\mu$, $Q_{\rm s}$,$\delta \langle r^2\rangle$ 	&		&	\cite{Au-radii1983,Au-radii1985,Au-moment1987,Au-radii1987,Au-radii1990,Au-radii1991,Au-radii1994,Au-radii1997}	\\
	&		&	180,182,198,199,$\rm 187^{m}$,$\rm 189^{m}$	&	$\mu$,$\delta \langle r^2\rangle$ 	&\cite{Au-radii2020-1,Au-radii2020-2}&	\cite{Au-radii1990,Au-radii2020-1,Au-radii2020-2}	\\
	&		&	176,177,179	&	$\mu$ 	&\cite{Au-moment2021,Au-moment2020,Au-moment2018}&	\cite{Au-moment2021,Au-moment2020,Au-moment2018}	\\	
\hdashline[0.2pt/2pt]

\multirow{1}{*}{Hg}	&	\multirow{1}{*}{80}	&	177-207 ($\rm 185-199^{m}$) (odd-$A$)	&	$I$, $\mu$, $Q_{\rm s}$,$\delta\langle r^2\rangle$ 	&	\cite{Hg-radii2018,Hg-radii2019,Hg-radii2021_PRL,Hg-radii2021-PRC}	&	\cite{Otten1989,Hg-radii2018,Hg-radii2019,Hg-radii2021_PRL,Hg-radii2021-PRC}	\\
	&		&	178-208(even-$A$)	&	$\delta\langle r^2\rangle$ 	&\cite{Hg-radii2018,Hg-radii2019,Hg-radii2021_PRL,Hg-radii2021-PRC}		&	\cite{Otten1989,Hg-radii2018,Hg-radii2019,Hg-radii2021_PRL,Hg-radii2021-PRC}	\\
\hdashline[0.2pt/2pt]

\multirow{1}{*}{Tl}	&	\multirow{1}{*}{81}	&	179-185,203,205,207,208,$\rm 182^{m}$-$\rm 187^{m}$,$\rm 195^{m1,m2,m3}$,$\rm 197^{m1,m2,m3}$	&	$I$, $\mu$,$\delta \langle r^2\rangle$ 	&	\cite{Tl-radii2017}	&	\cite{Tl-radii2013,Tl-radii2017,Tl-moment1985-1,Tl-radii1992-2}	\\
	&		&	190-194,196,$\rm 188^{m}$-$\rm 194^{m}$,$\rm 196^{m}$	&	$\mu$, $Q_{\rm s}$,$\delta \langle r^2\rangle$ 	&		&	\cite{Tl-moment1985-2,Tl-radii1987,Tl-radii1992-1}	\\
	&		&	187,188	&	$\mu$, $Q_{\rm s}$	&		&	\cite{Tl-moment1992}	\\
\hdashline[0.2pt/2pt]

\multirow{1}{*}{Pb}	&	\multirow{1}{*}{82}	&	182-194,196-212,214,$\rm 195^{m}$,$\rm 197^{m}$,$\rm 202^{m2}$	&	$\mu$, $Q_{\rm s}$,$\delta \langle r^2\rangle$ 	&		&	\cite{Pb-radii1983,Pb-radii1986,Pb-radii1987,Pb-radii1991,Pb-radii2007,Pb-radii2009}	\\
	&		&	$\rm 185^{m1,m2}$	&	$I$, $\mu$	&		&	\cite{Pb-moment2002}	\\
\hdashline[0.2pt/2pt]

\multirow{1}{*}{Bi}	&	\multirow{1}{*}{83}	&	187,$\rm 188^{g,m}$,189,191,202-213,$\rm 210^{m}$	&	$\mu$, $Q_{\rm s}$,$\delta \langle r^2\rangle$ 	&\cite{Bi-radii2017,Bi-radii2018,Bi-radii2018-2,Bi-radii2021}		&	\cite{Bi-radii1996,Bi-radii1997,Bi-radii2000,Bi-radii2017,Bi-radii2018,Bi-radii2018-2,Bi-radii2021} \\
	&		&	192$\rm ^{g,m}$-195$\rm ^{g,m}$,$\rm 197^{g,m}$,$\rm 198^{m}$	&	$\mu$,$\delta \langle r^2\rangle$	& \cite{Bi-radii2016,Bi-radii2017}&	\cite{Bi-radii2016,Bi-radii2017}	\\
\hdashline[0.2pt/2pt]

\multirow{1}{*}{Po}	&	\multirow{1}{*}{84}	&	200,202,204-211,216-218,$\rm 217^{m}$	&	$\mu$, $Q_{\rm s}$,$\delta \langle r^2\rangle$ 	&		& \cite{Po-radii2015}	\cite{Po-radii1991,Po-radii2011,Po-radii2015}	\\
	&		&	191-199,201,203,$\rm 197^{m}$,$\rm 199^{m}$,$\rm 201^{m}$,$\rm 203^{m}$	&	$\delta \langle r^2\rangle$ 	&		&	\cite{Po-radii2011,Po-radii2013}	\\
\hdashline[0.2pt/2pt]

At	&	85	&	195-211,217-219,$\rm 195^{m}$,$\rm 197^{m}$-$\rm 199^{m}$,$\rm 200^{m1,m2}$,$\rm 202^{m}$	&	$\mu$, $Q_{\rm s}$,$\delta \langle r^2\rangle$ 	&\cite{Bi-radii2016,Bi-radii2017}&	\cite{At-radii2018,At-radii2019}	\\
\hdashline[0.2pt/2pt]

Rn	&	86	&	202,204-212,218-223,225,$\rm 203^{m}$	&	$I$, $\mu$, $Q_{\rm s}$,$\delta \langle r^2\rangle$ 	&		&	\cite{Rn-moment1988,Rn-moment1987}	\\
\hdashline[0.2pt/2pt]

\multirow{1}{*}{Fr}	&	\multirow{1}{*}{87}	&	203-213,219-228,$\rm 204^{m1,m2}$,$\rm 206^{m1,m2}$	&	$I$, $\mu$, $Q_{\rm s}$,$\delta \langle r^2\rangle$ 	& \cite{Fr-moment2015,Fr-moment2017,Fr-radii2015-1,Fr-radii2016-1}		&	\cite{Fr-radii1987-2,Fr-moment2014,Fr-radii2014-2,Fr-moment2015,Fr-radii1985,Cs-radii1987,Fr-moment2017}	\\
	&		&	202,$\rm 202^{m}$,$\rm 218^{m}$,229,231	&	$I$, $\mu$, $\delta \langle r^2\rangle$ 	& &	\cite{Fr-radii2013,Fr-radii2013-cris, Fr-radii2014-1,Fr-moment2014,Fr-radii2015-2}	\\
	&		&	214	&	$I$, $\delta \langle r^2\rangle$ 	&\cite{Fr-radii2016-2}	&	\cite{Fr-radii2016-2}	\\
\hdashline[0.2pt/2pt]

Ra	&	88	&	208-214,220-230,232,233	&	$I$, $\mu$, $Q_{\rm s}$,$\delta \langle r^2\rangle$ 	&	\cite{Ra-radii2018}	&	\cite{Ra-moment1983,Ra-moment1987,Ra-radii1988-1,Ra-radii1988-2,Ra-radii2011,Ra-radii2018}	\\
\hdashline[0.2pt/2pt]

\multirow{1}{*}{Ac}	&	\multirow{1}{*}{89}	&	212-215,227	&	$I$, $\mu$, $Q_{\rm s}$,$\delta \langle r^2\rangle$ 	&	\cite{Ac-moment2017,Ac-moment2017-2}	&	\cite{Ac-moment2017,Ac-moment2017-2}	\\
	&		&	225,226,228,229	&	$\mu$, $\delta \langle r^2\rangle$ 	&	\cite{Ac-radii2019}	&	\cite{Ac-radii2019}	\\
\hdashline[0.2pt/2pt]

\multirow{1}{*}{Th}	&	\multirow{1}{*}{90}	&	227,228,230,232	&	$\delta \langle r^2\rangle$ 	&		&	\cite{Th-radii2012}	\\
	&		&	229, $\rm 229^{m}$	&	$\mu$, $Q_{\rm s}$,$\delta \langle r^2\rangle$ 	&	\cite{229mTh-moment2018,229mTh-radii2018}	&	\cite{Th-moment2011-1,Th-moment2011-2,Th-radii2012,229mTh-moment2018,229mTh-radii2018}	\\
\hdashline[0.2pt/2pt]

Pa	&	91	&	231	&		&\cite{Pa-levels2018}		&	\cite{Pa-levels2018}	\\
\hdashline[0.2pt/2pt]

\multirow{1}{*}{U}	&	\multirow{1}{*}{92}	&	234,236,238	&	$\delta \langle r^2\rangle$ 	&		&	\cite{U-radii1992}	\\
	&		&	233,235	&	$\mu$, $Q_{\rm s}$,$\delta \langle r^2\rangle$ 	&		&	\cite{U-radii1992,U-moment1973}	\\
\hdashline[0.2pt/2pt]

Np	&	93	&	237	&	$I$	&		&	\cite{Np-spin1976}	\\
\hdashline[0.2pt/2pt]

Pu	&	94	&	238-242,244	&	$\delta \langle r^2\rangle$ 	& \cite{Pu-radii2017} &	\cite{Pu-radii2017,Pu-radii1987}	\\
\hdashline[0.2pt/2pt]

Am	&	95	&	$\rm 240^{f}$,$\rm 242^{f}$,$\rm 244^{f}$	&	$Q_{\rm s}$,$\delta \langle r^2\rangle$ 	&		&	\cite{Am-radii1998,Am-radii2000}	\\
\hdashline[0.2pt/2pt]

Cm	&	96	&	242,244-246,248	&	$\delta \langle r^2\rangle$ 	&		&	\cite{Cm-radii1986,Cm-radii2006}	\\
\hdashline[0.2pt/2pt]

Bk	&	97	&	249	&	$I$	&		&	\cite{Cm-radii1986,Cm-radii2006}	\\
\hdashline[0.2pt/2pt]

Cf	&	98	&	249-252	&	$I$	&		&	\cite{Cf-spin1995}	\\
\hdashline[0.2pt/2pt]

\multirow{1}{*}{Es}	&	\multirow{1}{*}{99}	&	253	&	$\delta \langle r^2\rangle$ 	&		&	\cite{Es-radii2005}	\\
	&		&	253-255	&  $I$, $\mu$, $Q_{\rm s}$	&	\cite{Es-moment2022}	&	\cite{Es-moment1974,Es-moment2022}	\\
\hdashline[0.2pt/2pt]

Fm	&	100	&	255	&	$\mu$, $Q_{\rm s}$	&		&	\cite{Fm-moment2005}	\\
\hdashline[0.2pt/2pt]

No	&	102	&	252-254	&	$\mu$, $Q_{\rm s}$,$\delta \langle r^2\rangle$ 	&	\cite{No-radii2018}	&	\cite{No-radii2018}	\\
\hline
\end{longtable}
}
\end{center}

\section{Nuclear properties from atomic spectroscopy\label{sec:atoms}}
As the atomic nucleus is not point-like, its basic properties, such as spin, charge and current distribution (magnetic dipole moment, electric quadrupole moment, and charge radius), affect the electrons that are bound to it.
These effects lead to shifts in the electron energies and/or splittings in them, known as isotope shifts and hyperfine splittings, which are typically of the order $10^{-3}-1$~cm$^{-1}$ ($10^{-7}-10^{-4}$~eV). This is equivalent to a few parts per million relative to the atomic transition energy. The interaction between the atomic nucleus and its surrounding electrons can be described in a general form by using an electromagnetic multipole ($k$) expansion:
\begin{equation}
H_{\rm {hf}} =\sum_{k}\hat{\rm T}_{\rm N}^{k}\cdot \hat{\rm T}_{\rm e}^{k}
\label{eq01}
\end{equation}
where $\hat{\rm T}_{\rm N}^{k}$ and $\hat{\rm T}_{\rm e}^{k}$ are spherical tensor operators of rank $k$ acting on the nucleus and surrounding electrons, respectively. Here, even values of $k$ correspond to symmetry-conserving electric interactions, and odd ones are for symmetry-conserving magnetic interactions. Note that the $k=0$ electric monopole interaction corresponds to an exactly spherical nuclear charge distribution and results in a shift in the energy of atomic energy levels. The $k=1$ magnetic dipole and the $k=2$ electric quadrupole interactions are the dominant terms responsible for the hyperfine structure. The effect of higher-$k$ terms (i.e. $k=3$ magnetic octupole interaction) are several orders of magnitude smaller than that of $k=1,2$, and are practically out of the reach of current precision of laser spectroscopy techniques used for the study of radioactive isotopes. To our knowledge, the effect of $k=3$ interaction has only been observed in some stable isotopes~\cite{chi92,ger03,Sc-moment2020-45Sc,173Yb-octupole2021}. Constraining the parity-, and time reversal-violating electric dipole moment ($k=1$), and magnetic quadrupole moments ($k=2$), is a major open topic of research, with profound implications in our understanding of fundamental symmetries~\cite{Saf18,Chu19}.  

Therefore, with the achievable precision of current laser spectroscopy techniques, the majority of experimental studies at RIB facilities are focused on the investigation of the hyperfine interaction arising from the $k=0,1,2$ terms in the expansion~\cite{JPG2010,PPNP2016,PPNP2021}. This enables the routine extraction of nuclear magnetic dipole moments ($\mu$), electric quadrupole moments ($Q_{\rm s}$) in addition to the changes in mean-square charge radii ($\delta\langle r^2\rangle$). With the aim of probing more subtle effects, e.g. the $k=3$ interaction resulting from the nuclear magnetic octupole moment, efforts are underway to develop higher-resolution laser spectroscopy methods that maintain a sufficient sensitivity for studying radioactive species~\cite{Sc-moment2020-45Sc,173Yb-octupole2021}.

\subsection{\it Magnetic dipole and electric quadrupole hyperfine interactions\label{sec:hfs}}
In general, the interaction of the magnetic dipole moment $\bm{\mu_{I}}=g_{I}\mu_{\rm N}\bm{I}$ ($I$: nuclear spin) of the atomic nucleus with the magnetic field $\bm{B}_{e}$ at the location of the nucleus generated by the motion of surrounding electrons can be written as:
\begin{equation}
H_{\rm {m}} =-\bm{\mu_{I}}\cdot{\bm{B}_{e}}=A\bm{I}\cdot\bm{J}
\label{eq02}
\end{equation}
where $A$ is the magnetic dipole HFS parameter $A =\frac{\mu_{I}B_{e}}{IJ}$. Here, $\bm{B}_{e}$ is known to be proportional to the total angular momentum of the electrons ($\bm{J}$). 

\begin{figure*}[t!]
\begin{center}
\includegraphics[width=0.99\textwidth]{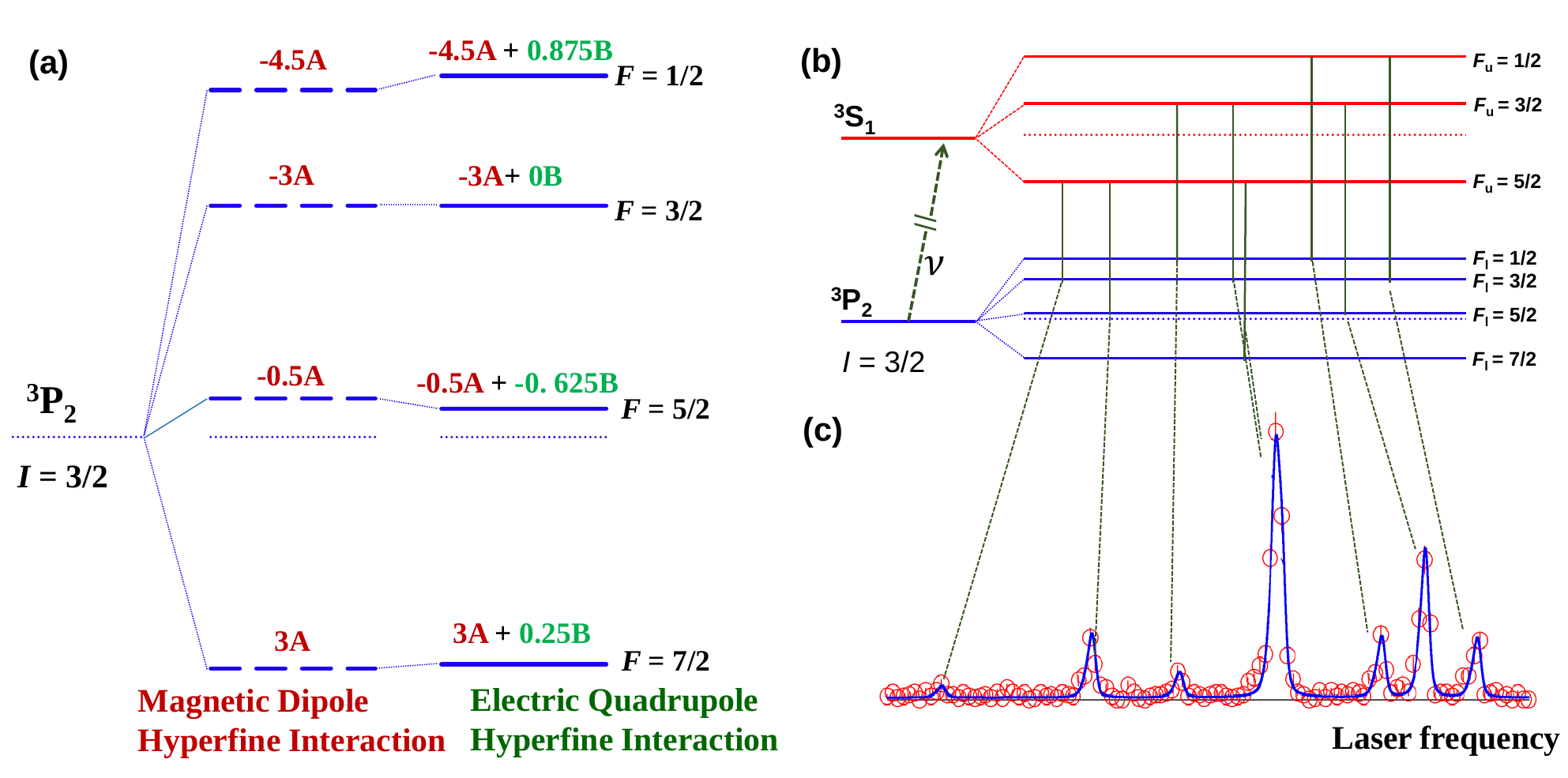}
\caption{\label{fig:fig2.1}\footnotesize{An example of a hyperfine structure (a) and spectrum (b) for the \mbox{$4s4p$ $^{3}P_{2}$} $\rightarrow$ \mbox{$4s5s$ $^{3}S_{1}$} atomic transition of $^{63}$Zn with nuclear spin of $I=3/2$ measured using the collinear laser spectroscopy (COLLAPS) setup ISOLDE-CERN~\cite{Zn-moment2017}.}}
\end{center}
\end{figure*}

Starting from this, the general expression describing the energy of each substate resulting from the magnetic dipole interaction is: 
\begin{equation}
\Delta E_{\rm m}/h =\frac{1}{2}AK=\frac{1}{2}A(F(F+1)-I(I+1)-J(J+1)).
\label{eq03}
\end{equation}
Here, $F$ is a quantum number $\bm{F}=\bm{I}+\bm{J}$ arising from the coupling of $I$ and $J$. The number of split substates is equal to $2I+1$ for the case of $I\le J$, and $2J+1$ for $J\le I$. Figure~\ref{fig:fig2.1}(a) presents an example of the four hyperfine substates ($F$) split from the $^{3}P_{2}$ state, due to the magnetic dipole hyperfine interaction between the atomic nucleus (with spin $I = 3/2$) and its bound electrons.

An electric quadrupole moment is expected for atomic nuclei with spin $I>1/2$ and a non-spherical charge distribution. The interaction between the quadrupole moment ($Q_{\rm s}$) of the nucleus and the average electric field gradient ($V_{zz}=\langle\frac{\partial^{2}V_{e}}{\partial z^{2}}\rangle$) at the location of nucleus produced by the surrounding electrons is given by:
\begin{equation}
H_{\rm {e}} = B\frac{6(\bm{I}\cdot\bm{J})^{2}+3(\bm{I}\cdot\bm{J})-2\bm{I^2}\cdot\bm{J^2}}{2I(2I-1)2J(2J-1)}.
\label{eq04}
\end{equation}
This leads to an additional energy shift of each substate $F$:
\begin{equation}
\Delta E_{\rm e}/h =\frac{B}{4}\frac{\frac{3}{2}K(K+1)-2I(I+1)J(J+1)}{I(2I-1)J(2J-1)}
\label{eq05}
\end{equation}
where $B = eQ_{\rm s}\langle\frac{\partial^{2}V_{e}}{\partial z^{2}}\rangle$ is the electric quadrupole HFS parameter. These additional shifts for the $F$ substates from the electric quadrupole hyperfine interaction are also illustrated in Fig.~\ref{fig:fig2.1}(a).

Therefore, for a general case  with $I,J>1/2$, the total energy shift of each hyperfine level $F$ is:
\begin{equation}
\Delta E_{\rm {hfs}}/h =\Delta E_{\rm m}/h+\Delta E_{\rm e}/h=\frac{1}{2}AK+\frac{B}{4}\frac{\frac{3}{2}K(K+1)-2I(I+1)J(J+1)}{I(2I-1)J(2J-1)}
\label{eq06}
\end{equation}
which combines the effects of both the magnetic dipole and electric quadrupole hyperfine interactions (Fig.~\ref{fig:fig2.1}(a)). 

In principle, by measuring the $\Delta E_{\rm {hfs}}$ of all hyperfine substates, the nuclear spin $I$ and HFS parameters ($A$ and $B$) can be determined, allowing the nuclear magnetic dipole and electric quadrupole moments to be extracted. For this, knowledge of the magnetic field $B_{e}$ and average electric field gradient $V_{zz}$ at the nucleus is required. This often needs to be calculated by atomic theory. However, as $B_{e}$ and $V_{zz}$ depend only on the electronic structure if one assumes an ideal point-like nucleus, they are expected to be a constant for all isotopes of a given element. Thus, the magnetic dipole and electric quadrupole moments can be deduced in a simpler and more elegant way, by using the known moments ($\mu$ and $Q_{s}$) and HFS parameters ($A$ and $B$) of a reference isotope: 
\begin{equation}
\mu=\mu_{\rm {ref}}\frac{IA}{I_{\rm{ref}}A_{\rm{ref}}},
\label{eq07}
\end{equation}
\begin{equation}
Q_{s}=Q_{s,\rm{ref}}\frac{B}{B_{\rm{ref}}}.
\label{eq08}
\end{equation}
In most cases, the reference nuclei are naturally occurring stable or long-lived isotopes, and their moments can be measured with high precision by other independent experimental methods, such as nuclear magnetic resonance (NMR) and nuclear quadrupole resonance (NQR) spectroscopy.

The assumption of an ideal point-like nucleus is clearly not sound for all real systems due to effects from the finite distribution of electric charge and nuclear magnetism. To take these into account, two additional corrections~\cite{HFA} need to be added to the expression of magnetic dipole HFS parameter $A$ such that: 
\begin{equation}
A=\frac{\mu_{I}B_{e}}{IJ}(1+\epsilon_{\rm{BR}})(1+\epsilon_{\rm{BW}})
=A_{\rm{point}}(1+\epsilon_{\rm{BR}})(1+\epsilon_{\rm{BW}})
\label{eq09}
\end{equation}
where $\epsilon_{\rm{BR}}$ is the Breit-Rosenthal-Crawford-
Schawlow (BR) effect caused by the extended nuclear charge~\cite{BR-effect}, and $\epsilon_{\rm{BW}}$ is the Bohr-Weisskopf effect (BW) arising from the non-uniform distribution of magnetization over the nucleus~\cite{BW-effect}. A direct experimental measurement of $\epsilon_{\rm{BR}}+\epsilon_{\rm{BW}}$ is not possible and precise theoretical calculations of $A_{\rm{point}}$ are challenging~\cite{PR1979,HFA2}. Thus, a commonly used method is to take the ratio of magnetic dipole HFS parameters ($A$ and $A'$) for two isotopes:
\begin{equation}
\frac{A}{A'}=\frac{\mu I'}{\mu'I}\frac{(1+\epsilon_{\rm{BR}})(1+\epsilon_{\rm{BW}})}{(1+\epsilon_{\rm{BR}}')(1+\epsilon_{\rm{BW}}')}\approx\frac{\mu I'}{\mu'I}(1+^{A}\Delta_{\rm{BW}}^{A'})(1+^{A}\Delta_{\rm{BR}}^{A'})\approx\frac{\mu I'}{\mu'I}(1+^{A}\Delta_{\rm{BW}}^{A'}).
\label{eq10}
\end{equation}
Here, the parameter $^{A}\Delta^{A'}$ is defined as the differential hyperfine anomaly of the two isotopes. As $^{A}\Delta_{\rm{BR}}^{A'}$ is expected to be very small~\cite{HFA} and $^{A}\Delta_{\rm{BW}}^{A'}$ from the BW effect is usually the dominant contribution to the differential hyperfine anomaly. Therefore, Eq.~\ref{eq07} becomes 
\begin{equation}
\mu=\mu_{\rm {ref}}\frac{IA}{I_{\rm{ref}}A_{\rm{ref}}}(1+{^{A}\Delta^{A'}}).
\label{eq11}
\end{equation}
Experimental determination of the hyperfine anomaly requires precise, independent measurements of the magnetic HFS parameters $A$ and magnetic moment $\mu$, as it is typically an effect on the order of 10$^{-3}$ with respect to $A$. In the majority of nuclei (those in light- and medium-mass elements), the hyperfine anomaly is assumed to be negligible by systematic studies and treated as an error contribution~\cite{K-moment2014}. However, for heavier systems, the magnitude of this effect can be significant~\cite{HFA2,HFA-Fr,HFA-Fr2,HFA-newphys}. Studying the hyperfine anomaly in these systems therefore could offer new insights into certain aspects of nuclear structure such as the nuclear magnetization radius~\cite{HFA-Fr,HFA-Fr2}.

\subsection{\it Isotope shifts\label{sec:IS}}

\begin{figure*}[t!]
\begin{center}
\includegraphics[width=0.99\textwidth]{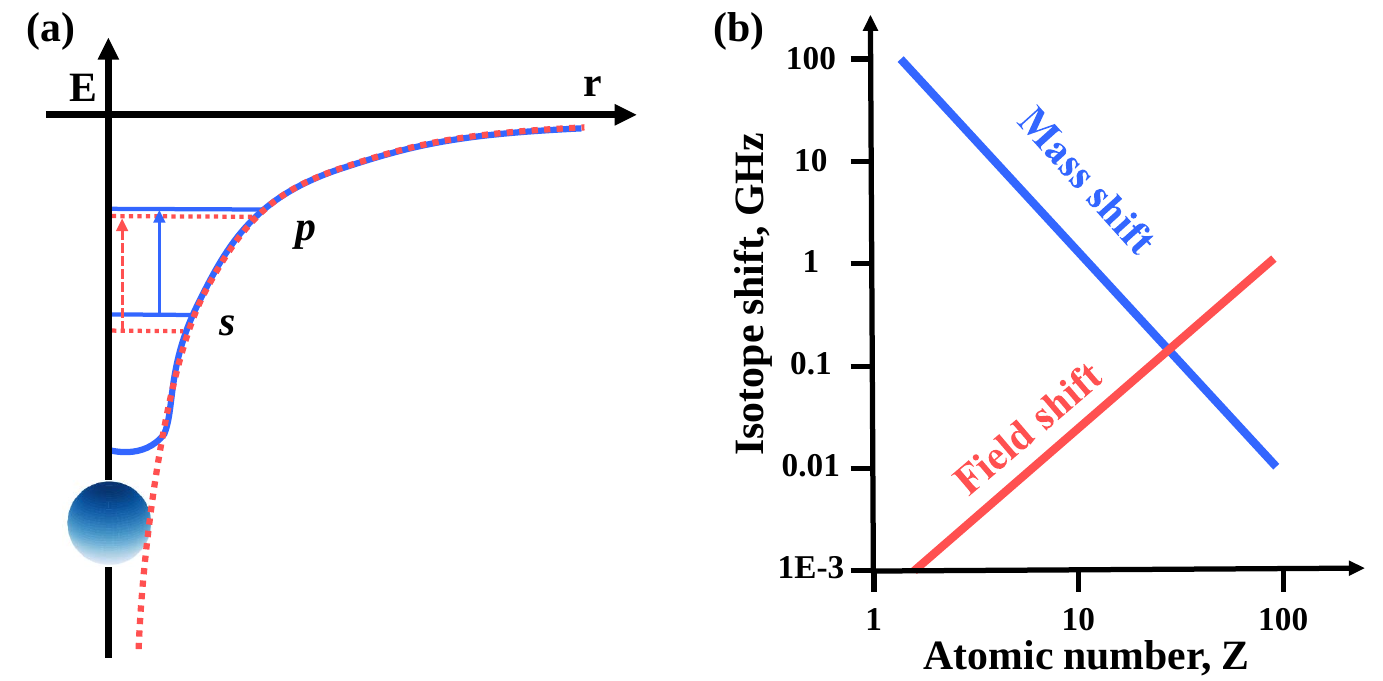}
\caption{\label{fig:fig2.2}\footnotesize{(a) Schematic diagram for the energy shift of $s-$ and $p$-electronic levels due to a nucleus with a finite size. This figure is reproduced from a figure in Ref.~\cite{Lu13}. (b) Qualitative illustration of the approximate magnitude of the mass shift and field shift as a function of the atomic number $Z$. The figure is reproduced from a figure in Ref.~\cite{Blaum_2013}}}
\end{center}
\end{figure*}

Starting with a simple example, as illustrated in Fig.~\ref{fig:fig2.2}(a), the finite size of atomic nucleus leads to a minute shift in the transition frequency of the two atomic energy levels (e.g. $s$ and $p$) when compared to that of a point nucleus. 
Therefore, for two nuclear systems (isotopes with mass numbers of $A$ and $A'$) of the same element, due to the difference in their mass and volume, there is a small difference ($\delta\nu$) in their atomic transition frequencies ($\nu^{A'}$, $\nu^{A}$) between the two electronic states. This is known as the isotope shift and is defined as:
\begin{equation}
\delta\nu_{\rm{IS}}^{AA'}=\nu^{A'}-\nu^{A}.
\label{eq12}
\end{equation}
For isotopes possessing non-zero nuclear spin, hyperfine structures for the two atomic energy level exist (given $J>0$), as shown in Fig.~\ref{fig:fig2.1} (b) for the atomic \mbox{$4s4p$ $^{3}P_{2}$} and \mbox{$4s5s$ $^{3}S_{1}$} states of zinc. Thus, the isotope shift between the transition frequencies of the two isotopes refers to the difference in the center of gravity (COG or centroid) of their hyperfine structures.

There are two major effects that contribute to the isotope shift~\cite{king}:
\begin{equation}
\delta\nu_{\rm{IS}}^{AA'}=\delta\nu_{\rm{MS}}^{AA'}+\delta\nu_{\rm{FS}}^{AA'}.
\label{eq13}
\end{equation}
The first part is the mass shift, that comes from the movement of a nucleus with finite mass $M$. It can be decomposed into two components, known as the Normal Mass Shift (\textsl{NMS}) and Specific Mass Shift (\textsl{SMS}). The \textsl{NMS} is equivalent to the case of hydrogen (one-electron case), and is simply proportional to the transition frequency ($\nu$) and to $\frac{M^{A'}-M^{A}}{M^{A'}M^{A}}$. The \textsl{SMS}, on the other hand, originates from correlation effects between any two electrons within a multi-electron system. 
This therefore renders the \textsl{SMS} very difficult to evaluate theoretically. A calibration based on known data (known as a \lq King-plot'~\cite{Fricke2004,king} approach, as will be briefly introduced in Sec.~\ref{sec:extraction}) is preferred in cases where it is possible. Nevertheless, the total mass shift can be factorized and written as
\begin{equation}
\delta\nu_{\rm{MS}}^{AA'}=\delta\nu_{\rm{NMS}}^{AA'}+\delta\nu_{\rm{SMS}}^{AA'}=(K_{\rm{NMS}}+K_{\rm{SMS}})\frac{M^{A'}-M^{A}}{M^{A'}M^{A}}=K_{\rm{MS}}\frac{M^{A'}-M^{A}}{M^{A'}M^{A}}.
\label{eq14}
\end{equation}
where $K_{\rm MS}$ is the mass shift factor. From Eq.~\ref{eq14}, one can see that, in light nuclei, the mass shift dominates the overall isotope shift, as schematically shown in Fig.~\ref{fig:fig2.2}(b). This is because that adding/removing one neutron to light nuclei will have a much larger fractional effect on $\delta\nu_{\rm{MS}}^{AA'}$ when compared to doing the same in heavier nuclei.

The second term in Eq.~\ref{eq13} is known as the field (or volume) shift arising from the change in electron energy due to the difference in spatial distribution of the nuclear charge, as shown in Fig.~\ref{fig:fig2.2}(a). This change in electron energy, due to the finite size of the nucleus, should then be proportional to the volume (mean-square charge radius $\langle r_{\rm {c}}^2\rangle$) of the atomic nucleus in addition to the non-relativistic probability density ($|\Psi_{e}(0)|^{2}$) of the involved electron at the centre of the nucleus. Thus, for an optical transition from state $i$ to state $f$, the shift in the transition frequency is proportional to the difference in the non-relativistic electron density between the two states ($\Delta|\Psi_{e}(0)|_{if}^{2}$). Assuming a constant electron density within the nuclear volume for all isotopes of an element in a non-relativistic regime, the field shift of the two isotopes ($A$ and $A'$) for an optical transition ($i\rightarrow f$) can be given by~\cite{king} 
\begin{equation}
\delta\nu_{\rm{FS}}^{AA'}=\frac{\pi a_{0}^{3}}{Z}\Delta|\Psi_{e}(0)|_{if}^{2}\delta\langle r_{\rm {c}}^2\rangle^{AA'}=F\delta\langle r_{\rm {c}}^2\rangle^{AA'}
\label{eq15}
\end{equation}
where $a_{0}$ is the Bohr radius, $Z$ is the proton number of the element, $\delta\langle r_{\rm {c}}^2\rangle$ is the change in the mean-square charge radii of two isotopes with mass numbers of $A$ and $A'$. Here, $F$ is the field shift factor. As discussed in Ref.~\cite{Blaum_2013}, in a rough approximation, a dependence of the field shift on the atomic number and mass number can be estimated to be 
\begin{equation}
\delta\nu_{\rm{FS}}^{AA'}\propto\frac{Z^{2}}{\sqrt[3]{A}}.
\label{eq16}
\end{equation}
Therefore, in heavier nuclei, the field shift dominates the overall isotope shift. Figure~\ref{fig:fig2.2}(b) plots the relative contributions of the field and mass shifts as a function of atomic number $Z$. Based on Eq.~\ref{eq15}, it is clear that the field shift carries information of nuclear charge radii and is therefore the part of interest in nuclear physics. In light nuclei, as the field shift is several orders of magnitude smaller than the mass shift, high-precision measurements of isotope shifts become crucial in order to precisely extract the $\delta\langle r_{\rm {c}}^2\rangle^{AA'}$.

Note that for a more proper treatment of the field shift, a relativistic correction factor $f(Z)$ may be considered, leading to a field shift factor $F(Z)=\frac{\pi a_{0}^{3}}{Z}\Delta|\Psi_{e}(0)|_{if}^{2}f(Z)$. In particular, for heavy nuclei, the assumption of a constant electron density over the nuclear volume is no longer valid and relativistic effects can not be neglected. The expression of the field shift then becomes $\delta\nu_{\rm{FS}}^{AA'}=F(Z)\lambda^{AA'}$. Here, $\lambda^{AA'}$ is the Seltzer moment, which can be written as a series of radial moments~\cite{Seltzer1969}:
\begin{equation}
 \lambda^{AA'}=\delta\langle r_{\rm {c}}^2\rangle^{AA'}+a\delta\langle r_{\rm {c}}^4\rangle^{AA'}+b\delta\langle r_{\rm {c}}^6\rangle^{AA'}+...
\label{eq17}
\end{equation}
As will be introduced in Sec.~\ref{sec:newphysics} and in Ref.~\cite{4th-radii}, high-precision measurements of isotope shifts may provide access to higher-order radial moments of atomic nuclei in addition to constraining potential new physics and phenomena.

\subsection{\it Extraction of nuclear observables\label{sec:extraction}}
Figure~\ref{fig:fig2.1}(b) and (c) show a typical example of the HFS spectrum for the atomic \mbox{$4s4p$ $^{3}P_{2}$} $\rightarrow$ \mbox{$4s5s$ $^{3}S_{1}$} transition in $^{63}$Zn ($I=3/2$) measured with the standard collinear laser spectroscopy method~\cite{Zn-moment2017}. The frequency ($\nu_{F_{\rm u}F_{\rm l}}$) of each allowed transition from the upper substate $F_{\rm u}$ to the lower substate $F_{\rm l}$ (selection rule: $\Delta F=0, \pm1$ with $F=0\rightarrow F=0$ forbidden) in Fig.~\ref{fig:fig2.1}(b), corresponding to each resonance peak in Fig.~\ref{fig:fig2.1}(c), can be described by the function~\cite{JPG2010}:
\begin{equation}
\nu_{F_{\rm u}F_{\rm l}}=\nu+\alpha_{\rm u}A_{\rm u}+\beta_{\rm u}B_{\rm u}-\alpha_{\rm l}B_{\rm l}-\beta_{\rm l}A_{\rm l}\\
\label{eq18}
\end{equation}
where $\nu$ is the centroid frequency of the \mbox{$^{3}P_{2}$} $\rightarrow$ \mbox{$^{3}S_{1}$} atomic transition. Here, $A_{\rm u}$ (and $B_{\rm u}$ ) and $A_{\rm l}$ (and $B_{\rm l}$) are the magnetic dipole (and electric quadrupole) HFS parameters for the upper (\mbox{$^{3}S_{1}$}) and lower (\mbox{$^{3}P_{2}$}) states, respectively. From Eq.~\ref{eq06}, one can deduce expressions for the coefficients $\alpha_{\rm u}$ ($\alpha_{\rm l}$) and $\beta_{\rm u}$ ($\beta_{\rm l}$), which are a function of the nuclear spin ($I$) and total angular momenta of the upper ($J=1$) and lower ($J=2$) atomic states:
\begin{equation}
\alpha=\frac{K}{2}, \hspace{2mm}
\beta=\frac{B}{4}\frac{\frac{3}{2}K(K+1)-2I(I+1)J(J+1)}{I(2I-1)J(2J-1)}
\label{eq19}
\end{equation}
with $K=F(F+1)-I(I+1)-J(J+1)$. Based on Eq.~\ref{eq18}, we can fit the HFS spectrum in Fig.~\ref{fig:fig2.1}(c) by using a $\chi^2$-minimization process (an example can be found in the Python package in SATLAS~\cite{SATLAS}). From this, the atomic parameters of $\nu$, $A_{\rm u}$, $B_{\rm u}$, $A_{\rm l}$, $B_{\rm l}$ can be obtained. With these parameters, we can derive magnetic moments~($\mu$), quadrupole moments~($Q_{\rm s}$) and changes in mean-square charge radii~($\delta\langle r^2\rangle$) without invoking any nuclear model, as will be detailed in the following. Note that the nuclear spin $I$ is predefined in the fitting process and an incorrect assignment will in most cases strongly affect the quality of the fit. \\

\hspace{-6mm}$\bullet{\textbf{ Nuclear spin}}$

One of the key strengths of high-resolution laser spectroscopy methods is that, in a large number of cases, the nuclear spins can be determined unambiguously in a nuclear model-independent way. The following approaches can be adopted to determine the nuclear spins from HFS spectra.
\begin{itemize}
\item [1)] \uline{Number of observed resonance peaks} \\
It has already been demonstrated in Eq.~\ref{eq03} that the number of hyperfine structure resonances is necessarily defined by the combination of the nuclear spin $I$ and the total angular momenta of the atomic states involved $J$, as $F=I+J, I+J-1, ..., \vert I-J\vert$. In cases where $I\le J$, the number of hyperfine substates $F$ is $2I+1$. Therefore, an unambiguous spin assignment can be immediately made by counting the number of the observed resonance peaks if they are all well resolved in the HFS spectrum. Taking again the \mbox{$^{3}P_{2}$} $\rightarrow$ \mbox{$^{3}S_{1}$} atomic transition of zinc as an example, three and eight HFS peaks are expected for zinc isotopes with $I=1/2$ and $I=3/2$, respectively (due to $J=1,2$). This method is exactly how the 1/2 spin was assigned to the isomeric state in $^{79}$Zn ~\cite{Zn-radii2016},
and what Fig.~\ref{fig:fig2.1}(b) presents for $I=3/2$ ground state of $^{63}$Zn. 
For all nuclear states of zinc isotopes with $I\ge 5/2$ (i.e. $I > J$), nine resonance peaks are expected in their HFS spectra, as has been observed for $^{65-79}$Zn (Fig.1 in Ref.~\cite{Zn-moment2017}). A direct spin determination in this manner is impossible for systems with $I> J$, however other approaches can be utilized. 
\begin{figure*}[t!]
\begin{center}
\includegraphics[width=0.75\textwidth]{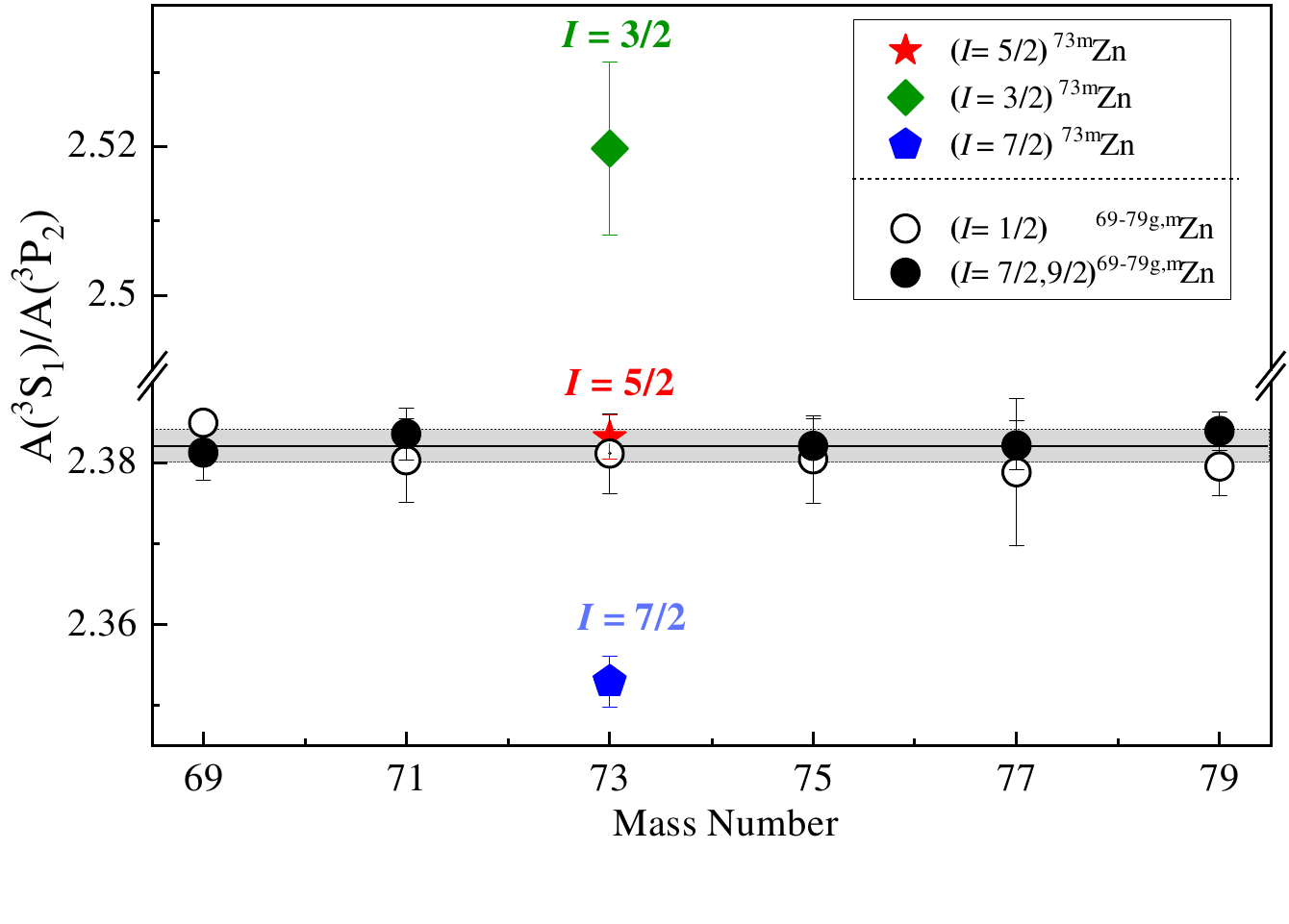}
\caption{\label{fig:fig2.3}\footnotesize{An example of spin determination for $^{73{\rm{m}}}$Zn based on the ratio between the HFS constants of the upper (\mbox{$^{3}S_{1}$}) and lower (\mbox{$^{3}P_{2}$}) states ($A(^{3}S_{1})/A(^{3}P_{2})$) extracted from the fitting of their HFS spectra~\cite{Zn-spin2018}. The black open and solid dots show the results for the 1/2 and high-spin (7/2, 9/2) states of $^{69-79}$Zn, respectively. For $^{73{\rm{m}}}$Zn, three possible spins are used in fitting its HFS spectrum, with a red star for $I=5/2$, a green diamond for $I=3/2$ and a blue pentagon for $I=7/2$. It is clear that the ratio $A(^{3}S_{1})/A(^{3}P_{2})$ for the case of $I=5/2$ consistent with that of other isotopes and isomers. This therefore determines the spin of $^{73{\rm{m}}}$Zn to be 5/2. The figure is re-plotted based on the results in Ref.~\cite{Zn-spin2018}.}}
\end{center}
\end{figure*}
\item [2)] \uline{Ratio of $A_{\rm{u}}$ and $A_{\rm{l}}$}\\
As discussed above in Sec.~\ref{sec:hfs}, the hyperfine anomaly, in the vast majority of cases, is very small and thus can be neglected. We can expect that the ratio of the magnetic dipole HFS parameters of the two atomic or ionic states probed by laser spectroscopy experiments, $R =A_{\rm{u}}/A_{\rm{l}}$, remains constant across the entire isotopic chain of a given element. This rule offers an indirect but effective and nuclear model-independent method to determine the spin of the studied isotope when fitting its HFS spectrum. In other words, if a correct spin is predefined when fitting the HFS spectrum of isotope with an unknown spin, the extracted $R =A_{\rm{u}}/A_{\rm{l}}$ will be the same, within experimental uncertainty, as that of the reference isotope. This approach has often been adopted for spin determinations of new isotopes, such as the neutron-rich $^{51}$Ca isotope~\cite{Ca-moment2015} and $^{75}$Cu~\cite{Cu-moment2009-1}. An experimental example is given in Fig.~\ref{fig:fig2.3}, where the spin of the long-lived isomeric state of the $^{73}$Zn is assigned unambiguously to be 5/2 instead of 3/2 or 7/2~\cite{Calvin-phd,Zn-spin2018}. This is because that the extracted $R$ with spin assumption of 5/2 perfectly agrees with that of other zinc isotopes.

\item [3)] \uline{Racah intensities of the resonance peaks}\\
Theoretically, the relative intensity of the resonance peaks in a typical HFS spectrum (Fig.~\ref{fig:fig2.1}(c)), corresponding to the allowed dipole transition strengths between two hyperfine states labeled with $F_{\rm u}$ and $F_{\rm l}$, can be calculated using the formula~\cite{Racah}:
\begin{equation}
I_{F_{\rm u}\rightarrow F_{\rm l}}\propto (2F_{\rm l}+1)(2F_{\rm u}+1)\left\{
\begin{matrix}
    J_{\rm u} & F_{\rm u} & I \\
    F_{\rm l} & J_{\rm l} & 1
\end{matrix}\right\}^{2}
\label{eq20}
\end{equation}
where the factor in brackets is a Wigner 6-$j$ symbol. This expression is also called the Racah intensity. Based on Eq.~\ref{eq20}, the nuclear spin ($I$) can be, in principle, determined from the relative intensities of resonance peaks in the measured HFS spectrum. However, in practice, the experimentally observed relative intensities of the HFS resonance peaks are often affected by the power of the probing laser. Thus a direct spin determination from Eq.~\ref{eq20} is not always possible, but can still be realized by a systematic comparison of the relative intensities of the HFS peaks of different isotopes measured under identical experimental conditions. This approach was employed  for the spin assignment of the neutron-rich $^{51}$K isotope~\cite{K-moment2013,K-radii2014-1}.
\end{itemize}

\hspace{-6mm}$\bullet{\textbf{ Nuclear magnetic ($\mu$) and quadrupole ($Q_{\rm s}$) moments}}$

The magnetic and quadrupole moments of a given isotope can be determined using Eq.~\ref{eq07} and Eq.~\ref{eq08}, respectively, from either the $A_{\rm u}$ and $B_{\rm u}$ of the upper state or $A_{\rm l}$ and $B_{\rm l}$ of the lower state, depending on their sensitivity to them. Taking once again the two atomic states of $^{63}$Zn in Fig.~\ref{fig:fig2.1}(b) as an example, $A_{\rm u}$ of \mbox{$^{3}S_{1}$} is two times more sensitive to the magnetic moment than $A_{\rm l}$ of \mbox{$^{3}P_{2}$}. The moments of zinc isotopes are therefore determined from the $A_{\rm u}$ of the \mbox{$^{3}S_{1}$} state and $B_{\rm l}$ of the \mbox{$^{3}P_{2}$} state~\cite{Zn-moment2017} ($B_{\rm u}$ of \mbox{$^{3}S_{1}$} is known to be nearly zero). In the case that HFS parameters for both states are comparable in magnitude, the final magnetic $\mu$ and quadrupole $Q_{\rm s}$ moments are often calculated as the weighted average of the two sets of values taking into account the correlation between the HFS parameters of the two states~\cite{Sc-moments2022,Ge-moment2020}.\\

\hspace{-6mm}$\bullet{\textbf{ Changes in mean-square charge radii $\delta\langle r^2\rangle$}}$

\begin{figure*}[t!]
\begin{center}
\includegraphics[width=0.90\textwidth]{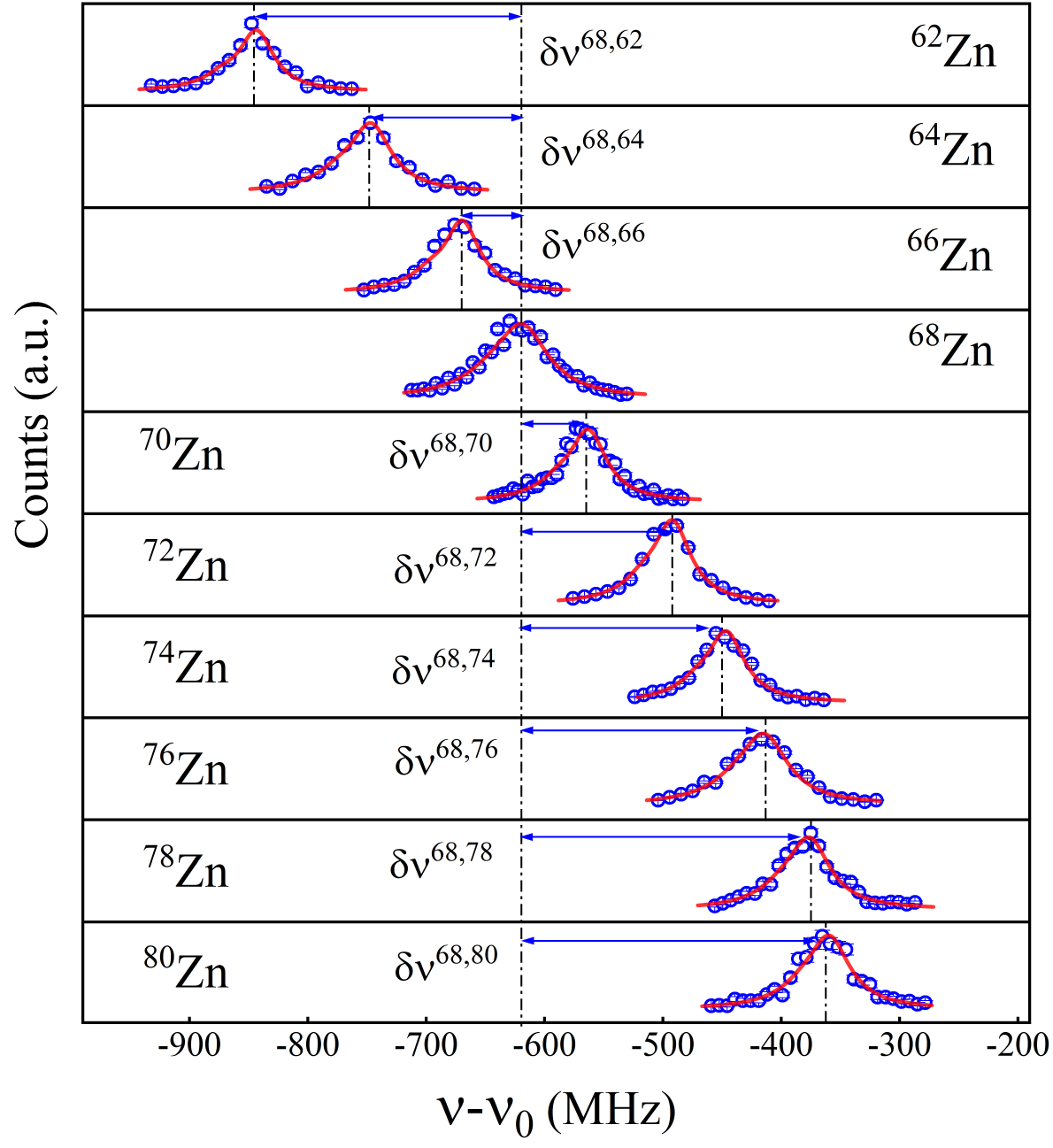}
\caption{\label{fig:fig2.4}\footnotesize{Optical spectra of even-even $^{62-80}$Zn isotopes (with $I=0$) measured in the \mbox{$4s4p$ $^{3}P_{2}$} $\rightarrow$ \mbox{$4s5s$ $^{3}S_{1}$} atomic transition by COLLAPS at ISOLDE-CERN~\cite{Zn-radii2019}, from which the isotope shifts, $\delta \nu=\nu^{A}-\nu^{68}$, are deduced. The figure is re-plotted based on the results in Ref.~\cite{Xie-phD}.}}
\end{center}
\end{figure*}

Based on Eq.~\ref{eq13} and the discussion in Sec.~\ref{sec:IS}, the total isotope shift $\delta\nu_{\rm {IS}}^{A,A'}$ for an optical transition between the isotopes $A$ and $A'$, can be summarized as: 
\begin{equation}
\delta\nu_{\rm {IS}}^{A,A'}= K_{\rm MS}\frac{M_{A}-M_{A'}}{M_{A}M_{A'}}+F\delta\langle r^2\rangle^{AA'}.
\label{eq21}
\end{equation}
Here, $K_{\rm MS}$ and $F$ are the mass shift and field shift factors, respectively, which are considered to be constant in a particular transition across an isotopic chain when relativistic effects are neglected.

Experimentally, we can deduce the isotope shift $\delta\nu_{\rm {IS}}^{A,A'}$ of two isotopes ($A,A'$) from the centroid frequencies determined from their HFS spectra. Figure~\ref{fig:fig2.4} presents the experimentally measured spectra of even-even $^{62-80}$Zn isotopes (with $I=0$) for the \mbox{$4s4p$ $^{3}P_{2}$} $\rightarrow$ \mbox{$4s5s$ $^{3}S_{1}$} atomic transition. As indicated in the figure, the centroid frequency of each isotope, defined through fitting the resonance, allows the isotope shifts of all isotopes relative to the reference $^{68}$Zn to be calculated as $\delta\nu^{68,A'}=\nu^{A'}-\nu^{68}$. Therefore, from the measured $\delta\nu^{A,A'}$, the changes in mean-square charge radii between the reference isotope $A$ and isotope $A'$, $\delta\langle r^2\rangle^{AA'}$, can be obtained. Further, the root-mean-square~(rms) charge radius of the isotope $A'$ can be determined from:
\begin{equation}
R^{A'}= \sqrt{\langle r^{2} \rangle^{A'}} = \sqrt{\delta \langle r^{2}\rangle^{A,A'} + \langle r^{2}\rangle^{A}},
\label{eq22}
\end{equation}
Here, $\langle r^{2}\rangle^{A}$ is the rms charge radius of reference isotope (e.g. $^{68}$Zn in Fig.~\ref{fig:fig2.4}), which is often the stable isotope and its charge radius known from other independent experiments, i.e. electron scattering experiments, and/or muonic and X-ray data~\cite{Fricke2004}.

In order to calculate $\delta\langle r^2\rangle^{AA'}$ using Eq.~\ref{eq21}, knowledge of the atomic factors $K_{\rm MS}$ and $F$ is required. In general, there are two standard approaches to determine these factors: atomic calculations using theoretical models and a calibration using a King-plot analysis. For elements with less than three stable isotopes, or in those where the charge radii of stable isotopes are unknown, theoretical calculation of atomic factors is the only possible approach. With advances in state-of-the-art theoretical atomic methods, the atomic mass- and field-shift factors are now calculable with much-improved accuracy and precision. These calculations have been widely used in the extraction of the $\delta\langle r^2\rangle$ in different mass regions of the nuclear chart~\cite{K-radii2021,Zn-radii2019,Sn-radii2020}. For elements that possess more than three stable isotopes whose charge radii are experimentally known from other techniques, the King-plot analysis~\cite{king} can be adopted to calibrate the atomic mass- and field-shift factors based on Eq.~\ref{eq21}, as detailed in Ref.~\cite{Fricke2004}.

\section{Nuclear global properties\label{sec:properties}}
Measurements of the basic properties of atomic nuclei allows us to gain critical insights into their structure and the complex nuclear force that binds them. As was detailed previously in Sec.~\ref{sec:atoms}, multiple fundamental properties of the atomic nucleus, such as the total angular momentum (nuclear spin), nuclear magnetic dipole moment, electric quadrupole moment and changes in mean-square charge radii, can be measured from the interaction with its bound electrons in a nuclear model-independent fashion. In this chapter, we discuss, using examples, how these observables allow us to better understand the nature of nuclear structure and how it evolves in unstable nuclei. It is becoming increasingly recognized that the properties of nuclei measured with laser spectroscopy experiments can act as stringent tests of current state-of-the-art nuclear models. This will be discussed in Sec.~\ref{physics} alongside results from laser spectroscopy techniques used to study exotic nuclei in different mass regions of nuclear chart. 

\subsection{\it Nuclear spins and electromagnetic moments\label{sec:moment}}
\subsubsection{Nuclear spins\label{sec:spin}}

The nuclear spin, ${I}$, is a quantum number assigned to each nuclear state and represents the total angular momentum of an atomic nucleus corresponding to the sum of the angular momentum of each of its $A$ nucleons including the orbital and `intrinsic' spin angular momentum. As a quantum many-body system with $Z$ protons and $N$ neutrons ($A$ nucleons) governed by short-range ($\sim$fm) nucleon-nucleon interactions, it is energetically favorable for the nucleons to be paired to give an angular momentum of 0$^+$ for each pair. Therefore, even–even nuclei will always possess a nuclear spin of 0$^{+}$ in their ground state, while odd-$A$ and odd-odd nuclei have half-integer spin and integer spin, respectively. In a simple nuclear shell-model picture, it is assumed that the constituent protons and neutrons in a nucleus move within a spherical mean field and the ground-state spin of the nucleus is defined by the angular momentum $j$ of the orbital in which the valence nucleon(s) reside. These nuclear shell-model concepts have played an essential role in our understanding of atomic nuclei since its conception~\cite{Mayer1949,Jensen1949} and have found particular success in describing nuclear ground-state spins of many near-magic nuclei. Most atomic nuclei are generally more complex and require models to account for pairing effects, multi-quasiparticle configurations and collectivity. In unstable nuclei with atypically large or small ratios of protons to neutrons, a change in magnitude of other features in the nucleon-nucleon interaction, e.g. the tensor force~\cite{SM-Otsuks2005} and three-body force~\cite{3N-many-body}, can lead to unexpected and sometimes dramatic changes in the structure of atomic nuclei. Prominent examples involve nuclear shell evolution~\cite{K-moment2013,Cu-moment2009-2}, intruder states~\cite{Mg-moment2005,Zn-radii2016}, disappearance of traditional magic numbers~\cite{Mg-moment2012} and the appearance of the new regions of magicity~\cite{Wie13,Ste13}. Experimentally, assigning the ground-state spin of odd-$A$ and odd-odd nuclei with other methods often relies upon invoking the conservation of total angular momentum or selection rules during the nuclear decays or reactions. Although they are of great value for nuclear structure studies, direct and unambiguous spin assignments are experimentally challenging for these methods. Laser spectroscopy techniques, as discussed in Sec.~\ref{sec:atoms}, offer a direct means to unambiguously determine the nuclear spins of ground and long-lived isomeric states, as detailed in Sec.~\ref{sec:extraction}. These spin assignments have proven pivotal in characterizing different nuclear structure phenomena.

\begin{figure*}[t!]
\centering
\includegraphics[width=0.99\textwidth]{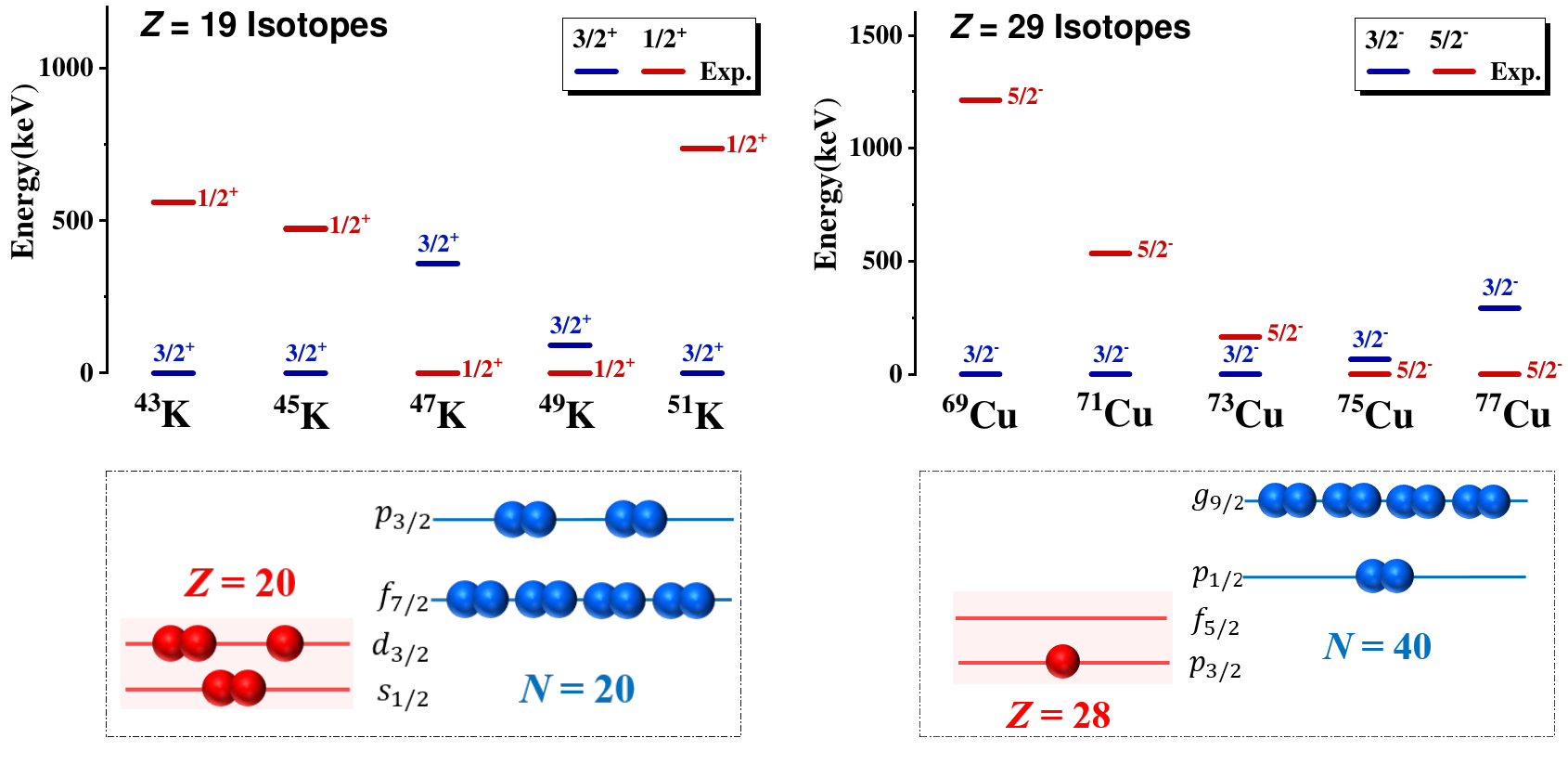}
\caption{\label{fig:fig3.1}\footnotesize{Example of proton shell evolution in potassium ($Z=19$) isotopes (left) and of copper ($Z=29$) isotopes (right), respectively. Information on the structural characteristics of these isotopes was obtained from measurements of the spins of their ground and first excited states~\cite{K-moment2013,Cu-moment2009-2}.}}
\end{figure*}

The importance of nuclear spin assignments for the structure study of the exotic nuclei has been demonstrated across different mass regions of the nuclear chart~\cite{K-moment2013,Mg-moment2005,Cu-moment2009-2,Zn-radii2016}, and will be discussed in Sec.~\ref{physics}. Here, we will give two examples, which shed light upon the (re)inversion of single-particle levels and the validation of effective nucleon-nucleon interactions in which the role of the tensor force is probed. Figure~\ref{fig:fig3.1} shows the energy of the low-lying states of potassium ($Z=19$) and copper ($Z=29$) isotopes. Potassium has 19 protons, with a single-proton hole in the $\pi 1d_{3/2}$ shell-model orbital below the $Z=20$ magic shell gap (left panel of Fig.~\ref{fig:fig3.1}). From a simple single-particle model perspective, a ground-state spin of 3/2$^+$ would be expected for all odd-$A$ potassium isotopes. However, as neutrons gradually fill the $\nu 0f_{7/2}$ and $\nu 1p_{3/2}$ orbitals, the ground-state spins of $^{47,49}$K were measured to be 1/2$^+$ by laser spectroscopy experiments. Progressing towards the more neutron-rich isotopes, the ground-state spin of $^{51}$K reverts back to 3/2$^+$ as the $\nu 1p_{3/2}$ orbital is fully occupied. This inversion and subsequent reinversion of the nucleon single-particle levels, demonstrated here with the ground-state spins of potassium isotopes, is one example of novel phenomena observed in exotic nuclei~\cite{K-radii1982,K-moment2013}. Theoretical interpretations for shell evolution, within the shell-model framework, attribute this phenomenon to result from the enhanced effect of the residual proton-neutron monopole interaction in weakly-bound nuclei~\cite{SM-Otsuks2005,K-moment2013}. By including the tensor contribution of the proton-neutron monopole interaction in the shell model, theoretical predictions showed that another inversion of proton $\pi 0f_{5/2}$ and $\pi 1p_{3/2}$ single-particle orbitals will occur when neutrons occupy the $\nu 1g_{9/2}$ orbital~\cite{SM-Otsuks2005}. This prediction was soon experimentally validated by the ground-state spins of neutron-rich copper isotopes (e.g. $^{75,77}$Cu) assigned by laser spectroscopy experiments~\cite{Cu-moment2009-2, Cu-moment2010-1}, as shown in right panel of Fig.~\ref{fig:fig3.1}. \\
\begin{figure*}[t!]
\begin{center}
\includegraphics[width=0.99\textwidth]{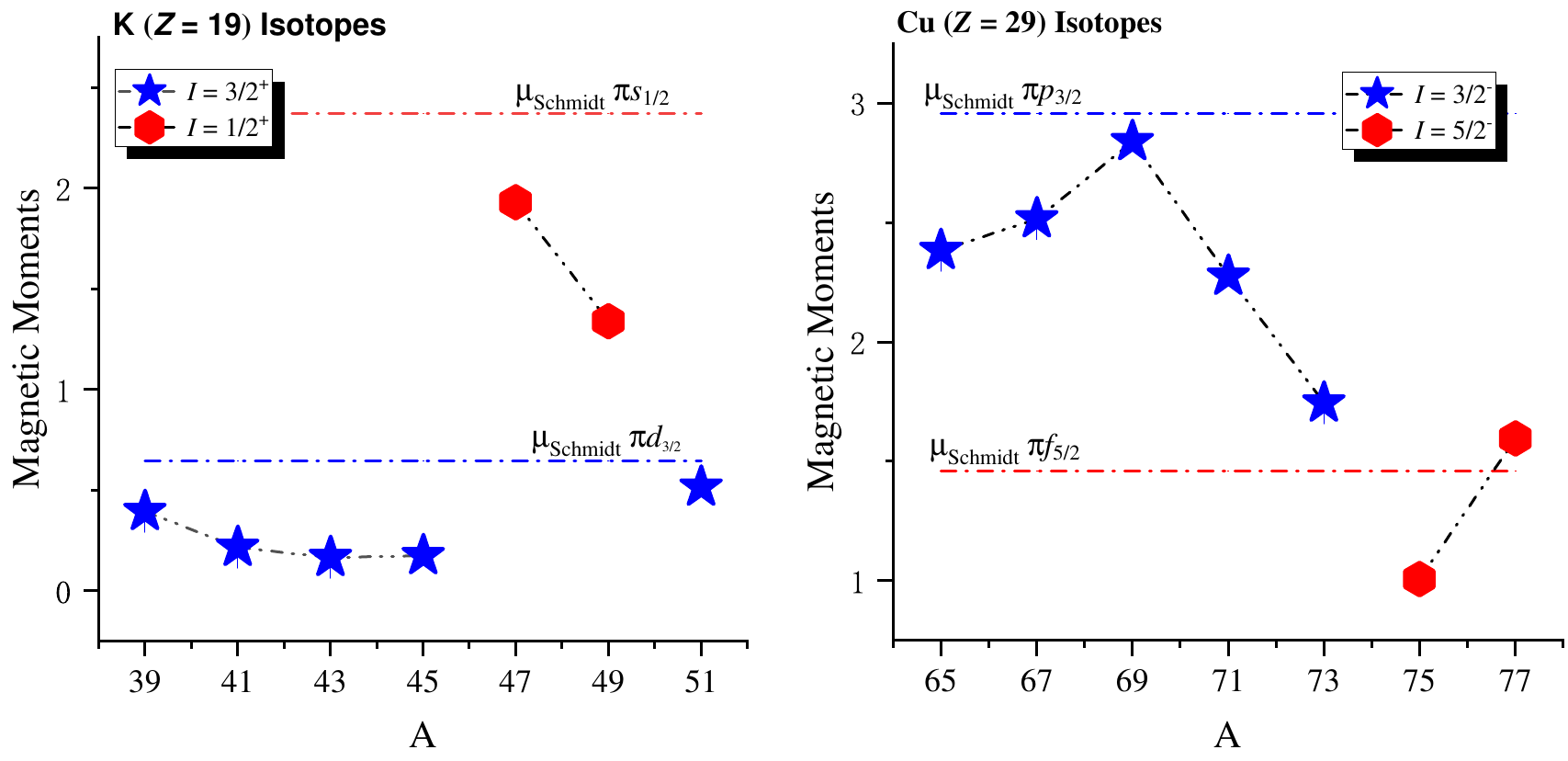}
\caption{\label{fig:fig3.2}\footnotesize{Experimental magnetic moments of potassium ($Z=19$) isotopes (left panel)~\cite{K-moment2013,K-moment2014} and of copper ($Z=29$) isotopes (right panel)~\cite{Cu-moment2009-2, Cu-moment2010-1}, compared to the Schmidt moments for a single proton occupying different shell-model orbitals. For potassium isotopes, Schmidt moments for proton in $\pi d_{3/2}$ and $\pi s_{1/2}$ orbitals are estimated using an effective spin $g$-factor of $g_{s}^{\rm eff} = 0.85g_{s}^{\rm free}$ and effective orbital $g$-factor of $g_{l}^{\pi} = 1.15$~\cite{K-moment2013}. While for copper isotopes, the Schmidt values are calculated using $g_{s}^{\rm eff} = 0.7g_{s}^{\rm free}$ for a single proton occupying $\pi p_{3/2}$ and $\pi f_{5/2}$ orbitals\cite{Cu-moment2009-2}.}}
\end{center}
\end{figure*}

\subsubsection{Nuclear magnetic moments\label{sec:u}}

The nuclear magnetic dipole moment arises from the magnetic field generated by the orbiting charged particle and the `intrinsic' spin ($s=1/2$) of a nucleon. In a semi-classical approximation (impulse approximation), the magnetic dipole operator of a nucleus can be expressed by the orbital and spin contributions such that
\begin{equation}
\bm{\mu}=\sum\limits_{i=1}^{i=A}g_{l}^i\bm{l}^i+\sum\limits_{i=1}^{i=A}g_{s}^i\bm{s}^i
\label{eq23}
\end{equation}
where $g_{l}$ and $g_{s}$ are known as the orbital and spin gyromagnetic factors ($g$-factors), $\bm{l}$ and $\bm{s}$ are the orbital and spin angular momentum operators~\cite{Castel1990}. The free proton and neutron $g$-factors are $g_{s}^{\pi}=5.587$, $g_{s}^{\nu}=-3.826$, $g_{l}^{\pi}=1$, $g_{l}^{\nu}=0$. The magnetic dipole moment is thus the expectation value of the $z$-component of the magnetic dipole operator, leading to the definition of $\bm{\mu}=g\bm{I}\mu_{\rm N}$, where $I$ is the non-zero spin of the nucleus, $\mu_{\rm N}$ is the nuclear magneton (the conventional unit of nuclear magnetic moments). 

As mentioned previously, in the nuclear shell-model picture, the ground-state spin of an odd-$A$ nucleus is primarily determined by that of its unpaired valence nucleon. Therefore, the single-particle magnetic moment for a valence nucleon occupying a particular shell-model orbital with total angular momentum $j=l\pm\frac{1}{2}$ is defined as~\cite{Neyens2003,Heyde1994}:
\begin{equation}
\mu (j) = \begin{cases} 
 g_{l}(j-\frac{1}{2})+\frac{1}{2}g_{s}, & j=l+\frac{1}{2}\\
\frac{j}{j+1}\left[I+\frac{3}{2}-\frac{1}{2}g_{s}\right], & j=l-\frac{1}{2}
\end{cases} 
\label{eq24}
\end{equation}
Using the known $g-$factors for free protons and neutrons, the single-particle magnetic moment for a valence proton or neutron in a shell-model orbital can be calculated in a straightforward manner. These are known as Schmidt moments. 

The fact is that experimental magnetic dipole moments of odd-$A$ (odd-$N$, even-$Z$ or odd-$Z$, even-$N$) nuclei deviate from the calculated Schmidt values, indicates that behavior of valence nucleons in a real nucleus differs to that of free nucleons. Two possible causes for such deviations are configuration mixing with other states and meson-exchange currents~(MEC) in the two-body magnetic moment operator~\cite{Arima1954,Castel1990}. To compensate for the influence of these two effects, ``effective” proton and neutron $g$-factors are often employed to calculate effective single-particle magnetic moments. For example, the dashed-and-dotted lines in Fig.~\ref{fig:fig3.2} are the calculated “effective” Schmidt moments using effective $g$-factors for a valence proton in $\pi d_{3/2}$ ($\pi s_{1/2}$) orbital (left panel) and in $\pi p_{3/2}$ ($\pi f_{5/2}$) orbital (right panel).

The magnetic moment of a composite nuclear state, e.g. a nuclear state described by the coupling between a valence proton and neutron, can be calculated through the addition theorem, using the known magnetic moments of its constituents as~\cite{Castel1990,Heyde1994}:
\begin{equation}
\mu(J) = \frac{J}{2}\left[\frac{\mu_{\pi}}{J_{\pi}}+\frac{\mu_{\nu}}{J_{\nu}}+\left(\frac{\mu_{\pi}}{J_{\pi}}-\frac{\mu_{\nu}}{J_{\nu}}\right)\frac{J_{\pi}(J_{\pi}+1)-J_{\nu}(J_{\nu}+1)}{J(J+1)}\right].
\label{eq25}
\end{equation}
For a largely deformed nucleus, calculation of its magnetic dipole moment requires accounting for both its collective and single-particle motions. Details and examples on the addition of nuclear magnetic moment of deformed systems can be found in Refs.~\cite{Castel1990,Heyde1994,Neyens2003} and will not be introduced here.

We see that nuclear magnetic moments are very sensitive to which shell-model orbitals are occupied by the valence nucleons (or holes). They therefore act as an excellent probe of the configuration and purity of the nuclear wavefunction. In addition, the nuclear magnetic moment offers a stringent test of nuclear shell-model calculations employing different effective interactions. Here, we will again take the potassium ($Z=19$) and copper ($Z=29$) isotopes as examples to demonstrate the sensitivity of the nuclear magnetic moment to the structural changes in nuclear shell ordering. In a naive shell-model picture, the ground-state magnetic moments of the odd-$A$ potassium and copper isotopes should remain constant, irrespective of their even-$N$ neutron number. We already know from the above discussion on nuclear spins in Fig.~\ref{fig:fig3.1} that the energy levels of 1/2$^+$ and 3/2$^+$ states in potassium isotopes and of 3/2$^-$ and 5/2$^-$ in copper isotopes migrate due to monopole interaction between the valence proton and increasing number of neutrons. As represented in Fig.~\ref{fig:fig3.2}, through comparing experimental magnetic moments to calculated effective Schmidt values of different specific proton orbitals, the dominant configuration involved in the ground-state wave function of each isotope can be elucidated. This provides compelling evidence that the shell-model orbitals in which the unpaired proton resides change with neutron number~\cite{K-radii1982,K-moment2013,Cu-moment2009-2, Cu-moment2010-1,SM-Otsuks2005}. 

\subsubsection{Nuclear quadrupole moments\label{sec:Q}}

The electric quadrupole moment arises from the non-spherical charge distribution of an atomic nucleus. It therefore carries information on nuclear deformation and acts as an indicator of collective effects. In the impulse approximation, the nuclear electric quadrupole moment can be expressed as the expectation value of the operator $\bm{Q}=e_i\sum\limits_{i=1}(3z_i^2-r_i^2)$ within the angular momentum state $|IM\rangle$ with the maximal projection $I=M$ or of $E2$ operator $\bm{Q}=e_i\sum\limits_{i=1}\sqrt{16\pi/5}r_i^2Y_2^0(\theta_i,\varphi_i)$, resulting in the expression~\cite{Heyde1994,Neyens2003}
\begin{equation}
Q_{\rm s}=\left(\frac{I(2I-1)}{2(I+1)(2I+3)(I+1)}\right)^{1/2}\langle I||\bm{Q}||I\rangle.
\label{eq26}
\end{equation}
From this, a zero, positive and negative quadrupole moment indicates a spherical, prolate and oblate charge distribution, respectively.

\begin{figure*}[t!]
\begin{center}
\includegraphics[width=0.70\textwidth]{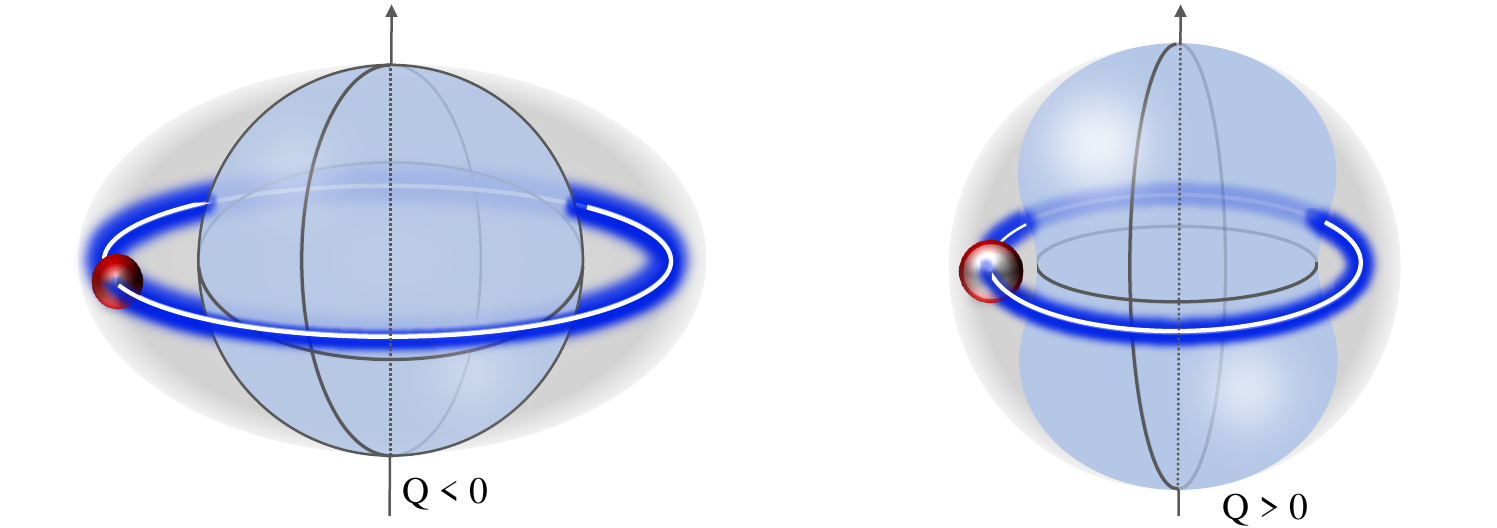}
\caption{\label{fig:fig3.3}\footnotesize{ (Left) A schematic for the nuclear shape with a single particle outside of a doubly magic nucleus, which induces an oblate charge distribution and thus a negative quadrupole moment. (Right) For a single hole within a doubly magic nucleus, a prolate charge distribution with a positive quadrupole moment is predicted.}}
\end{center}
\end{figure*}

In the extreme single-particle case, where a nucleon moves outside of a spherical core with no interaction between the two, the single-particle quadrupole moment can be calculated as~\cite{Castel1990,Heyde1994,Neyens2003}:
\begin{equation}
Q_{\rm{s.p.}}=-e\frac{2j-1}{2(j+1)}\langle r_j^2\rangle
\label{eq27}
\end{equation}
where $\langle r_j^2\rangle$ is the mean-square charge radius of the nucleon in the orbital $j$. Thus, for a single nucleon outside of a doubly magic nucleus, a negative quadrupole moment appears, which can be interpreted as oblate core polarization, as pictorially represented in the left panel of Fig.~\ref{fig:fig3.3}. The same model can be used for the case where a nucleon is missing (hole) from a closed shell, leading to a positive quadrupole moment $Q=-Q_{\rm{s.p.}}$, as shown in the right panel of Fig.~\ref{fig:fig3.3}. 
When $n$ particles are in the orbital $j$, the seniority scheme in the simple shell-model picture predicts 
\begin{equation}
Q_{\rm{s}}=\left(1-\frac{2n-2}{2j-1}\right) Q_{\rm{s.p}}(j)
\label{eq28}
\end{equation}
resulting in a linear increase of $Q_{\rm{s}}$ with $n$~\cite{Heyde1994,Castel1990}. 
From the above, we expect a vanishing quadrupole moment in the single-neutron case, as a neutron does not carry an overall charge. 
The quadrupole moment for a single-neutron outside of a closed shell however is finite due to the additional charge induced by the presence of other nucleons in the nucleus and the resulting interaction between them. Therefore, effective charges for protons and neutrons are often introduced, as $e^\pi_{\rm eff}=1+e_p$ and $e^\nu_{\rm eff}=e_n$ in analogy with the effective $g$-factors for magnetic moments. Thus, Eq.~\ref{eq27} becomes $Q_{\rm{s.p.}}=-e_{\rm {eff}}[(2j-1)/(2(j+1))]\langle r_j^2\rangle$.
Experimentally determined quadrupole moments however only partially agree with that of a single particle (or hole) outside of a spherical core at closed shells. 
In many cases, large values of quadrupole moments are observed, resulting from the motion of a large number of nucleons, for which a collective model needs to be introduced.

It is important to note that $Q_{\rm s}$ in Eq.~\ref{eq26} does not exist for nuclei with spin $I=0,1/2$. This does not necessarily mean that all $I=0,1/2$ nuclei possess a spherical charge distribution. It is the intrinsic quadrupole moment $Q_{\rm intr.}$, classically defined in the body-fixed axis system, that carries the most direct information relating to nuclear deformation. In the strong coupling limit, one can relate the spectroscopic quadrupole moment to the intrinsic quadrupole moment, through the relation:
\begin{equation}
Q_{\rm s}=\frac{3K^2-I(I+1)}{(2I+3)(I+1)}Q_{\rm intr.}
\label{eq29}
\end{equation}
with $K$ being the projection of the total angular momentum $I$ onto the symmetry-axis ($z$) of the deformed system. The nuclear quadrupole deformation parameter $\beta_2$, which represents to what extent the nucleus deviates from a spherical shape, can be linked to the intrinsic quadrupole moment ($Q_{\rm intr.}$)~\cite{Neyens2003}
\begin{equation}
Q_{\rm intr.}=\frac{3}{\sqrt{5\pi}}ZR_0^2\beta_{2}(1+0.36\beta_{2}).
\label{eq30}
\end{equation}
Experimentally, one can not determine the quadrupole deformation of nuclei with $I=0,1/2$ through measuring their spectroscopic quadrupole moment. This can instead be determined from measuring the $B(E2\uparrow)$, the probability of electric quadrupole $\gamma$-transition, but will not be detailed here. If nuclei possessing $I=0,1/2$ are deformed (a non-zero $Q_{\rm intr.}$), this deformation effect will be seen in the nuclear charge radii, which will be discussed in the following.

\subsection{\it Mean-square nuclear charge radii\label{sec:radii}}

The mean-square nuclear charge radius, $\langle r^2\rangle$, is defined as:
\begin{equation}
\langle r^2\rangle=\frac{\int r^2\rho_{\rm{ch}}(r)dV}{\int \rho_{\rm{ch}}(r)dV}.
\label{eq31}
\end{equation}
Here, $\rho_{\rm{ch}}(r)$ is the nuclear charge density distribution function, which can be determined from the nuclear form factor ($F(q)$) provided by electron scattering experiments:
\begin{equation}
F(q)=\frac{4\pi}{qZ}\int_{0}^{\infty}\rho_{\rm{ch}}(r)\sin(qr)rdr, 
\label{eqx}
\end{equation}
with
\begin{equation}
4\pi\int\rho_{\rm{ch}}(r)rdr=Z.
\label{eqy}
\end{equation}
In the above expression, $F(q)$ is the Fourier transform of $\rho_{\rm{ch}}(r)$ in the Born approximation. 
For low momentum transfer ($q$) in electron scattering experiments, Eq.~\ref{eqx} can be written as the expansions
\begin{equation}
F(q) =1-\frac{1}{3!}q^{2}\langle r^2\rangle+\frac{1}{5!}q^{4}\langle r^4\rangle.
\label{eqz}
\end{equation}
Therefore, electron scattering experiments also provide an independent means to access the mean-square charge radius and also offers the possibility to access the fourth radial moment $\langle r^4\rangle$~\cite{4th-radii}, an observable that can provide complementary information related to the nuclear surface~\cite{4th-radii2}.
From extensive studies of stable nuclei with electron scattering experiments~\cite{Fricke2004}, the charge density was found to be nearly constant within the nuclear volume. The trend of mean-square charge radii of stable nuclei was shown to approximately have the form
\begin{equation}
\langle r^2\rangle=\frac{3}{5}(r_{0}A^{1/3})^2.
\label{eq32}
\end{equation}
From this, the rms charge radius $R=\sqrt{\langle r^2\rangle}$ of stable nuclei was seen to scale with the nuclear mass number as $A^{1/3}$. It was found soon afterwards that this relation does not hold for radioactive isotopes, following the application of experimental methods which determined rms charge radii of unstable nuclei.
\begin{figure*}[t!]
\begin{center}
\includegraphics[width=0.99\textwidth]{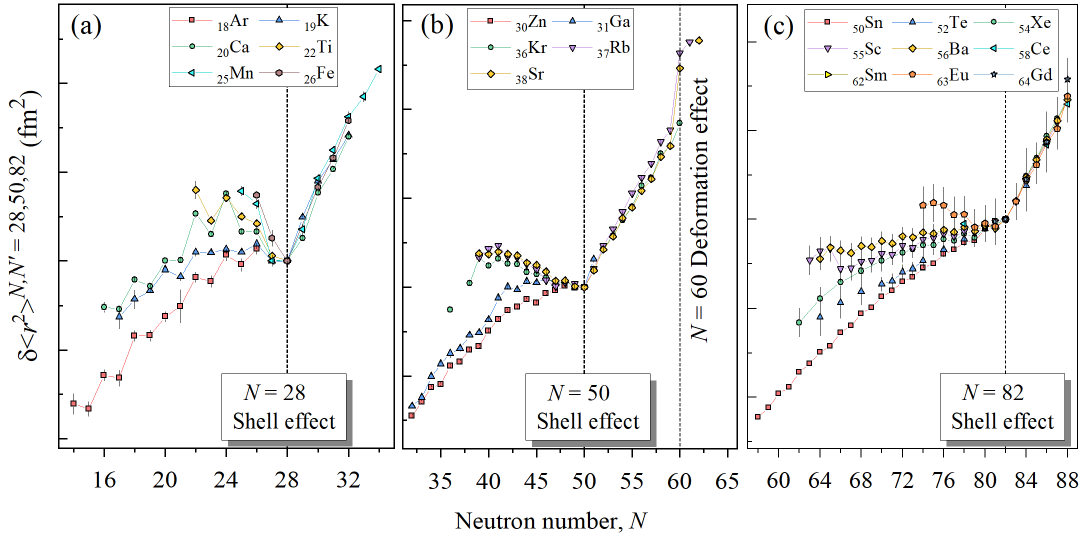}
\caption{\label{fig:fig3.4}\footnotesize{Changes in mean-square charge radii in the calcium ($Z=20$), nickel ($Z=28$), tin ($Z=50$) regions, as a function of the neutron number $N$, with respect to the reference isotope with a magic neutron number. A clear "kink" is commonly observed at neutron magic numbers. Interestingly, the charge radii of neutron-rich nuclei after these neutron magic numbers increase with a similar slope gradient for all proton numbers. This is in contrast to neutron-deficient isotopes, for which the charge radii exhibit a strong dependence upon the proton number.}}
\end{center}
\end{figure*}

\subsubsection{Nuclear charge radii trends}
The variations of rms charge radii of unstable nuclei in a given isotope chain behave very differently from Eq.~\ref{eq32}. Nuclear charge radii for different isotopic chains in the following regions of the nuclear chart are shown in Fig.~\ref{fig:fig3.4}: calcium ($Z=20$), nickel ($Z=28$) and tin ($Z=50$). Here, the changes in mean-square charge radii are defined as:
\begin{equation}
\delta \langle r^2\rangle^{A,A'}=\langle r^2\rangle^{A'}-\langle r^2\rangle^{A}
\label{eq33}
\end{equation}
with $A,A'$ the mass numbers of the two isotopes.
A striking pattern can be noticed in how the nuclear size evolves in all these regions. Namely, the size of the charge distribution appears to universally increase beyond the neutron closed-shells $N=28,50,82$, regardless of the proton number of the isotopes involved. This starkly contrasts the trends observed below the neutron magic numbers where the evolution of charge radii depends strongly upon the atomic number~($Z$). Reproducing and shedding light upon the mechanisms responsible for these phenomena is an ongoing area of theoretical and experimental effort~\cite{Per21,kor22}. 

The complex local variations in nuclear charge radii can be attributed to different aspects of nuclear structure and the interplay between them. In some cases, changes in mean-square charge radii exhibit dramatic changes with respect to neutron number.
Nuclear charge radii therefore act as sensitive probes of different facets of nuclear structure, for instance, shell effects~\cite{Ca-radii2016, K-radii2021}, pairing correlations~\cite{Cu-radii2016, Zn-radii2019}, nuclear deformation and shape staggering/coexistence~\cite{Zn-radii2016, Hg-radii2018}. High-precision charge radii of unstable nuclei have been critical in helping develop state-of-the-art nuclear theory~\cite{K-radii2021,Ca-radii2016,Cu-radii2020}. In addition, the charge radii of certain isotopes can also be related to the properties of the nuclear matter and may also aid in the search for new physics. Examples of these will be given in the following and in Sec.~\ref{physics}.

\subsubsection{For nuclear structure\label{sec:structure}}

$\bullet{\textbf{Benchmark for inter-nucleon interactions and many-body methods}}$

A consistent and accurate microscopic description of nuclear charge radii has been a long-standing challenge for nuclear theory~\cite{cau80,NNLOsat,Ca-radii2016}. As has been stated in Ref.~\cite{NNLOsat} and shown in Fig.1 therein, microscopic calculations, which routinely provide a relatively good description of nuclear binding energies, significantly underestimate nuclear charge radii. The discrepancy between theoretical calculations and experiment increases in magnitude in heavier elements.
As a variety of different many-body methods using the same nuclear interaction gives a similar description of experimental results, as shown recently in Ref.~\cite{Ni-radii2022}, this discrepancy is attributed to be probably due to an incomplete description of the underlying nuclear force. 
This problem was somewhat addressed with an interaction, NNLO$_{\rm{sat}}$, by including nuclear properties of selected light isotopes up to $A=25$ in the optimization process, necessary to determine low-energy coupling constants, of nuclear forces derived from chiral effective field theory ($\chi$EFT)~\cite{NNLOsat}. More recently, the $\Delta$NNLO$_{\rm {GO}}$ interaction, optimized with properties of $A\leq4$ nuclei and nuclear matter, was able to largely reproduce the magnitude of nuclear charge radii from $A=16$ up to $A=132$~\cite{NNLOgo}.\\ 

\hspace{-6mm}$\bullet{\textbf{ Nuclear shell effects}}$

\begin{figure*}[t!]
\begin{center}
\includegraphics[width=0.99\textwidth]{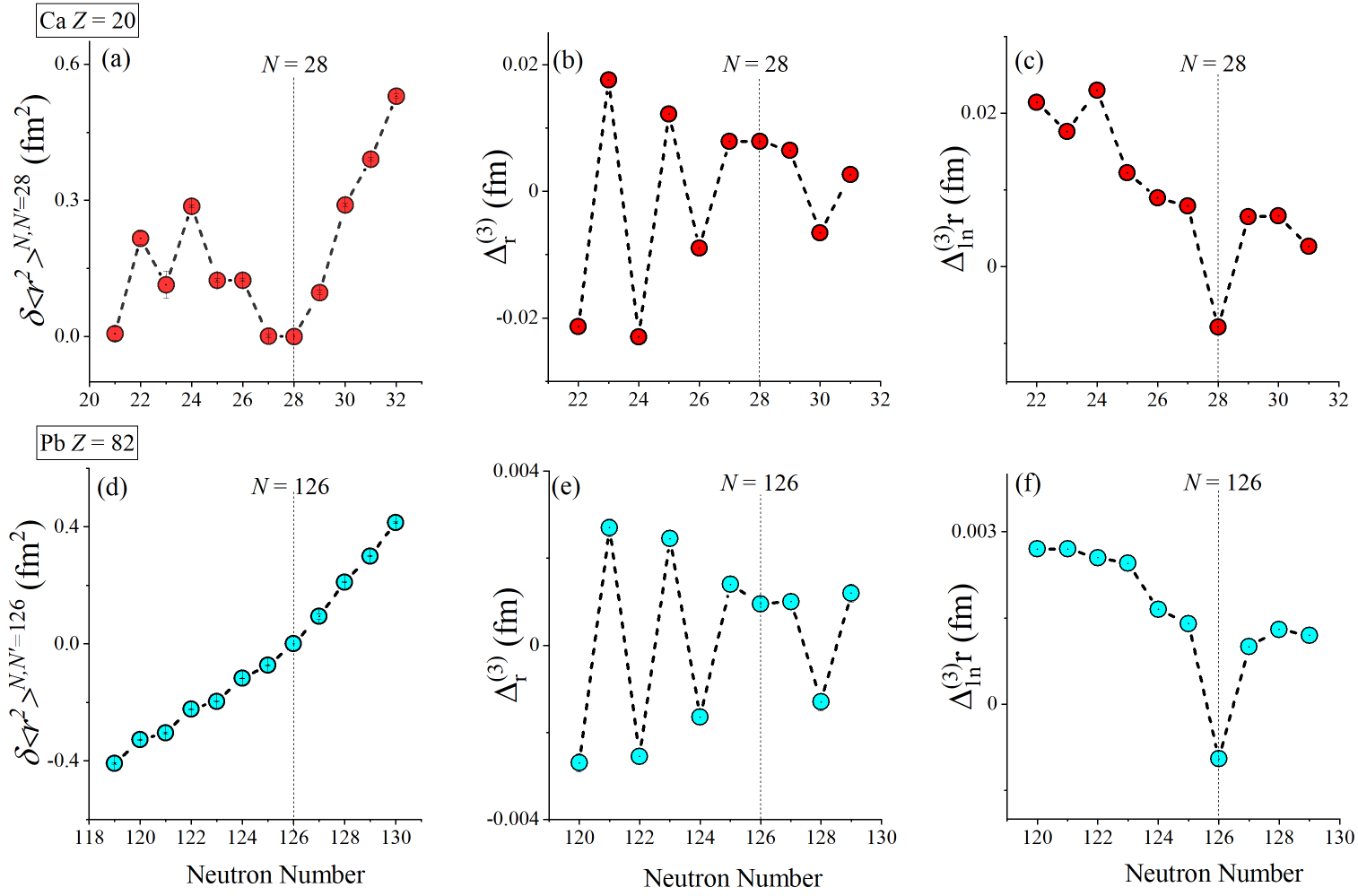}
\caption{\label{fig:fig3.5}\footnotesize{(a,d) Changes in mean-square charge radii of magic calcium and lead isotopes. (b,c,e,f) Odd-even staggering of nuclear charge radii of calcium and lead isotopes, obtained from the three-point radii differences $\Delta^{\rm (3)}r = \frac{1}{2}[r({N+1})-2r({N})+r({N-1})]$ and $\Delta_{\rm 1n}^{\rm (3)}r = \frac{1}{2}(-1)^{N+1}[r({N+1})-2r({N})+r({N-1})]$. Clear discontinuities are present at the magic numbers $N=28$ and 126.}}
\end{center}
\end{figure*}

Local minima in the trends of nuclear charge radii are often observed in isotopes with magic neutron numbers. This is due to the stabilizing effect of the shell closure which is correlated with a reduction of the nuclear radius. As shown in Fig.~\ref{fig:fig3.4} and Fig.~\ref{fig:fig3.5}(a)(d), the effect of neutron shell closures is often seen as an abrupt rise in the charge radii of isotopes after the neutron magic numbers, such as the well-known examples $N = 28, 50, 82$ and 126. Experimental charge radii therefore provide valuable input for characterizing new phenomena in exotic nuclei, such as the suggestion of new magic numbers~\cite{Ca-radii2016,K-radii2021} and the disappearance of conventional ones~\cite{Mg-moment2012,Al-radii2021}.\\

\hspace{-6mm}$\bullet{\textbf{ Pairing correlations}}$

A ubiquitous but intriguing feature of nuclear charge radii in a particular isotopic chain is the pattern of their local variations, known as odd-even staggering (OES). This effect describes the fact that the charge radii of most odd-$N$ isotopes are smaller than the average of their adjacent even-$N$ isotopes~\cite{Zn-radii2019,Cu-radii2020}. This OES effect of charge radii can be phenomenologically explained by the blocking effect of the odd nucleon. As explained in Refs.~\cite{OES-radii,Cu-radii2016,Zn-radii2019}, the additional unpaired neutron in an odd-$N$ isotope blocks a certain orbital and thus suppresses the scattering of neutron pairs to higher orbitals. This in turn reduces proton pair scattering and, as a result, leads to a smaller charge radius in the odd-$N$ isotope.

The OES effect is particularly pronounced in isotopes of calcium between $N=20$ and $N=28$ and is shown in Fig.~\ref{fig:fig3.4}(a) and Fig.~\ref{fig:fig3.5}(a). In other isotopic chains, for example the lead isotopes, this effect is barely visible in the $\delta \langle r^2\rangle$ (Fig.~\ref{fig:fig3.5}(d)). To better visualize and assess the OES effect in nuclear charge radii, a three-point radii difference is defined. This has the form~\cite{DFT-Fayan,Zn-radii2019,Cu-radii2020}:
\begin{equation}
\Delta^{\rm (3)}r = \frac{1}{2}[r({N+1})-2r({N})+r({N-1})].
\label{eq36}
\end{equation}
Figure~\ref{fig:fig3.5}(b) and (e) present the three-point radii differences in calcium and lead isotopes and show an identical pattern for the OES effect. Outlying points at $N=28, 126$ are attributed to effects from shell closures. In some cases, an abnormal OES (or outright inversion) can be caused due to sudden structure changes, for example increased collectivity~\cite{Ga-radii2017,Zn-radii2019,At-radii2019}.

If we modify the three-point radii differences in Eq.~\ref{eq36} by adding $(-1)^{N+1}$~\cite{Ga-radii2017}
\begin{equation}
\Delta_{\rm 1n}^{\rm (3)}r = \frac{1}{2}(-1)^{N+1}[r({N+1})-2r({N})+r({N-1})]
\label{eq37}
\end{equation}
the shell-closure effect can be examined more sensitively. As shown in Fig.~\ref{fig:fig3.5}(c) and (f), the pronounced dip in the $\Delta_{\rm 1n}^{\rm (3)}r$ clearly reflects the presence of shell closures at $N=28, 126$. This approach has been recently adopted to examine the closed-shell effect at the newly proposed $N=32$ magic number based on the charge radii of potassium isotopes~\cite{K-radii2021}. \\

\hspace{-6mm}$\bullet{\textbf{ Nuclear deformation}}$

Nuclear charge radii offer an additional means to probe nuclear deformation even for isotopes with $I=0,1/2$ which do not possess a spectroscopic quadrupole moment.
A notable example of this is the sudden increase in nuclear charge radii at $N=60$ around zirconium~($Z=40$)~\cite{Sr-radii1988,Zr-radii2002-1,Y-radii2007,Nb-radii2009}, which is partially shown in Fig.~\ref{fig:fig3.4}(b) and discussed in detail in the previous edition of this series~\cite{PPNP2016}. The origin of this behaviour was proposed to be due to a nuclear shape transition~\cite{PRL-otsuka2016}.

Assuming that the effects from higher order (e.g. octupole and hexadecapole) deformation can be neglected, the mean-square charge radius of a deformed nucleus, which retains its volume with respect to that of a spherical nucleus, can be expressed as:
\begin{equation}
\langle r^2\rangle=\langle r^2\rangle_{0}(1+\frac{5}{4\pi}\langle \beta_{2}^2\rangle)
\label{eq34}
\end{equation}
where $\beta_{2}$ is the quadrupole deformation parameter~\cite{PR1979}, $\langle r^2\rangle_{0}$ is the mean-square charge radius of the spherical nucleus, which is often calculated from the droplet model. The changes in the mean-square charge radii of two isotopes with mass numbers $A$ and $A'$ can then be written as:
\begin{equation}
\delta\langle r^2\rangle^{A,A'}=\delta\langle r^2\rangle_{0}^{A,A'}+\frac{5}{4\pi}\delta\langle\beta_{2}^2\rangle^{A,A'}\langle r^2\rangle_{0}.
\label{eq35}
\end{equation}
Therefore, information of nuclear deformation can be gained from measurements of $\langle r^2\rangle$ or $\delta\langle r^2\rangle^{A,A'}$. This contains not only the information on the static deformation $\beta_2$, linked to the quadrupole moment in Eq.~\ref{eq30}, but also the dynamic deformation. In particular, nuclear shape changes can be evaluated from their mean-square charge radii based on the Eq.~\ref{eq34} and Eq.~\ref{eq35}. For example, for a nucleus with a spherical ground state and a deformed isomer, the deformation of the isomer can be calculated from the measured isomer shift~($\delta\langle r^2\rangle^{\rm{g,m}}$). This was used previously to demonstrate the phenomena of shape coexistence~\cite{Hg-radii1979,Zn-radii2016}.

\subsubsection{Constraining properties of nuclear matter\label{sec:matter}}
\begin{figure*}[t!]
\begin{center}
\includegraphics[width=0.99\textwidth]{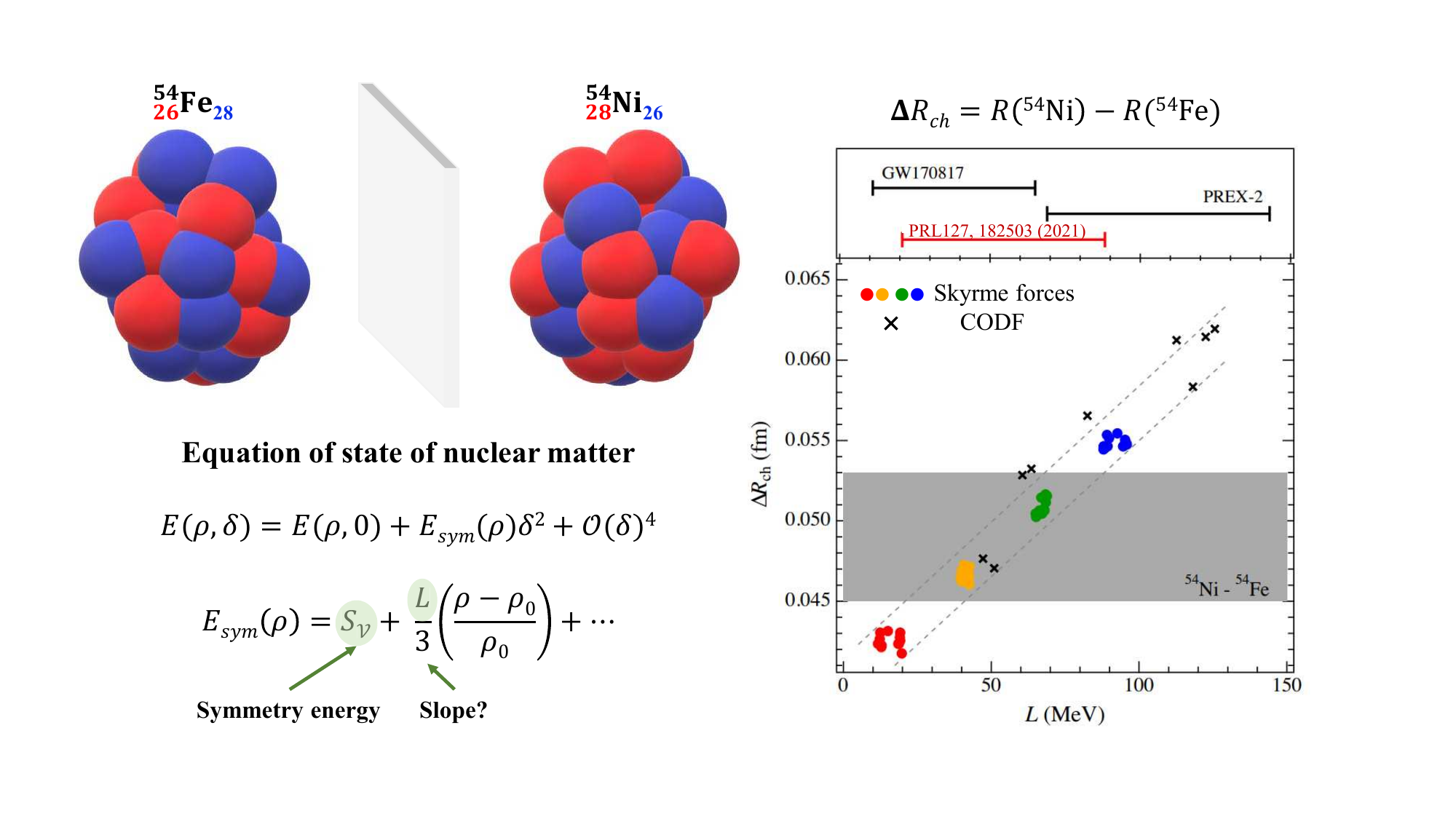}
\caption{\label{fig:fig3.6} Correlation of the charge radii difference between mirror nuclei and the slope of the symmetry energy, $L$, of the equation of state of nuclear matter. The correlation for the mirror pair $^{54}$Fe-$^{54}$N are obtained from mean field calculations with different Skyrme forces and covariant density function theory, which can be used to constrain the $L$. These experiments are complementary to the values extracted from gravitational wave observations (GW170817), and parity-violation electron scattering (PREX-II). The figure has been adapted from Ref.~\cite{Ni-radii2021}. }
\end{center}
\end{figure*}

Recently, it has been shown that nuclear charge radii can provide powerful constraints to the properties of nuclear matter such as the radii of neutron stars~\cite{weak-charge-distributions-of-the-48Ca} and to parameters in the equation of state~\cite{Brown17,Ni-radii2021}. Remarkably, the differences in charge radii of mirror nuclei are strongly correlated with the slope of the symmetry energy ($L$) in the equation of state~\cite{Wan13,Brown17}. An example of the correlation between the charge radii difference of the mirror pair $^{54}$Ni-$^{54}$Fe and $L$, is shown in Fig.~\ref{fig:fig3.6}. The different colors in the figure display the results of calculations using different Skyrme forces. 

By measuring these charge radii differences, strict limits could be set on the slope of the symmetry energy~\cite{Brown17,Ni-radii2021}. This approach is complementary to neutron-skin measurements performed by the parity-violation electron scattering experiment PREX-II at Thomas Jefferson
National Accelerator Facility (J-LAB) in the US~\cite{PREXII}. 

Only a handful of charge radii measurements are available for mirror nuclei. Moreover, large theoretical uncertainties on the structure of proton-rich nuclei can limit the uncertainty of extracted $L$ \cite{rei22}. 
The nuclear charge radius of the neutron-deficient $^{54}$Ni isotope recently measured at BECOLA setup of MSU~\cite{Ni-radii2021}, together with the known radius of its mirror isotope $^{54}$Fe, have provided a new constraint of $21\leq L\leq88$ MeV. This allowed neutron-skin thickness of $^{48}$Ca isotope to be predicted. The constraint on $L$ is consistent with that deduced from the binary neutron star merger GW170817, as shown in Fig.~\ref{fig:fig3.6}, indicating a soft neutron matter equation of state. 

\begin{figure*}[t!]
\begin{center}
\includegraphics[width=0.99\textwidth]{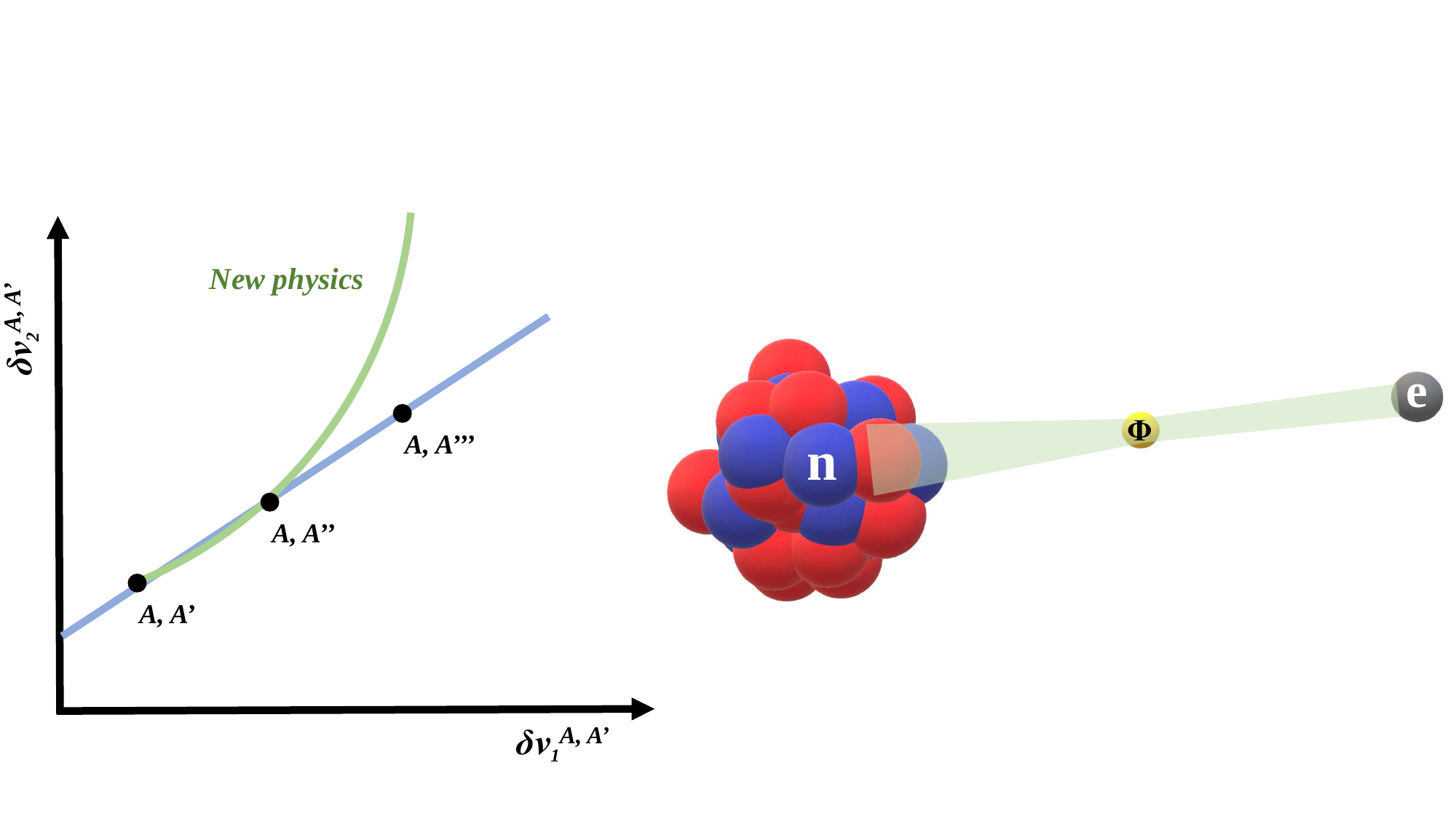}
\caption{\label{fig:fig3.7}Isotope shifts for dark matter searches. A non-linearity observed in the relation between isotope-shift measurements of two atomic transitions can be used to constrain the existence of new forces. A sketch of a non-linearity between the isotopes of masses $A$, $A^{'}$, and $A^{''}$, is exaggerated with the green curve.}
\end{center}
\end{figure*}
\subsubsection{Isotope shifts for new physics searches\label{sec:newphysics}}
In addition to their marked importance for nuclear structure, high-precision measurements of isotope shifts have proven to be powerful probes in the study of diverse physical phenomena, ranging from low-energy to high-energy nuclear physics~\cite{Ca-radii2016,Geb15,Saf18,Ber18}. Isotope shift measurements are sensitive to changes in the nuclear density distribution~\cite{Ca-radii2016,PPNP2016,Ca-radii2019,Sn-radii2019}. However, if the nuclear properties are fully understood, a precise measurement of these energy shifts can reveal subtle details of the electron-nucleus interaction~\cite{Ber18,Sta18,Del17}. Recently, several theoretical proposals have highlighted that high-precision isotopes shift measurements have the potential to constrain the existence of new hypothetical forces in addition to possible dark matter particles with unprecedented sensitivity~\cite{Saf18,Ber18,Sta18,Del17,Del17b,Fru17,Fla18}. This procedure is illustrated in Fig.~\ref{fig:fig3.7}. Based on Eq.~\ref{eq21}, there exists a linear relation at first order between isotope shifts and the changes in the mean-square charge radii. Thus, the isotope shifts measured in two different atomic transitions, $\delta \nu$ and $\delta \nu'$, follow a linear relation. This two-dimensional plot ($\delta \nu$ vs $\delta \nu'$) is also referred as a King plot~\cite{king}. A non-linearity observed in a King plot indicates the need to include additional terms in Eq.~\ref{eq21}, such as those in Eq.~\ref{eq17}. These may eventually be attributed to a new force between electrons and neutrons, or higher-order nuclear effects~\cite{4th-radii}.
Searches for a King-plot non-linearity have stimulated considerable developments in high-precision experimental techniques, which are now able to achieve sub-kHz precision~\cite{Geb15,Bra19,Man19}. Such experimental progress is not only promising for the search of new physics beyond the Standard Model of particle physics, but could also offer a means to access elusive nuclear observables such as the higher-order radial moment, $\langle r^4 \rangle$, and the nuclear dipole polarizability, $\alpha_{\rm D}$~\cite{Pap16,Fla18}. Precise measurements of these nuclear properties would have a marked impact on our knowledge of nuclear structure and nuclear matter~\cite{Ros13,Bir17,Zha18,Roc18,Roc18b,4th-radii}. Moreover, our understanding of these nuclear properties is critical for the interpretation of any new physics searches.

\section{Laser spectroscopy techniques\label{sec:hfsmethod}}
\begin{figure*}[t!]
\begin{center}
\includegraphics[width=0.83\textwidth]{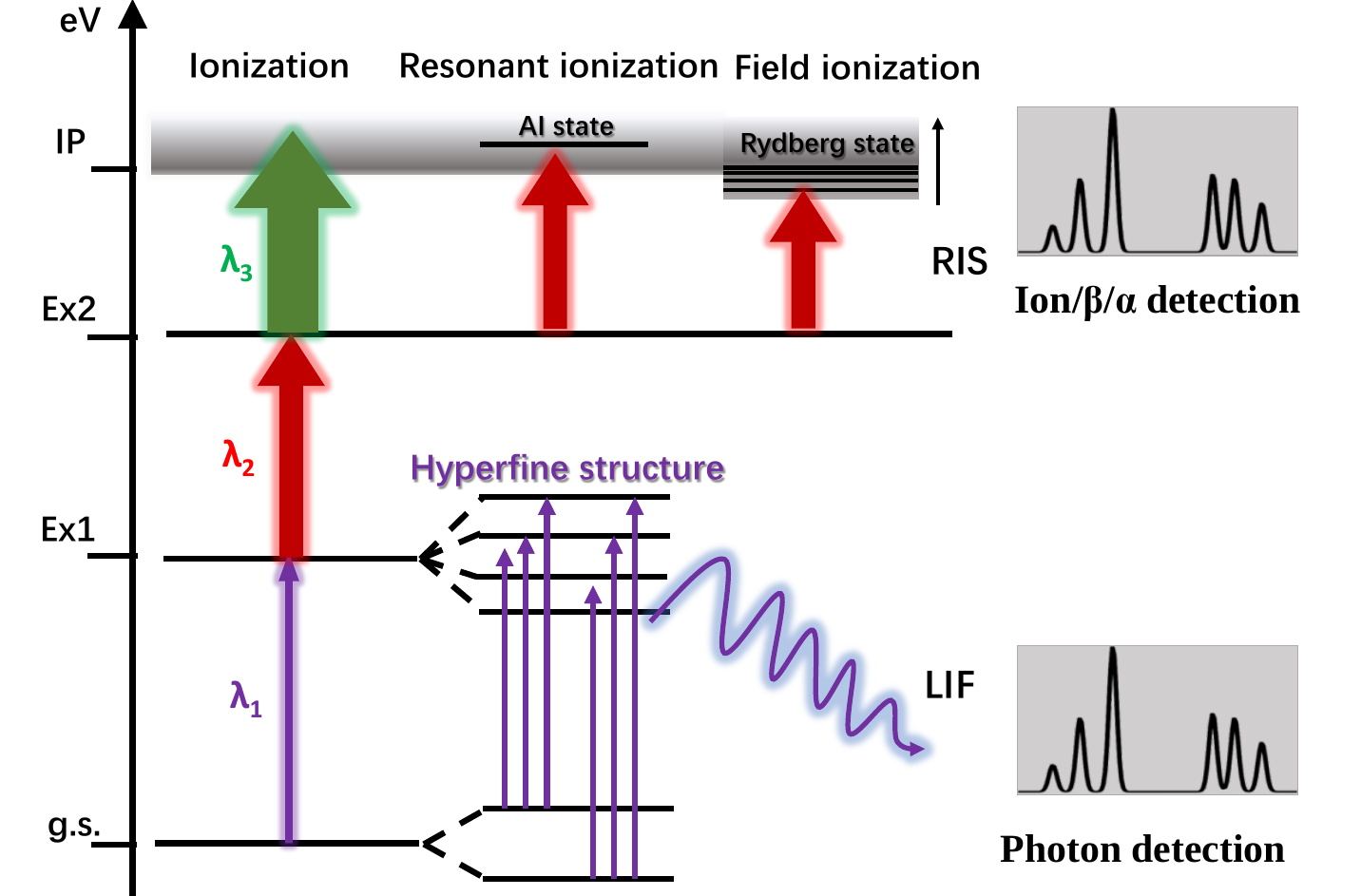}
\caption{\label{fig:fig4.1}\footnotesize{Approaches used to measure the HFS spectrum (see Fig.~\ref{fig:fig2.1}). The HFS spectrum can be measured by laser excitation and fluorescence detection, known as laser-induced fluorescence (LIF), or by subsequent resonance ionization and ion (or decay) detection, which is known as laser resonance ionization spectroscopy (RIS).}}
\end{center}
\end{figure*}
The structure of atoms is often probed by lasers whereby resonant excitation and/or ionization are used as experimental pathways. As schematically depicted in Fig.~\ref{fig:fig4.1}, atoms or ions of interest are resonantly excited from a lower state (commonly the ground state labelled as g.s.) to an excited state (labelled as Ex1) by a frequency-tunable laser. Detection of optical resonances of hyperfine transitions between the g.s. and Ex1 can be realized by various approaches. In this review, we will mainly focus on two broadly used approaches: laser-induced fluorescence (LIF) and laser resonance ionization spectroscopy (RIS). In the former case, the HFS spectrum can be measured by detecting fluorescent photons emitted from Ex1 as a function of probe laser frequency. The RIS technique involves the sequential absorption of multiple photons to step-wise excite and ionize a particular atom. Measurements of HFS spectrum are then achieved by detecting the positively charged resonant ions, or by detecting their decay products (e.g. $\alpha$-particles or $\beta$-particles) of ionized species, as a function of the frequency of the probe laser. 
Ionization from an excited level can be realized by three different processes shown in Fig.~\ref{fig:fig4.1}: 1) non-resonant ionization - in which the valence electron residing in an excited state (labelled as Ex2) is excited through brute force above the ionization potential non-resonantly; 2) auto-ionization - where the valence electron undergoes further resonant excitation to an auto-ionizing state; 3) Rydberg/field ionization - in which the valence electron is resonantly excited to a high-lying Rydberg state and subsequently ionized by an electric field or thermally.

\begin{figure*}[t!]
\begin{center}
\includegraphics[width=0.90\textwidth]{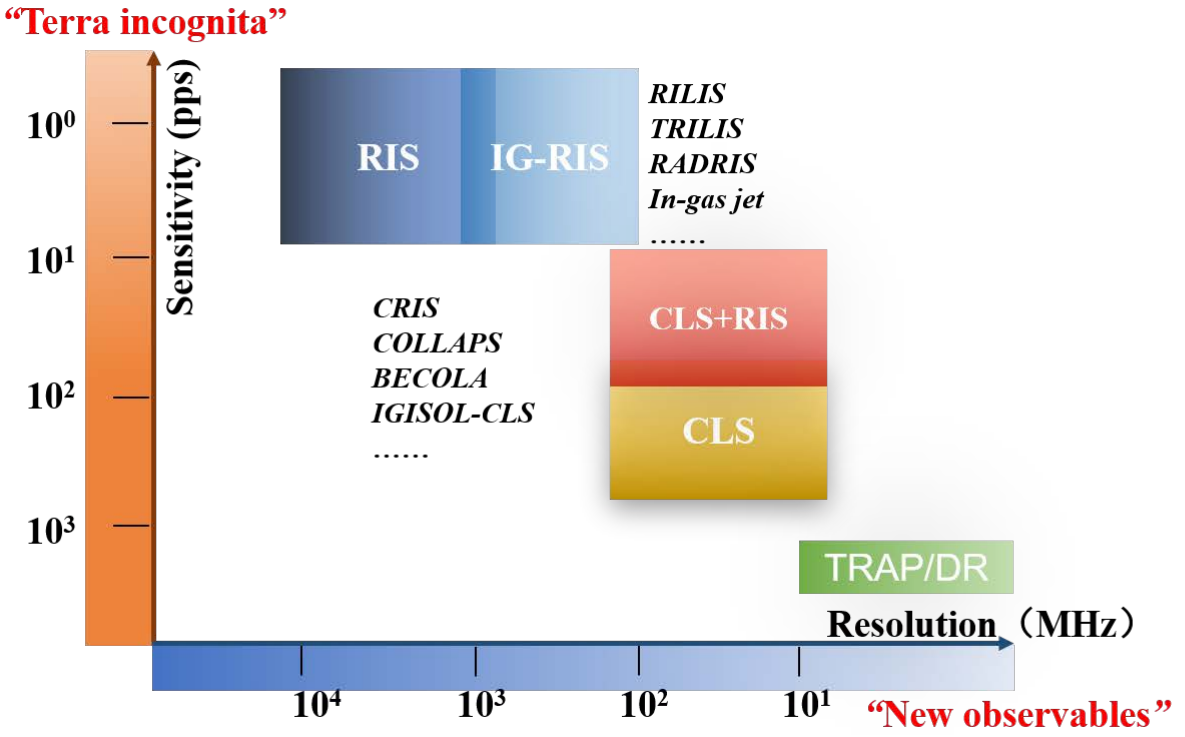}
\caption{\label{fig:fig4.2}\footnotesize{A qualitative presentation of the sensitivity and the resolution that can be achieved for different techniques used for the study of short-lived isotopes: collinear laser spectroscopy, in-source laser spectroscopy, and trap-assisted laser spectroscopy. Details related to each variant can be found in the text of this section~(Sec.~\ref{sec:hfsmethod}).}}
\end{center}
\end{figure*}

Different experimental techniques have been developed for performing HFS spectrum measurements. The study of exotic isotopes at RIB facilities requires highly sensitive techniques as the isotopes of interest are typically produced in small quantities. Three main categories of techniques are commonly applied in these studies: collinear laser spectroscopy, in-source laser spectroscopy, and trap-assisted laser spectroscopy. Each technique possesses its own combination of sensitivity and resolution, which are the two main factors that define the applicability of these techniques in different regions of the nuclear chart. Figure~\ref{fig:fig4.2} presents a qualitative comparison of the sensitivity and spectral resolution of these three classes of methods. In this section, we will give an overall introduction of these three categories of laser spectroscopy techniques, together with their respective characteristics.

\subsection{\it Collinear laser spectroscopy\label{sec:cls}}

The concept of using a collinear geometry of the laser and fast ion beam was proposed in 1970s. This was envisaged as a means to overcome the loss of sensitivity when performing spectroscopy perpendicularly on a collimated thermal atomic beam without sacrificing resolution~\cite{CLS0,CLS01}. This is possible due to the fact that the original energy spread $\delta E$ of ions is preserved during electrostatic acceleration. Thus, the velocity spread ($\delta v$) of an ion beam along its axis of motion is significantly reduced when the ion velocity, $v=\sqrt{2E/m}=\sqrt{2eU/m}$, is increased. This reduction is given by
\begin{equation}
\delta E = \delta(\frac{1}{2}mv^2)=mv\delta v \Rightarrow \delta v=\frac{1}{\sqrt{2mE}}\delta E.
\label{eq38}
\end{equation}
At an acceleration voltage of a few tens of kV (e.g. 30-60~kV), the velocity spread of the ion beam is narrowed by three orders of magnitude compared to a thermal beam. As a result, the linewidth of the Doppler-shifted laser frequency observed by a fast beam is given by:
\begin{equation}
\delta \nu = \nu_{0}\frac{\delta E}{\sqrt{2eUmc^2}}
\label{eq39}
\end{equation}
where $\nu_{0}$ is the rest-frame transition frequency. This ensures an experimental linewidth that is comparable with the typical natural linewidth (few or few tens MHz) of the transition involved.

The collinear laser spectroscopy (CLS) experimental technique was realized for the measurement of HFS spectrum in 1978 by probing the D$_{2}$ line of stable sodium and caesium, demonstrating the potential of the method for high-resolution and high-sensitivity optical measurements~\cite{CLS1} for the time. Soon after, this method was applied to perform measurements of unstable neutron-rich caesium isotopes at the TRIGA reactor at Mainz~\cite{CLS2,CLS2-2}. 
This approach enabled the measurement of the HFS and isotope shifts of unstable isotopes in long isotopic chains after its implementation at the ISOLDE facility at CERN~\cite{COLLAPS1,COLLAPS2} whereby a new era in the study of short-lived isotopes far from stability began. The majority of the isotopes summarized in Fig.~\ref{fig:fig1.2} and Table~\ref{tab:table1} were investigated by CLS experiments, yielding a wealth of data across different mass regions of the nuclear chart, some of which will be detailed in Sec.~\ref{physics}. After continuous developments, many technical innovations have been realized to greatly improve the capabilities of CLS, such as the use of high-quality, bunched ion beams produced by gas-filled radiofrequency quadrupole (RFQ) ion trap (also named as cooler-buncher)~\cite{Hf-radii2002,Zr-radii2002-1,RFQ}, and the combination of CLS with the highly sensitive RIS technique~\cite{CRIS-NIM2013,CRIS-NIM2016,CRIS-NIM2020}. 

Collinear laser spectroscopy, and extensions of it, play an indispensable role in the study of the unstable nuclei today. Setups employing this class of techniques include COLLAPS, CRIS, Versatile Ion-polarized Techniques On-line (VITO) and Multi Ion Reflection Apparatus for Collinear Laser Spectroscopy (MIRACLS) at ISOLDE-CERN~\cite{COLLAPS1, CRIS-NIM2020,sta15,VITO,MIRACLS}, CLS at IGISOL of JYFL~\cite{Y-radii2018,IGISOL-CLS}, BECOLA and RISE at FRIB~\cite{BECOLA}, CFBS at ISAC-TRIUMF~\cite{CFBS}, CLS at BRIF-CIAE~\cite{BRIF-CLS}, CLS at ALTO~\cite{ALTO}, LaSpec at FAIR~\cite{Laspec}, CLS at RISP-ROAN~\cite{RISP-RAON} and at RIKEN-RIBF~\cite{RIKEN-CLS}. These existing and planned CLS setups are summarized in Fig.~\ref{fig:fig1.1} and Table~\ref{tab:table2}, together with the RIB facilities where they are based. In the following, details of the standard CLS technique based on detection of LIF, as well as the collinear resonance ionization spectroscopy (CRIS) technique will be introduced.

\subsubsection{Collinear laser spectroscopy with LIF detection\label{sec:collaps}}
\begin{figure*}[t!]
\begin{center}
\includegraphics[width=0.99\textwidth]{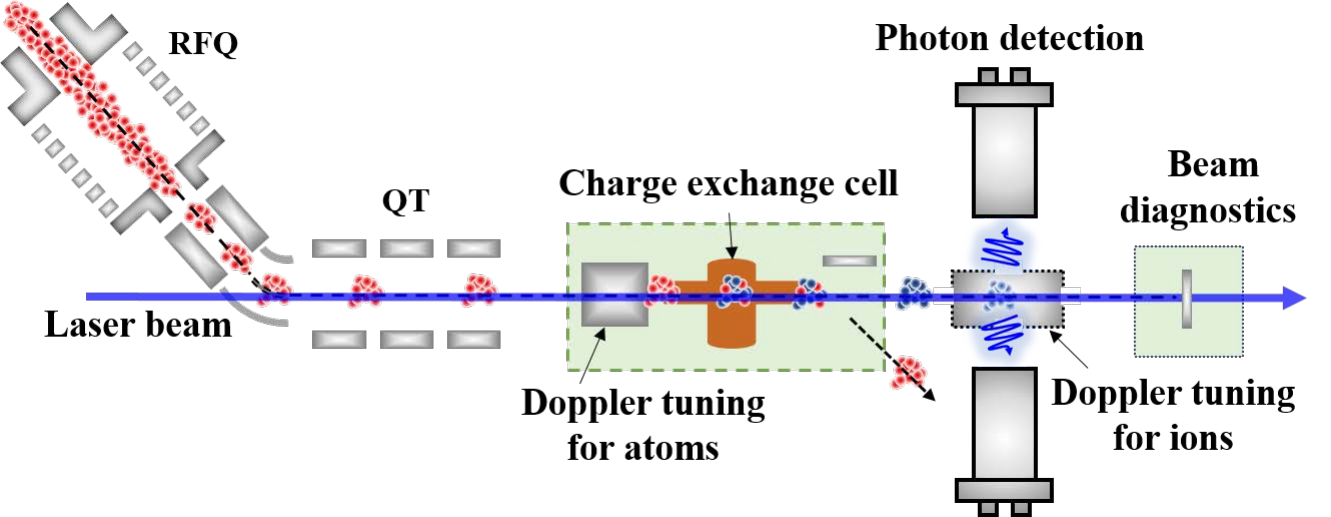}
\caption{\label{fig:fig4.3}\footnotesize{A typical schematic of a collinear laser spectroscopy setup based on detecting laser-induced fluorescence. A bunched ion beam extracted from an RFQ trap is guided into a laser spectroscopy beam line where it is (anti-)collinearly overlapped with a narrow-linewidth continuous-wave laser beam. A quadrupole triplet (QT) lens is used to shape the ion beam profile. The velocity of the ion beam is tuned before being neutralized into atoms in the charge exchange cell. The energetic atoms with a variable velocity are then resonantly excited by the laser beam in the region where the photon detection system are located. When an ionic transition is probed, the velocity of the ion beam is tuned with an electrode in the photon detection region. The emitted photons from the excited atom or ions are collected by the photon detection system as a function of the atom/ion velocity to obtain their HFS spectra. A beam diagnostics system is installed at the end of the beamline to assess the ion beam intensity as well as the neutralization efficiency.}}
\end{center}
\end{figure*}

Figure~\ref{fig:fig4.3} presents an illustration of the main components of current CLS setups in which a laser beam is superimposed along the ion beam path.

\textbf{Charge-exchange process:} 
For the majority of elements, the best-suited transitions for HFS spectrum measurements exist in the neutral atomic system. In these cases, the ion beam needs to be neutralized before it can interact with the laser light. Neutralization of the singly charged ions into atoms is usually realized in-flight by passing them through a charge-exchange cell filled with an alkali-metal vapor. The most commonly used alkali metals are sodium and potassium. At different CLS experimental setups worldwide, two main designs of charge-exchange cell, which differ in their geometry, are used; a horizontal design as shown in Fig.~\ref{fig:fig4.3}~\cite{COLLAPS1,BRIF-CLS,VITO2} and a vertical one~\cite{BECOLA,CFBS}. 

\textbf{Doppler tuning:} 
In a CLS experiment, a convenient alternative to scanning the laser frequency directly to measure the HFS spectrum of a particular species is to instead alter the incoming ion beam velocity. 
In this approach, the laser frequency remains fixed and is stabilized using an external reference cell and/or a high-precision wavelength meter.
Changing the ion beam velocity therefore changes the frequency experienced by the atoms or ions in their rest frame due to the changing Doppler shift. 
In a collinear (or anti-collinear) geometry, the Doppler-shifted laser frequency experienced by the ion or atom beam with a varying total kinetic energy $E$ is
\begin{equation}
\nu=\nu_{0}\times\frac{\sqrt{1-\beta^2}}{{1\pm\beta}}, \\
\beta=\sqrt{1-\frac{m^2c^4}{(E+mc^2)^2}}
\label{eq40}
\end{equation}
where $\nu_{0}$ is the fixed laser frequency and $m$ is the mass of the atom or ion studied. The positive sign~($+$) or negative sign~($-$) in Eq.~\ref{eq40} corresponds to a collinear (parallel direction of laser and ion beams) or anti-collinear (anti-parallel direction of laser and ion beams) geometry, respectively. An advantage of the Doppler-tuning method is that it can result in a more stable and controllable frequency experienced by the atoms/ions when compared to scanning the laser frequency directly. Furthermore, Doppler tuning has a better ability to perform fast HFS spectrum measurements which can save valuable beamtime and to some extent mitigate fluctuations in the detected background stemming from instability in beam-related operating parameters at RIB facilities. As shown in Fig.~\ref{fig:fig4.3}, for experiments on neutral atoms, the velocity of the beam must be tuned by an electrode in the region preceding charge exchange. For ions, Doppler tuning can be realized by varying the voltage applied to an electrode residing directly in the photon detection region. A commonly used electrode design for voltage scanning consists a series of ring electrodes~\cite{COLLAPS1,BECOLA}. In recent years, a crown-shaped electrode was used for voltage scanning of CLS experiments to reduce downstream beam-steering effects at high applied Doppler-tuning voltages~\cite{VITO2,BRIF-CLS}.

\textbf{Photon detection:} 
During a CLS experiment, atoms or ions are resonantly excited in the interaction region (photon detection region in Fig.~\ref{fig:fig4.3}) after absorption of laser photons. Usually, strong transitions with transition rates of $A=10^{7-8}$~s$^{-1}$, corresponding to short excited-state lifetimes (few or few tens of ns), are preferred. The resonantly excited atoms or ions will then spontaneously decay and emit photons. By detecting fluorescent photons as a function of the scanning voltage (and therefore laser frequency), the HFS spectrum of the studied isotope in atomic or ionic form, can be obtained~(e.g. Fig.~\ref{fig:fig2.1}). To optimize the solid angle of the photon collection process, dedicated photon detection systems are required~\cite{K-radii2014-1,COALA,COALA2}, which usually consist of photo-multiplier tubes (PMTs) and telescopes (lens system) for imaging photons from the interaction region onto the photocathode of the PMT. Significant care has to be taken in order to reduce the background experienced by the PMT, which mostly originates from scattered laser light, PMT dark counts, and non-resonant photons from collisional excitations during and after the charge-exchange process. The overall experimental efficiency of CLS experiments depends on several factors, such as the laser-atom(-ion) interaction probability, the transition rate of the chosen optical transition, the solid angle of the detection system, as well as the wavelength-dependent quantum efficiency of the PMT. In the case where CLS experiments are performed on a continuous beam, the general sensitivity of the technique, defined as the required incoming ion rate to detect one laser-induced fluorescent photon, is about $10^{3-4}$~s$^{-1}$ for measuring ionic species~\cite{BECOLA,COALA} and $10^{4-5}$~s$^{-1}$ for measuring atoms~\cite{BECOLA, ALTO,BRIF-CLS}. These limits are mainly determined by the background signal rate caused by scattered laser light.

\textbf{Use of bunched ion beams:} 
The rather limited sensitivity of the CLS technique when employed in studying continuous ion beams limits its applicability to isotopes which are produced with a sufficiently high yield. The introduction of devices which efficiently bunch ion beams for CLS experiments is regarded as possibly the most important innovation in the 2000s~\cite{JPG2010}. Doing this results in a simultaneous increase in the instantaneous fluorescent photon rate and a significant reduction in the laser-related background.

As shown in Fig.~\ref{fig:fig4.3}, continuous ion beams from RIB facilities can be cooled and then accumulated by using an RFQ cooler-buncher~\cite{Hf-radii2002,Zr-radii2002-1,RFQ}. The accumulation time ($T$) used to trap the ions in the RFQ can range from a few to a few hundred milliseconds, depending on several factors e.g. the half-life of the studied unstable isotope and the ion beam intensity.
The ions are released from the RFQ in bunches with a typical temporal width ($\Delta T$) of a few $\mu$s. Only signals that originate when a bunch of atoms or ions is present in the light-collection region are accepted, resulting in a significant suppression of the background from scattered laser light and PMT dark counts. This drastic improvement significantly extended the reach of CLS experiments to more exotic nuclei as demonstrated in Refs.~\cite{Zr-radii2002-1,Cu-moment2009-2,Ga-moment2010-1,K-moment2013, Ca-moment2015}. Additional details on the initial nuclear structure studies undertaken after the introduction of ion beam cooler-bunchers can be found in the reviews~\cite{JPG2010,PPNP2016,JPG2017}.

\textbf{Versatility of CLS:} 
For some specific cases, the standard CLS technique can be modified or extended into variations that enable a higher sensitivity and/or a higher-precision measurements. These variations of CLS are briefly summarized in the following:
\begin{itemize}
\item [1)] \uline{Collisional ionization detection}~\cite{Kr-moment1995,Kr-radii1996} \\
The combination of the conventional CLS method and collisional ionization detection can be used for higher-sensitivity measurements. This approach was used for the study of exotic krypton isotopes, reaching a sensitivity of 10$^3$ atoms/s even with a continuous ion beam delivered at  ISOLDE-CERN~\cite{Kr-moment1995,Kr-radii1996}.  

\item [2)] \uline{Collinear and anti-collinear CLS}~\cite{Be-radii2009,Be-radii2012,8B-CLS} \\
As discussed in Sec.~\ref{sec:IS} and shown in Fig.~\ref{fig:fig2.2}, for light-mass nuclei, the field shift, which is sensitive to the changes in mean-square nuclear charge radii, is orders of magnitude smaller than that in heavy nuclei. Therefore, in order to deduce the nuclear charge radii of light nuclei with sufficient precision, isotope shifts need to be measured with a higher precision than what is typical. To achieve this, two independent laser beams with precisely known frequencies in a combined collinear and anti-collinear geometry can be used. This enables the uncertainty resulting from the acceleration voltage to be negated. Concurrently, the absolute frequencies of the lasers can be determined with high precision with a frequency comb. This approach was successfully used for the on-line study of the neutron-halo nucleus $^{11}$Be~\cite{Be-radii2009}, and will be used in the near future to study the proton-halo nucleus $^{8}$B~\cite{8B-CLS}.

\item [3)] \uline{Optical pumping and $\beta$-NMR}~\cite{Mg-moment2008,Mg-moment2009,K-radii2015} \\
Spin-polarization of RIBs can be achieved at CLS setups through optical pumping, enabling the $\beta$-NMR technique to be applied for determining the properties of unstable nuclei. Such experiments have been performed previously at COLLAPS at ISOLDE, CERN~\cite{Mg-moment2008}, BECOLA at NSCL, MSU~\cite{K-radii2015}, and CFBS at ISAC, TRIUMF~\cite{Li-moment2014}. These examples yielded high-precision measurements of spins, moments and radii of the magnesium isotopes, which have provided insights into the \lq island of inversion\rq~\cite{Mg-moment2008,Mg-moment2009}, the quadrupole moments of $^{9,11}$Li isotopes~\cite{Li-moment2014} and the charge radii of $^{36,37}$K~\cite{K-radii2015}. Polarized RIBs also unlock the potential for studying phenomena beyond nuclear structure such as fundamental interactions, material science and life sciences. For this purpose, a dedicated setup (named VITO) has been established at ISOLDE-CERN, in order to perform $\beta$-NMR/NQR experiments using spin-polarized RIBs~\cite{VITO,VITO2}.

\item [4)] \uline{In-cooler optical pumping}~\cite{Mn-moment2016,Mn-radii2016} \\
Optical pumping can also be used to enhance the population of a selected ionic metastable state before it is delivered to a CLS experiment.
This can allow a certain state to be populated from which transitions that are better suited to nuclear properties studies exist. Optically pumping ions in a RFQ cooler-buncher was first demonstrated for a CLS experiment at IGISOL, enabling the measurement of the moments and charge radii of niobium isotopes~\cite{Nb-radii2009}. It was also successfully applied in the ISOLDE cooler-buncher to enhance the population of ionic manganese in a selected metastable state with nuclear quadruple moment sensitivity, allowing for its determination in neutron-rich manganese isotopes at COLLAPS~\cite{Mn-moment2016, Mn-radii2016}.

\item [5)] \uline{ROC}~\cite{ROC-COLLAPS}\\
A method for ultra-sensitive radiation detection of decaying isotopes after optical pumping and state-dependent neutralization (ROC) was developed and validated in the 1990s~\cite{Ca-radii1992-2}. This yielded an improvement in sensitivity able to measure $^{50}$Ca produced at 5$\times$ 10$^{4}$ ions/s (continuous beam) in the presence of an isobar with a rate of around 10$^{8}$ ions/s. In order to study more exotic calcium isotopes, produced at rates of less than 10 ions/s, an improved ROC setup was developed based on a reconfiguration and extension of the COLLAPS experimental apparatus~\cite{ROC-COLLAPS}. The first commissioning of this new setup was performed on $^{51}$Ca and $^{52}$Ca~\cite{ROC-COLLAPS}, displaying a large increase of sensitivity by more than an order of magnitude with respect to a previous CLS experiment~\cite{Ca-radii2016} and the earlier ROC setup~\cite{Ca-radii1992-2}. Further optimization of the setup is ongoing, with the aim to study the neutron-rich $^{53,54}$Ca isotopes.
\end{itemize}

\subsubsection{Collinear resonance ionization spectroscopy\label{sec:cris}}
\begin{figure*}[t!]
\begin{center}
\includegraphics[width=0.99\textwidth]{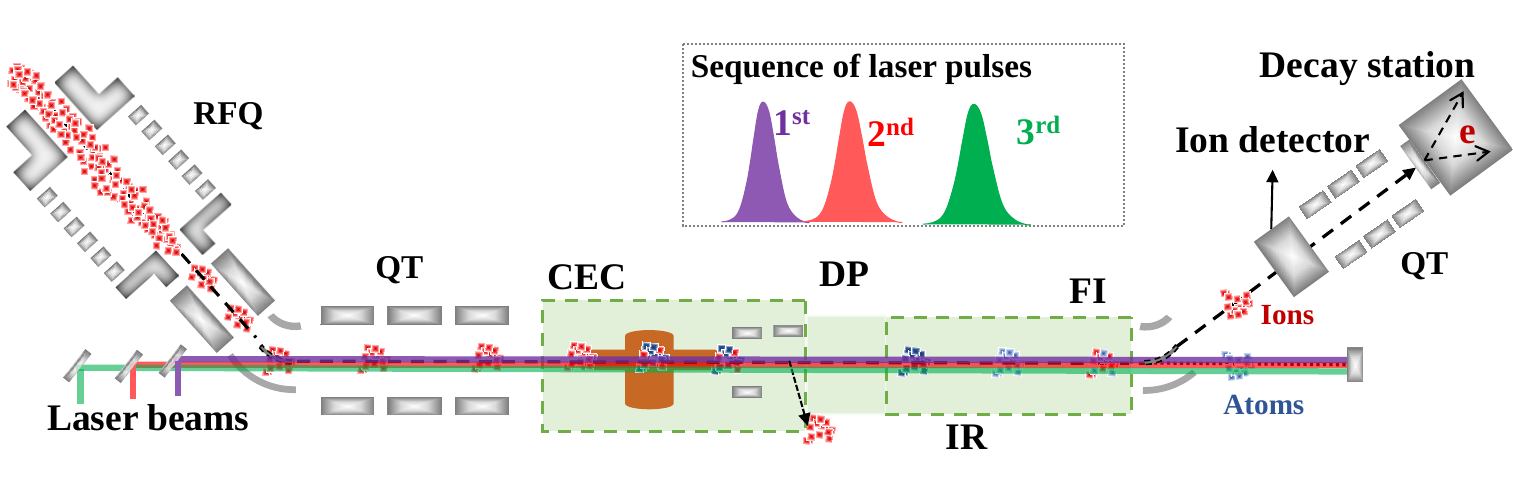}
\caption{\label{fig:fig4.4}\footnotesize{A typical schematic of a collinear resonance ionization spectroscopy~(CRIS) setup. A bunched ion beam extracted from an RFQ trap enters the beamline, where its profile is shaped by a quadrupole triplet (QT) lens. The ion beam is neutralized in a charge-exchange cell (CEC). Non-neutralized ions are deflected away from the neutral beam using an electrostatic deflector after the CEC. Neutral atoms are then overlapped in space and time with multiple pulsed laser beams (figure in the inset) in the interaction region (IR). The atoms in the IR region are resonantly excited and subsequently ionized with multiple pulsed laser beams or resonantly excited to Rydberg states and then field ionized via a field ionizer~(FI). The resulting ions are guided and counted by an ion detector as the function of the first step laser frequency to obtain the HFS spectrum. The resulting ions can also be guided to a decay station, and their decay particles can be detected as a function of the laser frequency. In this way, the HFS spectrum can be measured in a more sensitive way, as detailed in the text of Sec.~\ref{sec:cris}.}}
\end{center}
\end{figure*}

Even with bunched ion beams, the required ion rate for fluorescence-detected CLS experiments is generally around 10$^{3-4}$ ions/s. Studying species produced at rates less than this, especially when delivered alongside large isobaric contaminants, remains highly challenging. RIS techniques combined with ion- or decay-particle detection (Fig.~\ref{fig:fig4.1}), often used by in-source RIS (which will be described in Sec.~\ref{sec:insource}), are orders of magnitude more sensitive allowing isotopes produced at rates below 1 particle per second to be studied~\cite{RILIS2013,Hg-radii2018,Ac-moment2017,No-radii2018}. However, a general shortcoming of these techniques is their low spectral resolution, typically around a few GHz, which limits their applicability to heavy-mass regions and mostly to measurements of magnetic moments and charge radii.

An approach combining the high resolution of CLS and the high sensitivity of RIS, named collinear resonance ionization spectroscopy (CRIS), is a natural extension to CLS. The original proposal for the CRIS method dates back to 1982 which was initially conceived to enable detection of rare, radioactive isotopes with ultra-low abundance~(e.g.$^{26}$Al)~\cite{CRIS0}. The first experiment using the CRIS method was realized at ISOLDE-CERN by Schulz \text{et al.},~\cite{CRIS01} in 1991 to measure the HFS of unstable ytterbium isotopes using two-step excitation to a Rydberg state and subsequent field ionization. Although the efficiency achieved in this experiment was measured to be just 10$^5$~s$^{-1}$, owing to duty-cycle losses performing spectroscopy with pulsed lasers on a continuous atom beam, the promise of the method was confirmed.
Following the introduction of RFQs for ion beam cooling and bunching, a subsequent demonstration of the CRIS technique was performed off-line on stable aluminium~\cite{Kieran-phd}.

The first dedicated CRIS experimental setup was established at ISOLDE-CERN~\cite{CRIS-NIM2013}, following the installation of ISCOOL (ISolde COOLer-buncher)~\cite{RFQ}. 
This setup aimed to study the nuclear properties of isotopes produced in smaller rates, which lay out of reach for conventional CLS and/or are in mass regions inaccessible to in-source RIS. 
The first experimental campaign of CRIS measured magnetic moments and charge radii of neutron-deficient francium isotopes, reaching a high experimental efficiency of 1:100~\cite{Fr-radii2013-cris,Fr-moment2014}. In addition, a decay-assisted CRIS experiment was also realized during the same period, allowing an unambiguous determination of the ground- and isomeric-state HFS in $^{202,204}$Fr~\cite{Fr-radii2014-1}. These initial experiments were already covered in the 2016 review~\cite{PPNP2016}. 
During these campaigns, a broadband pulsed Ti:sapphire laser was used, resulting in a low spectral resolution~\cite{Fr-radii2013-cris,Fr-moment2014}. This was mainly due to the lack of available narrow-linewidth pulsed lasers and that using continuous-wave (cw) lasers would induce optical pumping as the neutral francium atoms transit between the charge-exchange cell and ionization region. The latter would result in a significantly reduced experimental sensitivity~\cite{Ivan-phd}. These problems were solved by generating short pulses by \lq chopping' cw light~\cite{Fr-moment2015} or by utilizing an injection-seeded Ti:sapphire laser system~\cite{Laser-injec}. These systems enabled high-resolution HFS spectra to be obtained with a comparable linewidth to those measured with conventional CLS~\cite{Fr-moment2015, Cu-moment2017,Fr-moment2017}.

A schematic of CRIS is depicted in Fig.~\ref{fig:fig4.4}. Ion bunches with a temporal length of a few $\mu$s are delivered into the CRIS setup usually every 10~ms (determined by the laser system used with lowest repetition rate) and neutralized in-flight using a potassium or sodium vapor in the charge-exchange cell. The non-neutralized ions are deflected away from the neutral atoms using an electrostatic deflector plate located afterwards. Neutral atoms populating long-lived high-lying states after charge exchange can be ionized and removed from the neutral beam with the same deflector. The remaining neutral beam is then overlapped in space and synchronized in time with multiple pulsed laser beams within the 1.2-m long interaction region. To reduce non-resonant collision ionization of the neutral beam, this region is maintained at ultra-high vacuum ($\sim10^{-10}$~mbar). In the interaction region, laser resonant excitation and subsequent ionization occurs. As shown in Fig.~\ref{fig:fig4.1}, in most cases, it is the first transition step that is probed by a high-resolution laser system. Resulting ions are then guided towards an ion detector and counted as a function of laser frequency allowing the HFS to be measured. The relative timings of the laser pulses (see the inset of Fig.~\ref{fig:fig4.4}) are not only essential for optimal ionization efficiency (as is the case for in-source RIS), but can also have a profound influence on the linewidth and lineshape of the observed resonances in the scanning transition. These effects were systematically investigated in Refs.~\cite{Fr-moment2015,CRIS-PRA2017,K-radii2019,CRIS-NIM2020-2}.

After years of development, the CRIS technique~\cite{CRIS-NIM2016,CRIS-NIM2020} is now a proven method for making measurements that simultaneously achieve a high resolution and sensitivity. This has allowed it to routinely measure isotopes produced at rates of a few tens or hundreds per second across most regions in the nuclear chart.

\textbf{\uline{Decay-tagged CLS spectroscopy}:} For many cases, the sensitivity of CRIS can be further enhanced by implementing the decay-tagging approach~\cite{K-radii2021}. The feasibility of $\alpha$-decay-tagged CRIS method was demonstrated previously~\cite{Fr-moment2014,Fr-moment2015}. More recently, in order to study very challenging cases, such as $^{52}$K, the $\beta$-tagged CRIS method was employed~\cite{K-radii2021}. As shown in Fig.~\ref{fig:fig4.4}, a decay station can be placed at the end of the CRIS setup. Resonant laser ions produced in the interaction region are implanted into a thin aluminum plate. The emitted $\beta$-particles are then counted by the surrounding detectors as a function of the laser frequency to obtain a HFS spectrum. This development was motivated by the intense stable contaminant $^{52}$Cr which prevents a conventional CRIS measurement. As isobaric contaminants are often stable or long-lived, they will not produce a significant background when using $\beta$-counting. A dedicated decay station with the possibility of combining $\beta$- and $\gamma$-detection is under development. This will enable measurement of short-lived isotopes produced at even smaller rates in addition to performing decay studies on pure isomeric beams produced through CRIS.

\textbf{\uline{Field ionization}:} The high sensitivity of CRIS is due in part to the favorable characteristics of direct ion (or decay) detection. There however exists some cases where the non-collisional background rate from intense isobaric contaminants hinders or outright prevents a successful measurement.
Field ionization following multi-step resonant excitation to Rydberg states as shown in the Fig.~\ref{fig:fig4.1}, has shown potential to further enhance the experimental sensitivity of CRIS~\cite{CRIS01,CRIS-FI}. The benefits of using this approach are two-fold. First, it circumvents the requirement for a high-power laser needed for efficient non-resonant ionization. Second, it can prevent collisional ions originating from the region preceding the field ionization unit from being detected, reducing the background further~\cite{CRIS01,CRIS-FI}.

The improved combination of sensitivity and resolution of CRIS enabled a pioneering experimental program to study the short-lived radioactive molecules (e.g. RaF)~\cite{Gar20,Udre21}. This promises to open up new opportunities for research in nuclear structure, also in fundamental symmetries and astrophysics, as will be detailed in Sec.~\ref{sec:Molecular}. 

\subsection{\it In-source resonance ionization for spectroscopy and ion production\label{sec:insource}}

The exceptional efficiency and selectivity of the resonance ionization process has given rise to a range of techniques which utilize it both as a spectroscopic method and for radioactive ion beam production.
A review of the applications of resonance ionization in the field of nuclear physics is given in Ref.~\cite{RIS2012}.
A class of ultra-sensitive methods, which generally involve performing resonance ionization spectroscopy of radioactive atoms in regions close to where they are produced, have been developed, known as in-source RIS techniques.
These methods have proven to be extremely sensitive, allowing radioactive species produced at rates far below 1 ion per second to be measured~\cite{Hg-radii2018,Po-radii2013}.

This section will start with a brief mention of resonance ionization laser ion sources (Sec.~\ref{sec:rilis}) before an overview of in-source RIS techniques where both hot-cavity (Sec.~\ref{sec:hot-cavity}) and in-gas variants (e.g. in-gas cell in Sec.~\ref{sec:kul1} and in-gas jet in Sec.~\ref{sec:kul2}) will be discussed.
Finally, another resonance ionization-based technique, RADRIS, used for spectroscopy of the heavy actinide (e.g. nobelium), will be shown in Sec.~\ref{sec:RADRIS}.

\subsubsection{Resonance ionization laser ion sources\label{sec:rilis}}
The unique combination of high efficiency and selectivity of resonance ionization motivated the development of ion sources based on the process. 
The first realization of such an ion source was at the IRIS facility in Gatchina where resonantly ionized isotopes of ytterbium, neodynium and holmium were produced~\cite{ALKHAZOV1991400}. 
Shortly after, the Resonance Ionization Laser Ion Source (abbreviated to RILIS) was implemented at ISOLDE-CERN~\cite{1993550} and also at ISAC-TRIUMF, known as TRILIS~\cite{TRILIS}. 
A recent review of the ISOLDE-RILIS is given in Ref.~\cite{RILIS2017}. 
In these examples, the ionization process occurs in an ion source, typically a hot cavity surface source, coupled to a thick target where the nuclear reaction products of interest are produced - known as `hot cavity' RILIS.

In addition to these examples, laser ion sources have been implemented at other RIB facilities which utilize gas cells to catch recoils from other types of nuclear reactions, such as fusion evaporation. An operational example is FURIOS at IGISOL in Jyv\"askyl\"a~\cite{FURIOS}.

A powerful use of laser ion source infrastructure involves employing it for HFS measurements.
By performing spectroscopy on species of interest close to their site of production where they are most abundant, resulting resonant ions can then be delivered to any part of a facility for means of detection. 
Examples include direct ion detection with a Faraday cup or a single ion detector, nuclear decay detection, in addition to detecting ions after mass selection performed in ion traps.

This experimental combination, which is illustrated schematically in Fig~\ref{fig:fig4.2}, enables extremely sensitive measurements that can access isotopes that lie far beyond the reach of standard, fluorescence-detected collinear laser spectroscopy methods.

\subsubsection{Hot-cavity in-source resonance ionization spectroscopy}\label{sec:hot-cavity}
At thick-target ISOL facilities, isotopes of interest are produced through a variety of nuclear reactions resulting from energetic light ions (usually protons) impinging upon a thick target.
The reaction products are stopped within the target, heated to high temperatures (often exceeding 2000$^\circ$C) to aid diffusion and effusion of short-lived species which pass into a transfer line as a vapor. 
Afterwards, they enter an ion source where they undergo ionization.

\begin{figure*}[t!]
\begin{center}
\includegraphics[width=0.99\textwidth]{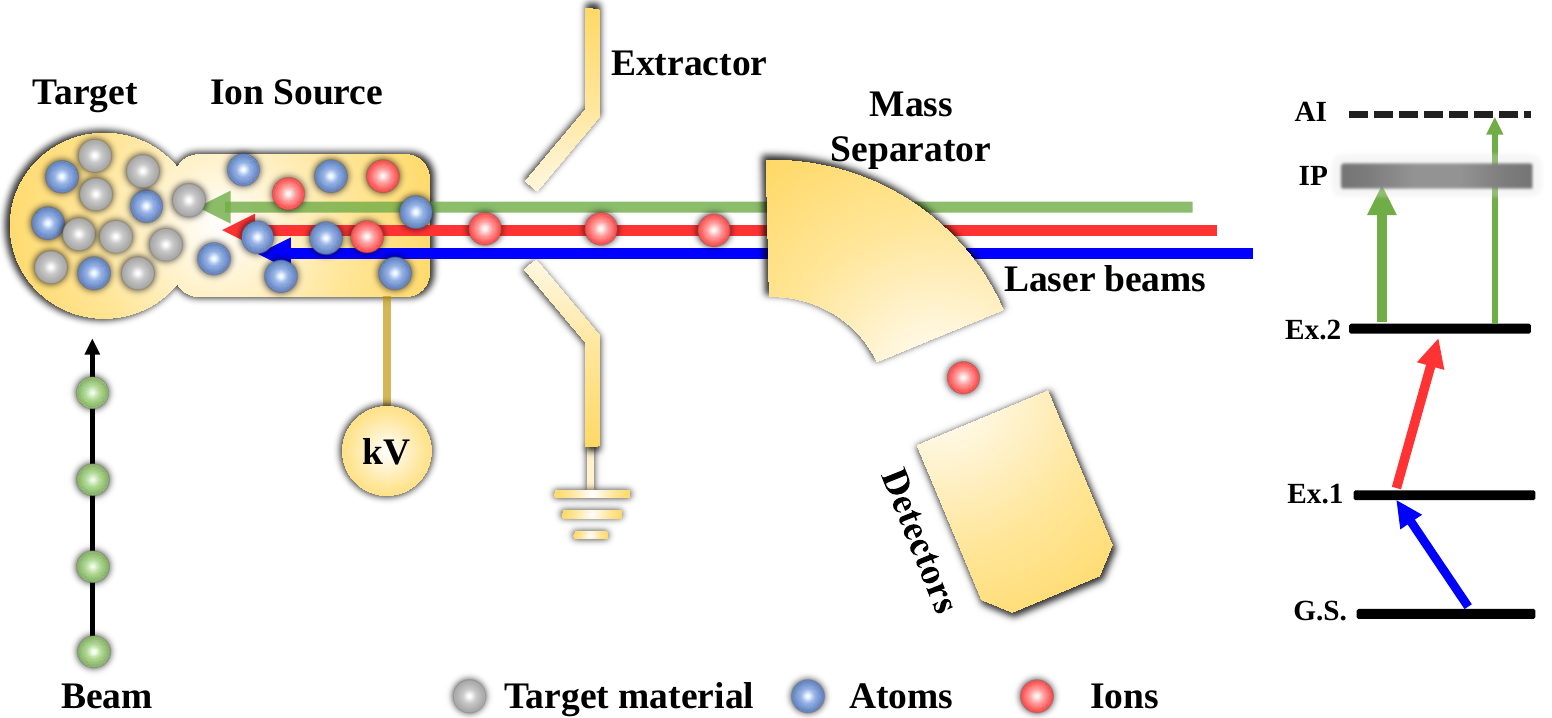}
\caption{\label{fig:fig4.5}\footnotesize{A typical schematic of the in-source laser spectroscopy setup at ISOLDE CERN~\cite{RILIS2013}. The figure is modified based on that in Ref.~\cite{At-IP2013}.}}
\end{center}
\end{figure*}

Figure~\ref{fig:fig4.5} shows a schematic of the hot-cavity in-source spectroscopy technique used at thick-target ISOL facilities.
Resonance ionization is achieved through sending the necessary pulsed laser light into the exit aperture of the ion source where element-selective ionization produces singly charged ions of the element of interest.
To maximize the likelihood that at least one laser-atom interaction occurs as species transit through the transfer line and ion source, a high repetition rate, usually 10~kHz, is required for the pulsed lasers. The ion source environment provides a degree of ion confinement which greatly aids ion survival.
The laser ionized species are electrostatically extracted and then mass-separated and detected as a function of laser frequency.

A key advantage of performing spectroscopy at the source of production is that resulting resonant ions can be detected at any experimental setup situated at the facility. 
For high-yield cases, resonant ions can be detected by simply using a Faraday cup or single ion detector. In species produced at smaller rates and/or in cases where significant isobaric contamination masks the resonant ions of interest, a Multi-Reflection Time-Of-Flight Mass Spectrometer~(MR-ToF—MS) can be utilized.
The ultimate sensitivity of the technique is usually achieved through detecting resonant ions through their characteristic nuclear decay, which in most cases is a unique identifier of a given isotope or isomer.
Another highly sensitive detection method that was recently demonstrated for detecting resonant ions harnessed the exceptional mass-resolving power of the Phase Imaging Ion Cyclotron Resonance technique (PI-ICR)~\cite{Ag-radii2021}.

Utilizing detection methods that can distinguish between different nuclear states in the same isotope for counting of resonant ions, allows for the HFS of each individual state to be ascertained, even in cases where they largely overlap. This is known as decay-assisted laser spectroscopy.
This allows for the determination of frequency region of the HFS spectrum that selectively ionizes a single nuclear state. 
The ions in a selected nuclear state can then be delivered to perform laser-assisted decay spectroscopy.
This symbiotic link resulting from simultaneous nuclear state-selective ionization and detection capabilities has proven to be a powerful experimental tool in the lead region which exhibits rich isomerism~\cite{Hg-radii2018}.

However, the spectral resolution of HFS spectra measured with hot-cavity in-source RIS is limited by the high-temperature environment that is required for efficient extraction of short-lived isotopes at thick-target ISOL facilities.
The full width at half maximum (FWHM) of the resonance peaks in a HFS spectrum resulting from the Doppler broadening of a transition of interest with frequency $f_{0}$ in a given isotope is given by
\begin{equation}
\Delta f_{\rm{FWHM}} = \sqrt{\frac{8kT}{mc^{2}}}f_{0}
\label{eq:Doppler}    
\end{equation}
where $m$ is the mass of the isotope, and $T$ is the temperature of the ion source.
By substituting typical values for $T$ into Eq.~\ref{eq:Doppler}, broadened resonance widths in the GHz range are obtained, depending on $m$ and $f_{0}$.
This is significant enough to completely obscure the hyperfine structure in the majority of elements accessible at thick-target ISOL facilities.

The applicability of hot-cavity in-source RIS for nuclear structure studies is therefore limited to elements which possess atomic transitions that are highly sensitive to nuclear properties.
This is typically the case for heavy elements around and beyond the lead region in addition to some exceptional cases in medium-mass regions such as copper ($Z=29$)~\cite{Cu-moment2011-2} and silver ($Z=47$)~\cite{Ag-radii2021}.
Even in these elements, reliable extraction of nuclear spins and quadrupole moments can be challenging using hot-cavity in-source RIS where the hyperfine structure is often not fully resolved.

However, in elements in which no stable isotopes exist, the extraction of the nuclear observables of interest in most cases relies upon atomic mass- and field-shift factors and/or field gradients calculated with theoretical methods.
The uncertainty stemming from these calculated quantities often dominates the overall uncertainty on the resulting nuclear observables.
Therefore, hot-cavity in-source RIS is able to in some cases provide measurements of nuclear properties with a similar overall uncertainty to collinear laser spectroscopy despite the larger statistical uncertainty on isotope shifts and HFS parameters measured with the technique.

In addition to HFS spectroscopy, hot-cavity in-source RIS is often employed to identify new allowed transitions and/or ionization schemes in elements of interest.
Such studies are sometimes motivated by the requirement of a particular experiment for a beam of a given isotope with a higher specified intensity/purity than what is available currently.
In some cases, it is often employed as an initial means of constructing a detailed electronic fingerprint of an element for which there exists little experimental data.
The large Doppler broadening inherent to hot-cavity in-source RIS does not hinder such efforts in these experiments.

In-source laser ionization has been used for the study of radioactive isotopes present in different ion source types.
The majority of studies to date have utilized the interior volume of a standard surface ion source as the region to perform spectroscopy. 
A recent demonstration of an element-selective resonance ionization mode of an electron-impact ion source at ISOLDE was reported~\cite{VADLIS}, allowing spectroscopy of very neutron-deficient mercury isotopes produced from a molten lead target to be achieved~\cite{Hg-radii2018}.
A later development of this ion source (named Versatile Arc Discharge and Laser Ion Source: VADLIS) modifying the extraction electrode further improved the ion yield from laser resonance ionization~\cite{VADLIS2}.

At thick-target ISOL facilities, surface-ionized isobaric contaminants sometimes prove too intense even for the most sensitive experimental techniques.
By coupling a positive ion repeller and radiofrequency ion guide to the standard surface source, a new ion source type is realized which can be used to strongly suppress surface-ionized isobars.
Laser ionization can still occur on atoms that propagate through the ion guide structure after they exit the surface source.
Resulting resonant ions experience a confining field produced by the ion-guide structure to aid their transport to be mass separated and delivered or detected.
This type of source is named the Laser Ion Source and Trap (abbreviated to LIST) at ISOLDE-CERN~\cite{LIST2015} and the Ion Guide Laser Ion Source (abbreviated to IG-LIS) at ISAC-TRIUMF~\cite{IGLIS-TRIUMF}.
The first on-line application of the LIST at ISOLDE-CERN for spectroscopy was used to measure the neutron-rich $^{217,219}$Po isotopes~\cite{Po-radii2015}.
Surface contamination suppression factors in excess of around 10$^{4}$ have been achieved which come at the cost of around a factor of 20-50 reduction in production yield of the isotope of interest.

The hot-cavity in-source RIS technique has been extensively used on-line at thick-target ISOL facilities such as ISOLDE-CERN in addition to ISAC-TRIUMF and IRIS.
It is also employed at the off-line separator facility RISIKO at JGU Mainz, where many laser and/or ion source developments that are implemented at on-line facilities were first initiated~\cite{Tc-radii2020,Si-2013}.
Recent developments using a hot-cavity catcher at IGISOL, Jyv\"askyl\"a enabled the technique to be used to study down to the very neutron-deficient $^{96}$Ag isotope~\cite{Ag-radii2021}. 

Efforts to expand the capabilities of hot-cavity RIS have been undertaken primarily through improving its resolution by circumventing or even eliminating Doppler broadening.
Achieving a reduction in spectral linewidth yields multiple benefits for the technique. 
It results in a higher precision on all measured isotopes shifts and HFS parameters allowing for more precise measurements of nuclear magnetic moments and charge radii as well as a more reliable extraction of nuclear spins and/or quadrupole moments in more elements.
Furthermore, entire new isotope chains in medium-mass regions become measurable using this high-sensitivity technique.\\

\hspace{-6mm}$\bullet{\textbf{{ Perpendicularly Illuminated LIST}}\label{PI-LIST}}$

One such approach to reduce the spectral linewidth of hot-cavity in-source RIS is to perform measurements where laser light is directed upon the atomic ensemble in a perpendicular geometry, largely circumventing Doppler broadening. 
Accessing the atoms in this manner is possible using the LIST (or IG-LIS) where laser light can be sent through the open space that exists between the rod electrodes used for ion confinement.

The technique, named Perpendicularly Illuminated Laser Ion Source Trap (abbreviated to PI-LIST), was first demonstrated on technetium and promethium isotopes at RISIKO~\cite{Tc-radii2020,Pm-radii2020}.
Using PI-LIST, spectral linewidths of less than 100~MHz were achieved, allowing the HFS of the long-lived $^{97,98,99}$Tc isotopes to be fully resolved.
The improvement in resolution resulting from the perpendicular interaction geometry comes with a loss in efficiency when compared to Doppler-limited collinear ionization in normal LIST operation.
This loss of efficiency is compounded by the additional factor between standard hot-cavity RILIS and LIST mode.

Despite the reduced experimental sensitivity, experimental campaigns at on-line facilities utilizing PI-LIST either directly as a spectroscopic tool or as a means to provide a higher degree of isomer selectivity for downstream experimental setups~\cite{PI-LIST-proposal}, are expected to yield new results in the coming years.\\

\hspace{-6mm}$\bullet{\textbf{ Doppler-free two-photon in-source RIS}}$

Another approach involves eliminating Doppler broadening almost entirely through utilizing atomic transitions where two counter-propagating photons are absorbed simultaneously.
These two-photon transitions are subject to different selection rules and their absorption cross section depends strongly on the laser linewidth and intensity in addition to the proximity of real states existing close to the virtual states through which the excitation occurs.
Two-photon laser spectroscopy was first demonstrated in a hot-cavity ion-source setup on stable silicon isotopes at RISIKO, Mainz~\cite{Si-2013}.
A more recent demonstration was able to perform two-photon laser spectroscopy on stable rubidium at ISOLDE, yielding a linewidth below 100~MHz in the hot-cavity environment~\cite{Rb-2photon-2020}.

\subsubsection{In-gas-cell resonance ionization spectroscopy\label{sec:kul1}}
\begin{figure*}[t!]
\begin{center}
\includegraphics[width=0.99\textwidth]{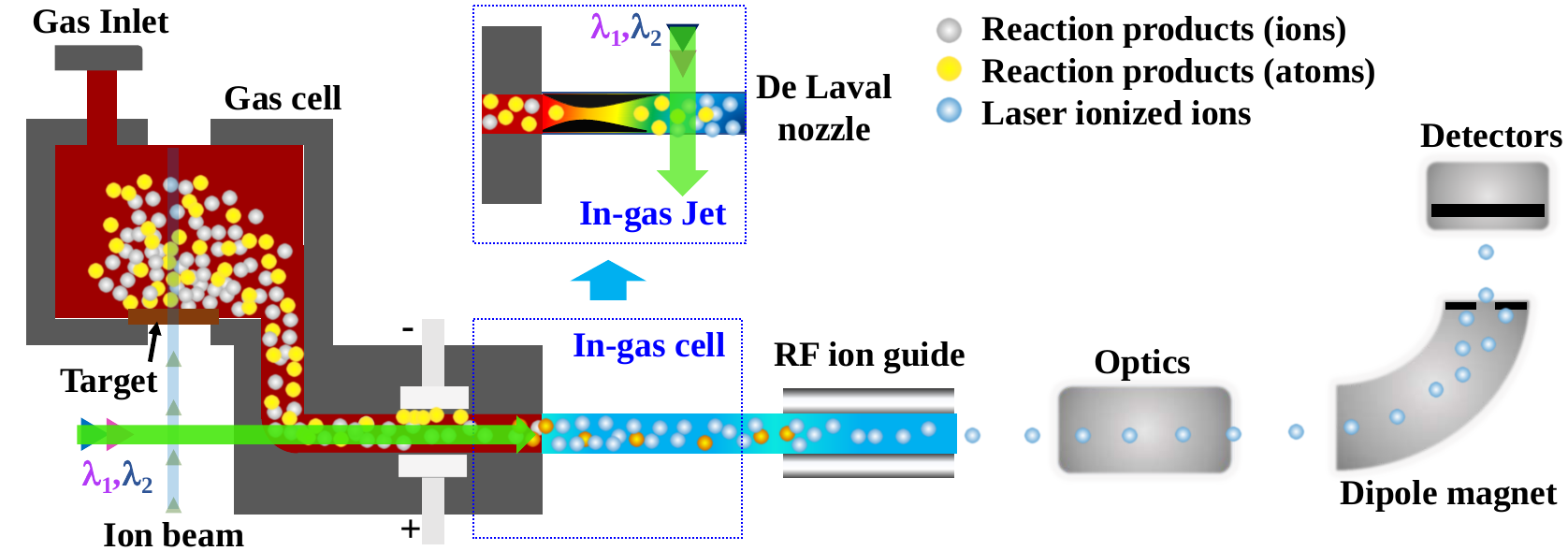}
\caption{\label{fig:fig4.6}\footnotesize{A typical schematic of in-gas-cell and in-gas-jet laser spectroscopy setups. The figure is modified from that in Refs.~\cite{Ac-moment2017}.}}
\end{center}
\end{figure*}

Resonance ionization can also be used for element-selective ion production and spectroscopy at facilities where nuclear recoils produced through a variety of mechanisms are caught within a gas cell.
At these facilities, a range of nuclear reactions can be specifically targeted through consideration of projectile, projectile energy and target composition to produce isotopes of interest.
The resulting radioactive isotopes are initially produced with both a high recoil energy and energy spread. From interacting with a pure, inert buffer gas enveloping the production region, the reaction products can be thermalized and then transported away through flow of the gas. 
They are then directed out of the gas cell where they enter a high-vacuum region and undergo mass separation before being directed to an experimental station.
The key advantage of the facilities operating in such a manner is that they exhibit a significantly lower chemical dependence compared to the thick-target ISOL method where entire regions of the periodic table are rendered practically inaccessible.
Shorter-lived isotopes are also more widely producible at gas-cell facilities owing to the rapid and mostly chemically insensitive extraction time.
By directing laser light into the gas cell and optimizing the gas flow conditions for neutral atom formation and transport, laser resonance ionization can be achieved enabling RIS. Resulting resonant ions can then be guided out of the gas cell and transported with an ion guide, a mass separator, and detected as a function of laser frequency to enable HFS measurements. This technique is referred to as in-gas-cell RIS, a schematic for which is shown in Fig.~\ref{fig:fig4.6}.

The significantly lower operational temperatures of the gas-cell environment results in a reduced thermal broadening of atomic lines measured with in-gas cell RIS.
However, additional mechanisms resulting from collisions with the buffer gas, known as collisional or pressure broadening, increase the experimentally achievable spectral linewidth to the GHz range. Reduction of the pressure in the laser ionization region can reduce the pressure broadening but this, however, also affects the atom formation and transport processes. 
A compromise must therefore be made to balance these factors.
The applicability of the technique is therefore approximately equivalent to hot-cavity in-source RIS but is able to measure the elements that are exclusively available in atomic form at gas-cell facilities.

Recent examples of in-gas-cell RIS studies have taken place at different facilities including IGISOL at Jyv\"askyl\"a~\cite{IGISOL-gascell}, KISS at RIKEN-RIBF~\cite{Ir-radii2020,Os-radii2020} and LISOL~(Leuven Isotope Separator On-Line) in Louvain-La-Neuve~\cite{Ac-moment2017-2}.

\subsubsection{In-gas-jet resonance ionization spectroscopy\label{sec:kul2}}
If the flow of atoms exiting a gas cell is directed through a particular type of nozzle into a high-vacuum region, a collimated supersonic jet can be formed.
The pressure $P$ and temperature $T$ in a supersonic jet of Mach number, $M$ originating from an environment with initial pressure $P_{0}$, temperature $T_{0}$ are given by
\begin{equation}
T = T_{0} [1+\frac{\gamma-1}{2}M^{2}]^{-1},
\label{eq:T}    
\end{equation}
\begin{equation}
P = P_{0} [1+\frac{\gamma-1}{2}M^{2}]^{\frac{-\gamma}{\gamma-1}}
\label{eq:P}    
\end{equation}
where $\gamma$ is the ratio of specific heat capacities equivalent to 5/3 for monoatomic gases.
It can be seen from Eq.~\ref{eq:T} and Eq.~\ref{eq:P} that both the pressure and temperature scale inversely with respect to $M$, leading to a reduction in both temperature- and pressure-related broadening.
Therefore, by performing RIS on atoms within the supersonic jet, a significantly higher spectroscopic resolution can be obtained.
Resulting resonant ions are then transported by an ion guide through a mass separator to be detected. This is known as in-gas-jet RIS, as partly shown in Fig.~\ref{fig:fig4.6}. The first off-line investigations took place at LISOL~\cite{GasJet2013} with the first on-line demonstration performed at the same facility. This experiment was able to measure isotopes of actinium in the vicinity of the $N=126$ shell-closure~\cite{Ac-moment2017,Ac-moment2017-2}.
This initial on-line experiment, where $M\sim6$ was obtained, enabled an improvement in resolution exceeding an order of magnitude, yielding linewidths of around 400~MHz. Importantly, this was achieved without any appreciable loss of efficiency when compared to in-gas-cell studies of the same isotopes.
Ongoing developments at KU Leuven characterizing supersonic jets through imaging the fluorescence of copper atoms seeded within them will allow for further improvements in spectral resolution~\cite{GasJetPRX}.

As this in-gas-jet technique has proven to be both a high-efficiency and high-resolution method, experimental setups are being installed at other RIB facilities, such as the at S$^{3}$ low-energy branch (S3-LEB) under development at SPIRAL2-GANIL~\cite{S3-LEB} and at the in-flight separator SHIP at GSI~\cite{GasJetGSI}, where a variety of heavy elements will be delivered in the future.

\subsubsection{Radiation detected resonance ionization spectroscopy (RADRIS)}\label{sec:RADRIS}
A variant of gas-cell techniques is Radiation Detected Resonance Ionization Spectroscopy (abbreviated to RADRIS), which has been developed at the in-flight separator SHIP of GSI~\cite{No-atom2016}. 
In this approach, energetic (around 40~MeV) ions resulting from fusion evaporation reactions are delivered through a thin entrance window into a gas cell where they thermalize.
A large fraction of the stopped fusion products persist in a positive charged state enabling them to be collected by a catcher filament.
The filament is periodically heated, evaporating captured fusion products where they undergo laser resonance ionization.
Resulting ions are then detected through measuring their characteristic decay as a function of laser frequency.
This technique was successfully used to perform the first measurements in the nobelium ($Z=102$) isotopic chain, which is the heaviest element to be investigated by laser spectroscopy experiments so far~\cite{No-atom2016,No-IP2018,No-radii2018}. In these measurements, an exceptional experimental efficiency of 6.4(1.1)\% was recorded in the case of the even-even $^{254}$No isotope.
More details of this experimental technique and the associated experiments can be found in a recent review dedicated to laser spectroscopy of the actinides~\cite{PPNP2021}.

\subsection{\it In-trap studies\label{sec:trap}}
The technology for laser spectroscopy studies of trapped radioactive isotopes has experienced continuous developments in the last decades, with applications in a diverse range of fields in fundamental science and technology~\cite{Spr97,Beh09,Lu13,Huc17,Mar21}. Compared with collinear laser spectroscopy, measurements with trapped species enable longer interaction times, thus a significantly better precision can be achieved. 
A few selected examples of ongoing developments of trapped neutral and ionic radioactive atoms are summarized below. 

\subsubsection{Neutral traps\label{sec:mot}}

\begin{figure*}[t!]
\begin{center}
\includegraphics[width=0.99\textwidth]{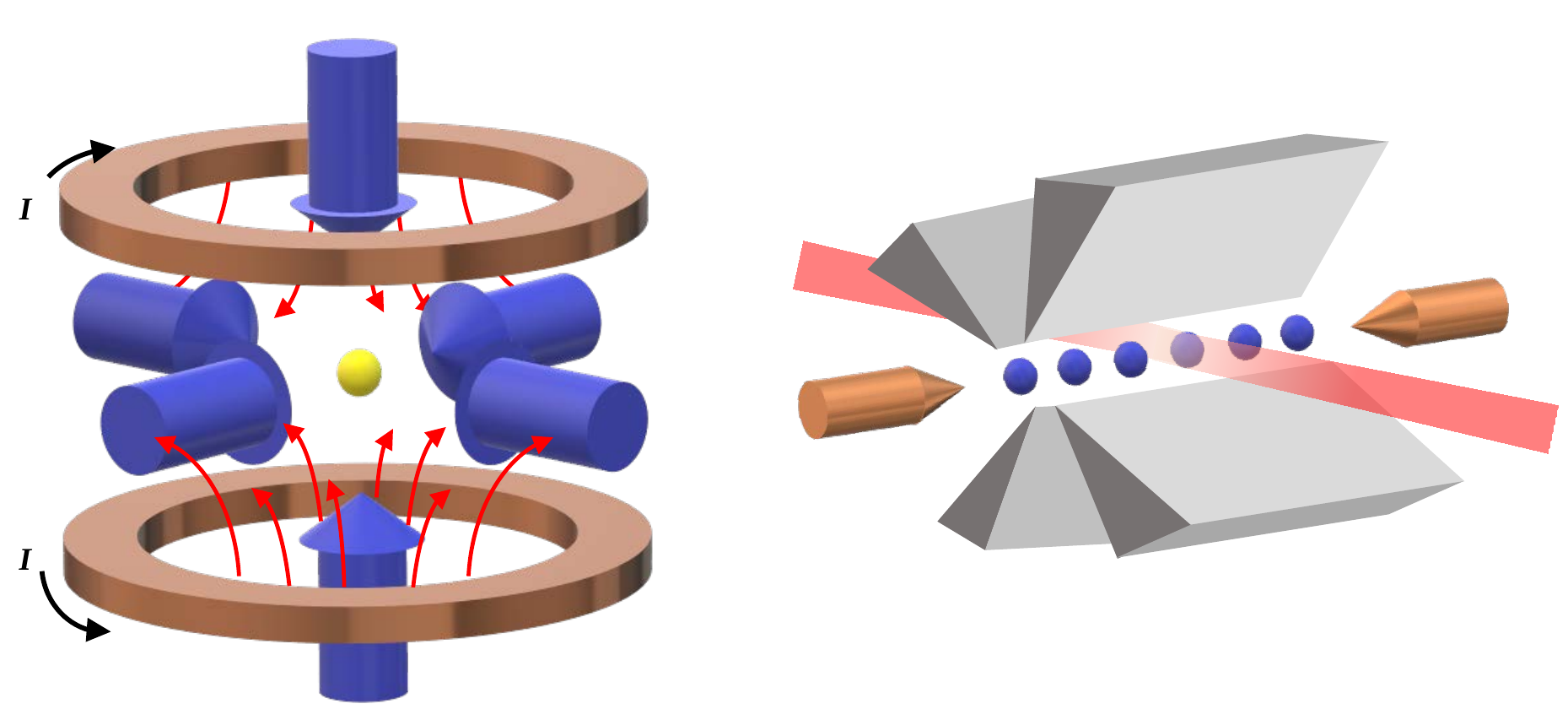}
\caption{\label{fig:fig4.7}\footnotesize{(Left) Cartoon schematic of a magneto-optical trap, MOT. It is used for trapping of neutral species. The current loops are used to produce a magnetic field, and multiple laser beams provide laser cooling of trapped atoms and molecules (shown in yellow color).
(Right) Cartoon schematic of a radiofrequency Paul traps. They are commonly used for trapping of charged species and enable further cooling by multiple laser beams.}}
\end{center}
\end{figure*}
The application of Magneto-Optical Trapping (MOT) to radioactive isotopes has allowed precise control and interrogation of different radioactive atoms. Some examples include $^{6-8}$He~\cite{Lu13}, $^{21}$Na~\cite{Lu94}, $^{37,38}$K~\cite{K-radii1997}, $^{79}$Rb~\cite{Gwi94}, $^{210,221}$Fr~\cite{Sim96,Tan13}, and $^{225}$Ra~\cite{Par15}. A schematic of a MOT trap is shown in Fig.~\ref{fig:fig4.7}(left). Multiple laser beams are used to reduce the kinetic energy of the atoms or molecules within a spatially varying magnetic field, producing an ensemble of cold neutral atoms or molecules. 
Experimental setups, developed at Argonne National Laboratory in the US, have allowed the measurement of the nuclear charge radii of neutron-rich $^{6,8}$He isotopes~\cite{He-radii2004,He-radii2007,Lu13}, with efforts now being dedicated to search for the electric dipole moment of $^{225}$Ra~\cite{Par15}. Recently, experiments have been performed for the $\beta$-recoil correlation measurement from $^{6}$He isotopes trapped a MOT to search for exotic tensor-type contributions to the charged weak current~\cite{6He-MOT}. Ongoing experimental programs at ISAC-TRIUMF in Canada are focused on precise studies of francium isotopes for party-violation studies~\cite{Beh09,Tan13} and $\beta$-decay of potassium isotopes for searches of new physics~\cite{Fen18}.
The HUNTER (Heavy Unseen Neutrinos from Total Energy-momentum Reconstruction) apparatus for radioactive atom trapping and high-resolution decay-product spectrometry, is being developed at Temple University in US to search for sterile neutrinos using precision beta-decay studies of trapped $^{131}$Cs atoms~\cite{Mar21,HUNTER1}. 

A novel technique (called OROCHI) is being developed at RIBF-RIKEN~\cite{RIKEN} and uses superfluid helium (He II) as a host matrix in which energetic radioactive isotopes produced at the facility can be trapped~\cite{OROCHI}. In this technique, the trapped atoms, occupying a well-defined location in the helium, are spin-polarized through optical pumping after repeated laser excitation and de-excitation cycles. Then, laser radiofrequency (RF) or microwave (MW) double resonance spectroscopy of the trapped isotopes can be performed in order to extract their nuclear spins and moments. The proof-of-principle demonstration of this technique was conducted off-line on stable isotopes of several elements (e.g. Rb, Cs, Ag, Au) and the first on-line commissioning experiment was successfully performed on stable and radioactive $^{84-87}$Rb isotopes delivered by RIPS~\cite{OROCHI}. Recent progress in constructing a new fluorescence detection system for OROCHI has been reported in Ref.~\cite{OROCHI2}. This provided an improvement in detection limits of around two orders of magnitude. New RF (and MW) double resonance spectroscopy experiments of radioactive isotopes are expected in the near future, owing to this progress.

\subsubsection{Ion traps\label{sec:iontrap}}
Ion traps are considered workhorses in the study of radioactive isotopes~\cite{dill18}. Gas-filled radiofrequency Paul traps (see Fig.~\ref{fig:fig4.3} and Fig.~\ref{fig:fig4.7}(right)) are commonly used for bunching and cooling of radioactive beams for secondary experiments~\cite{RFQ}. Furthermore, Penning traps and MR-TOF-MS are widely employed for performing highly precise and sensitive mass measurements of short-lived isotopes~\cite{dill18}. Employing the latter as a means of ion detection for laser spectroscopy experiments of radioactive beams has so far allowed them to operate at the sensitivity frontier studying species produced in vanishingly small quantities. An ongoing effort, the MIRACLS experiment at ISOLDE-CERN~\cite{sel21,MIRACLS}, aims to perform CLS on ions circulating in an MR-TOF-MS. The benefits from this stem from the massively increased number of laser-ion interactions attainable in such a scheme. This results in an increased sensitivity when compared to conventional CLS, for species that possess a suitable ionic transition and structure. The proof-of-principle experiment developing this approach has been demonstrated on stable $^{24,26}$Mg~\cite{sel21,MIRACLS}. On-line experiments studying neutron-rich and -deficient magnesium isotopes are planned~\cite{MIRACLS-Mg}.

Precision laser spectroscopy studies in Paul traps are well established for stable ions. These devices have also allowed the properties of long-lived isotopes of radioactive alkaline-earth elements such as $^{133}$Ba$^+$~\cite{Huc17} and $^{226}$Ra$^+$~\cite{Hol22} to be studied in relation to their possible application in quantum information technologies. Performing precision in-trap studies with short-lived isotopes remains a challenge as it demands highly efficient deceleration, trapping, and measurement techniques~\cite{Be-moment2008}. 

\begin{table*}[p!]
\caption{\label{tab:table2}\footnotesize{Summary of laser spectroscopy setups, which are in operation, in the process of planning/commissioning, at both current or new-generation radioactive ion beam facilities worldwide.}}
\vspace{2mm}
\renewcommand*{\arraystretch}{1.2}
\resizebox{\textwidth}{!}{
\begin{tabular}{llllllll}
\hline
Continent &Country & RIB Facility & Setup name  & Type &  Detection method & Status& Refs. \\
\hline
\multirow{11}{*}{\textbf{Europe}} 
 & Belgium     & LISOL                 & --        & RIS      & Ion/Decay         &Terminated & \cite{LISOL}\\
 & Finland     & IGISOL/JYFL                & --      & CLS      & Photon            &Operational &\cite{Y-radii2018,IGISOL-CLS}\\
 & France      & ALTO                  & --     & CLS+PB   & Photon/Decay      &Commissioning&\cite{ALTO}\\
 & France      & SPIRAL2/GANIL                 & LUMIERE & CLS      & Photon/Decay      &Planning&\\
 & France      & SPIRAL2/GANIL                 & S$^{3}$ & RIS      & Ion/Decay         &Commissioning&~\cite{S3-LEB}\\
 & Germany     & TUD\footnotemark[1]   & COALA   & CLS      & Photon            &Operational&\cite{COALA}\\
 & Germany     & MAINZ\footnotemark[1] & RISIKO      & RIS      & Ion               &Operational& ~\cite{Tc-radii2020}\\
 & Germany     & GSI                   & RADRIS  & RIS      & Ion               &Operational&\cite{PPNP2021}\\
 & Germany     & FAIR                  & LaSpec  & CLS+RIS  & Ion/Photon/Decay  &Planning&\cite{Laspec}\\
 & Switzerland & ISOLDE/CERN                & COLLAPS & CLS      & Photon            &Operational&\cite{JPG2017}\\
 & Switzerland & ISOLDE/CERN                & CRIS    & CLS+RIS  & Ion/Decay           &Operational&\cite{CRIS-NIM2020}\\
 & Switzerland & ISOLDE/CERN                 & RILIS   & RIS      & Ion/Decay         &Operational&\cite{RILIS2013}\\
 & Switzerland & ISOLDE/CERN                 & VITO    & CLS+PB   & Photon/Decay      &Operational&\cite{VITO}\\
 & Switzerland & ISOLDE/CERN                 & MIRACLS & CLS+TRAP & Photon            &Commissioning&\cite{MIRACLS}\\
 & Russia & IRIS                  & --      & RIS      & Ion/Decay           &Operational&\cite{IRIS}\\
 & Russia & JINR              & GALS       & RIS      & Ion/Decay           &Commissioning&\cite{GALS}\\
\hdashline
\multirow{5}{*}{\textbf{North America}}  
 & Canada      & ISAC/TRIUMF                & TRILIS  & RIS      & Ion               &Operational&\cite{TRILIS}\\
 & Canada      & ISAC/TRIUMF                & CFBS & CLS+PB   & Photon/Decay      &Operational&\cite{CFBS}\\
 & USA         & NSCL/MSU              & BECOLA  & CLS+PB   & Photon/Decay      &Operational&\cite{BECOLA}\\
 & USA         & FRIB/MSU              & RISE        & CLS+RIS  & Ion/Decay         &Commissioning&\\
 & USA         & CARIBU/ANL                & --        & CLS      & Photon            &Commissioning&~\cite{8B-CLS}\\
 & USA         & MIT\footnotemark[1]   &FRECIOSA & CLS+RIS  & Ion             &Commissioning&\\
\hdashline
\multirow{5}{*}{\textbf{Asia}}  
 & China       & PKU\footnotemark[1]  &  --       & CLS+RIS  & Photon/Ion   &Commissioning&\cite{PKU-CLS} \\
 & China       & BRIF/CIAE                 &  --        & CLS      & Photon      &Operational&\cite{BRIF-CLS} \\
 & China       & HIAF/IMP                 &  --        & CLS+RIS  & Photon/Ion/Decay  &Planning&\\
 & Japan       & RIBF/RIKEN                & KISS     & RIS      & Ion/Decay         &Operational&\cite{KISS}\\
 & Japan       & RIBF/RIKEN                & PALIS    & RIS      & Ion               &Commissioning&\cite{PALIS} \\
 & Japan       & RIBF/RIKEN                & OROCHI   & TRAP     & Photon            &Commissioning &\cite{OROCHI}\\
 & Japan       & RIBF/RIKEN                & --         & CLS      & Photon            &Commissioning &\cite{RIKEN-CLS}\\
 & South Korea & RAON/IBS                 & CLS      & CLS+RIS  & Ion/Photon        &Commissioning&\cite{RISP-RAON} \\
 \hline
\end{tabular}}
\vspace{2mm}
\footnotemark[1]{These are the offline (in-house) laser spectroscopy setups.}
\end{table*}
\subsection{\it Radioactive ion beam facilities\label{sec:RIB}}
\begin{figure*}[t!]
\begin{center}
\includegraphics[width=0.99\textwidth]{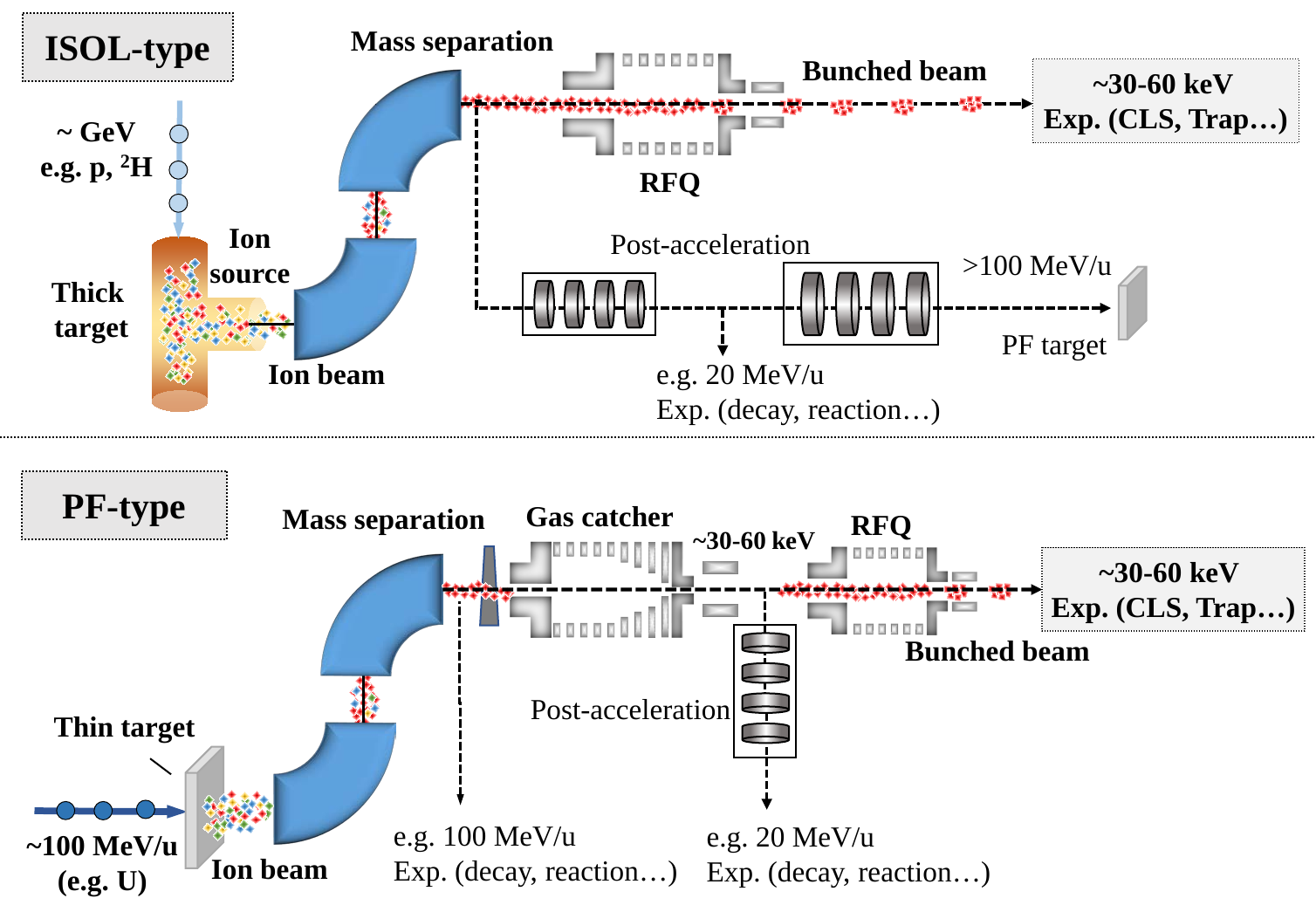}
\caption{\label{fig:fig4.8}\footnotesize{An typical schematic of the (a) ISOL (Isotope Separation On-Line and (b) PF (Projectile Fragmentation) types of radioactive ion beam facilities.}}
\end{center}
\end{figure*}

To study the properties of exotic nuclei, the above-discussed laser spectroscopy techniques need to be integrated into the RIB facilities where these species are produced. There are two main (and complementary) types of RIB facilities that have produced about 3000 unstable isotopes and isomers over the last half a century. These systems form the nuclear chart in its current form today (as partly shown in Fig.~\ref{fig:fig1.2}). These facilities are defined by their operating principle either utilizing Isotope Separation On-Line (ISOL) or in-flight separation (also named Projectile Fragmentation (PF))~\cite{RIB1,RIB2}. Initially, laser spectroscopy experiments of exotic isotopes were almost exclusively carried out at ISOL-type facilities, such as ISOLDE-CERN~\cite{ISOLDE2017}, IGISOL of JYFL~\cite{IGISOL2014} and ISAC-TRIUMF~\cite{ISAC-TRIUMF2014}. This is mainly due to the fact that RIBs are naturally delivered at ISOL-type facilities at suitable energies ($\sim$30-60~keV) and with favorable properties (e.g. low emittance and energy spread). This contrasts RIBs produced through in-flight PF which possess an energy of typically tens or hundreds of MeV per nucleon and with a significant energy spread, preventing direct application of laser spectroscopy techniques. With advanced beam stopping and cooling devices~\cite{Gas-cell,Gas-cell2}, low-energy RIBs suitable for precision experiments, can now be delivered at PF-type RIB facilities (e.g. NSCL at MSU~\cite{BECOLA} and RIBF at RIKEN~\cite{RIKEN,Gas-cell2,RIKEN-CLS}). An impressive series of results from laser spectroscopy experiments operating at PF RIB facilities has already been generated~\cite{Fe-radii2016,Ca-radii2019,Ni-radii2021,Ni-radii2022-2}. Figure~\ref{fig:fig4.8} presents a simplified schematic for ISOL- and PF-type RIB production techniques in addition to how laser spectroscopy setups are implemented at these facilities. In the following, an introduction to RIB production at ISOL and PF facilities as well as the characteristics of the generated RIBs will be given. A mention of a concept proposing to merge both PF and ISOL techniques in a single future RIB facility will also be given. 

\subsubsection{ISOL radioactive ion beam facility\label{sec:ISOL}}
The ISOL technique was initiated in Copenhagen~\cite{ISOL} in the 1950s. It was later expanded and developed at different laboratories where access to high-energy light-ion beams was available. The most prominent example of an ISOL facility is at CERN and named ISOLDE (Isotope Separator On-Line DEvice). This facility has produced radioactive isotopes since 1967 where it initially used the 600-MeV proton beam from the Synchro–Cyclotron~\cite{ISOLDE1971}. Since 1992, it has utilized the 1.4-GeV proton beam from the Proton-Synchrotron Booster~\cite{ISOLDE2017}. 

As shown in Fig.~\ref{fig:fig4.8}, RIBs at ISOL facilities are produced by bombarding a thick target with a high-energy proton beam or with neutrons produced through spallation or fission reactions in a neutron converter. The resulting reaction products diffuse and effuse out of the target, heated to high temperature to aid these processes, and enter a transfer line. They undergo ionization through either surface, plasma or laser ionization in an ion source. The singly charged ions are then extracted and electrostatically accelerated to few tens of keV (e.g. 30-60 keV). The isotope of interest can be selected using a mass separator and delivered to an experimental setup located downstream.

The ISOL method has major advantages for RIB production. The produced beams are often highly intense and are of high quality, possessing a small energy spread and low emittance. 
This makes them suitable for precision low-energy experiments aiming to perform mass measurements in ion traps or HFS measurements using laser spectroscopy. However, the production efficiency of the RIBs depends highly on the chemical properties of the element of interest. A result of this is that some regions of the periodic table are extremely difficult to extract in atomic form. Furthermore, the approach is also limited by the time required for reaction products to be released from the target, which prevents isotopes with half-lives less than a few ms to be delivered before they decay.

For experiments which require bunched ion beams, such as collinear laser spectroscopy experiments, RFQ cooler-bunchers are utilized~\cite{RFQ}. For experiments requiring a high energy, such as those using Coulomb excitation or transfer reactions, charge breeding and subsequent post-acceleration techniques are used.

\subsubsection{PF RIB facilities\label{sec:PF}}
The production of RIBs at PF facilities was first studied conceptually and then demonstrated in the 1970s~\cite{PF,PF2}. In PF reactions, isotopes of interest are produced in flight by impacting stable energetic heavy ions (with energies of few tens or few hundreds of MeV per nucleon) upon a thin target (e.g. beryllium). The resulting fragments propagate along the incident direction of the heavy ions with a similar velocity. As shown in Fig.~\ref{fig:fig4.8}, the fragment of interest can then be selected by a series of mass separators before being delivered to subsequent beam manipulation devices (e.g. storage ring, gas catcher) or experimental devices. 

RIBs produced in this manner are insensitive to the chemical properties of the element of interest and thus more species are available for study at this type of facility. The direct fragmentation of heavy ions upon the thin target also permits shorter-lived species to be produced. However, RIBs delivered at PF facilities are typically plagued by their poor beam quality and possess a large energy spread in addition to a substantial longitudinal and transverse emittance.  
Due to the distinct and complementary advantages of the PF approach, there exists a broad range of scientific opportunities at these facilities. Although not immediately suitable for precision low-energy experiments, the produced beams are suitable for many other kinds of experiments including transfer reactions, in-beam $\gamma$- and decay spectroscopy. In fact, the pioneering measurements of reaction cross sections of exotic nuclei, performed in the 1980s at Lawrence Berkeley Laboratory, which uncovered evidence for the neutron halo of $^{11}$Be and $^{11}$Li~\cite{PF-halo1,PF-halo2}, used RIBs produced through PF. These results helped kickstart and motivate a new era of RIB physics around the world~\cite{PF-halo3}. 

These RIBs can not be directly used for laser spectroscopy experiments. However, trap-related techniques, for example, the RF or MW double-resonance spectroscopy using OROCHI~\cite{OROCHI}, and SLOWRI~\cite{Be-moment2008} trap techniques at RIBF, RIKEN~\cite{RIKEN} have been used for the study of nuclear properties of exotic nuclei at PF facilities. Owing to advances in stopping cell technology (also called gas catchers or gas stoppers) which can condition PF RIBs to become high-quality, low-energy beams~\cite{Gas-cell,Gas-cell2}, as illustrated in the bottom panel of Fig.~\ref{fig:fig4.8}, significantly more opportunities have become available for laser spectroscopy experiments aiming to measure exotic nuclei.

\subsubsection{Next-generation RIB facilities: merging the PF and ISOL techniques\label{sec:PF-ISOL}}

With the currently operational RIB facilities worldwide, including both ISOL and PF facilities, as partly shown in Table~\ref{tab:table2}, about 3000 unstable nuclei have been artificially produced and identified. This number constitutes around one third of the total number of nuclei that are theoretically predicted to be bound. It is expected that more than 1,000 more isotopes could become accessible from the newly operational Facility for Rare Isotope Beams (FRIB) at Michigan State University (MSU), US. This PF facility has the current most powerful heavy ion accelerator and promises to give unprecedented access to exotic dripline nuclei.

To enable the exploration of nuclei in uncharted regions (\textit{terra incognita}) of the nuclear chart such as towards the neutron dripline in medium- and heavy-mass regions, next-generation RIB facilities, such as EURISOL~\cite{EURISOL} and BISOL~\cite{BISOL}, have been proposed. These facilities plan to combine both the ISOL and PF techniques. These facilities aim to produce hitherto unreachable neutron-rich nuclei through PF reactions utilizing a radioactive neutron-rich heavy-ion projectile (e.g. $^{132}$Sn), initially produced through the ISOL method and post-accelerated up to 150 MeV per nucleon. Exotic nuclei approaching the neutron dripline in the medium-mass region are projected to be producible in sufficient intensities to be experimentally investigated using this approach.

\section{Study of exotic nuclei\label{physics}}
Recently, major progress has been achieved in improving the technical capabilities of laser spectroscopy methods. For example, the CRIS experiment has performed highly sensitive and high-resolution HFS measurements~\cite{Cu-radii2020,K-radii2021} in addition to enabling studies of short-lived radioactive molecules \cite{Gar20,Udre21}. The in-gas-jet method has been realized for efficient, high-resolution HFS measurements~\cite{Ac-moment2017}. The heaviest element to be measured by laser spectroscopy experiment is now nobelium ($Z=102$)~\cite{No-atom2016} owing to recent campaigns using the RADRIS technique at GSI. 
The results made possible by these developments have contributed significantly to our understanding of nuclear structure and the nucleon-nucleon interactions that govern it, offering stringent tests that have stimulated rapid advances in atomic and nuclear theory.

In this section, selected results exemplifying the progress achieved in studying exotic nuclei across the nuclear chart since the last review in this series will be highlighted. In addition, a new avenue of research performing laser spectroscopy studies of molecules containing unstable nuclei will be introduced (Sec.~\ref{sec:Molecular}). Although much progress has been made in most mass regions of the nuclear chart, there exist some areas where none or few new measurements have been made with laser spectroscopy techniques (such as that around zirconium $Z=40$). In these cases, we refer to previous reviews~\cite{JPG2010,PPNP2016}.

\subsection{\it Light mass region\label{sec:light}} 
\begin{figure*}[t!]
\begin{center}
\includegraphics[width=0.99\textwidth]{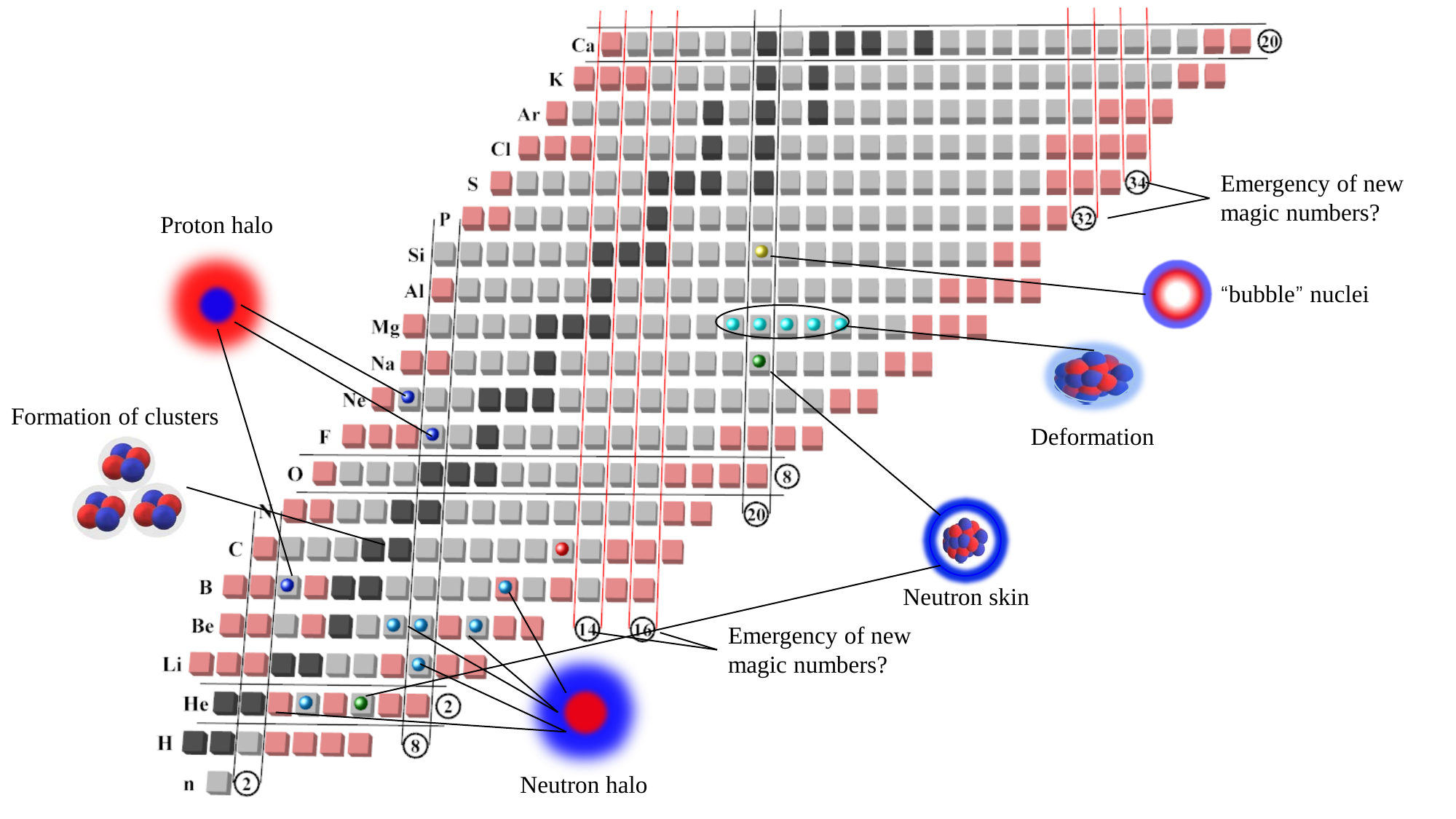}
\caption{\label{fig:fig5.1}\footnotesize{Lightest isotopes of the nuclear chart. Distinct physical phenomena such as halo nuclei, cluster formation, neutron skins, and magic numbers are highlighted for particular isotopes.}}
\end{center}
\end{figure*}
Light nuclei play a critical role in the development of microscopic nuclear theory~\cite{Ep09,Car15}. As illustrated in Fig.~\ref{fig:fig5.1}, by adding or removing just a few nucleons, light nuclear systems can exhibit distinctive nuclear structure phenomena such as the clustering of nucleons, halo nuclei, and the emergence of new shell-structures~\cite{Lu13}. Some of these exotic behaviours have been thoroughly examined in the past by high-precision laser spectroscopy experiments. Examples of this include the confirmation of the neutron halo in neutron-rich helium~($Z=4$)~\cite{He-radii2007}, lithium~($Z=5$)~\cite{Li-radii2006} and beryllium~($Z=6$)~\cite{Be-radii2009,Be-moment2014} isotopes and are detailed in a series of review articles~\cite{JPG2010,Lu13,JPG2017,PPNP2016}. Recent efforts have also been devoted towards realizing a high-precision laser spectroscopy study of the charge radius of $^{8}$B~($Z=5$)~\cite{8B-CLS} at ATLAS, Argonne National Laboratory, US. This isotope is proposed to possess a one-proton halo supported by its large quadrupole moment~\cite{8B-Qmoment} as well as the extended spatial distribution of its loosely bound proton~\cite{8B-halo}.  
In recent years, significant progress has been achieved towards understanding these nuclei in connection with the underlying theory of the strong force, quantum chromodynamics (QCD)~\cite{bea14,Lon18}.
In addition to their importance for nuclear physics, laser spectroscopy measurements of few-body electron systems provide precision tests of atomic physics and many-body QED~\cite{Nor15,Li20}. Some recent highlights include spectroscopy of C$^{3+}$ ions in a storage ring~\cite{Win21}, and the developments towards the study of boron isotopes~\cite{B-radii2019}. 

Owing to developments in chiral effective field theory and many-body methods~\cite{Car15,Lon18,ham20}, nuclear theory can now provide accurate \textit{ab initio} calculations of the properties of isotopes in the vicinity of carbon and oxygen isotopes~\cite{Lon18}. There however is a distinct lack of data about the evolution of nuclear size in this region of the nuclear chart.

\begin{figure*}[h!]
\begin{center}
\includegraphics[width=0.99\textwidth]{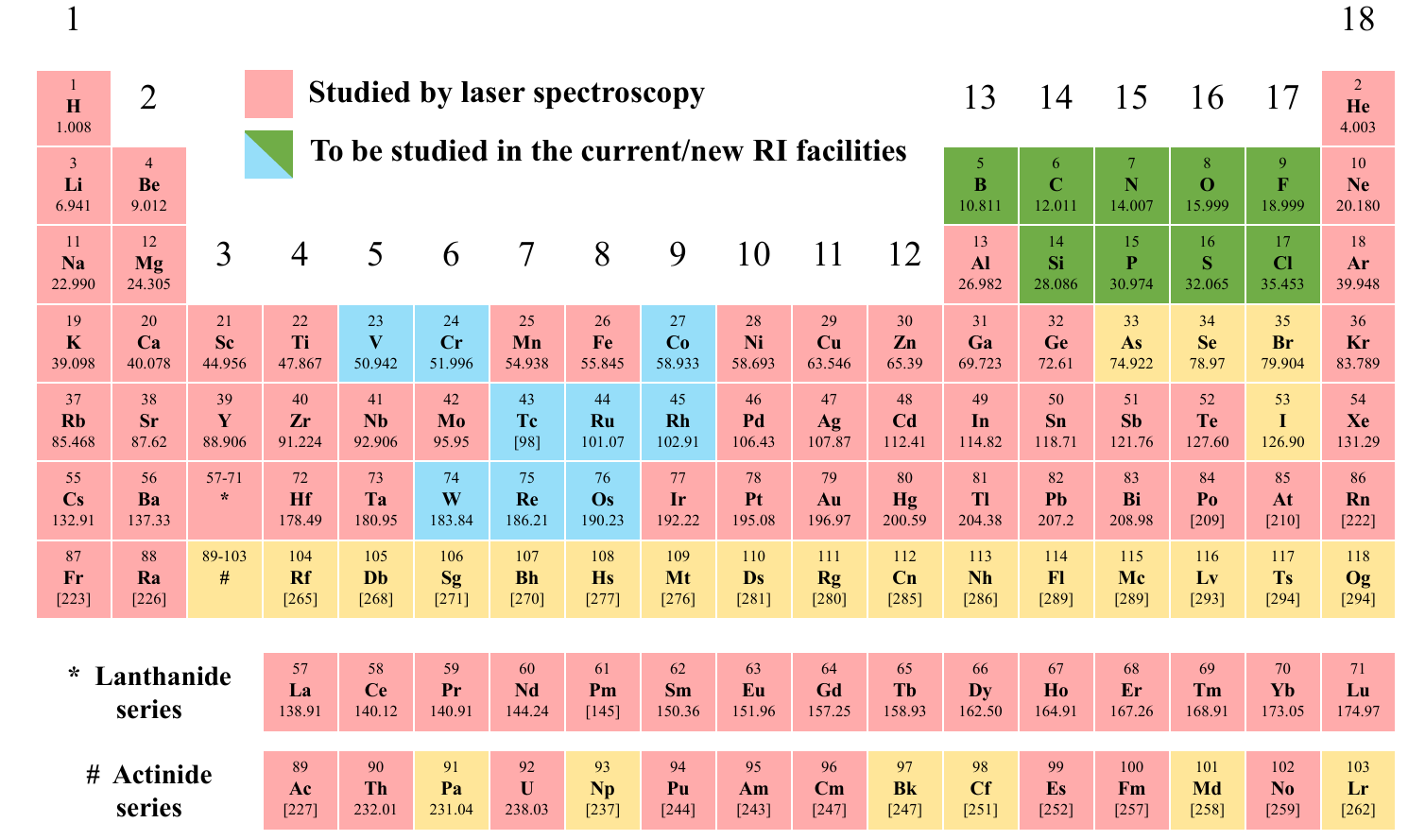}
\caption{\label{fig:fig5.2}\footnotesize{Elements for which laser spectroscopy has been performed on exotic isotopes (red squares), and whose exotic isotopes are expected to be measured with existing techniques in RIB facilities (blue squares). A gap in our knowledge around the region of carbon and oxygen isotopes is noticeable  (green squares). These light nuclei are of particular importance to understand the emergence of new nuclear phenomena from inter-nucleon interactions.}}
\end{center}
\end{figure*}
As shown in Fig.~\ref{fig:fig5.2}, exotic isotopes of a large variety of elements have already been studied using laser spectroscopy methods. Clear gaps however exist in our current knowledge including the region containing carbon and oxygen isotopes. Although exotic isotopes of these elements are already produced at existing RIB facilities, there are four major challenges that have so far prevented the extension of laser spectroscopy studies to these elements and those surrounding them:  
\begin{itemize}
\item [1)] Atomic or ionic transitions from the ground state lie mainly in the vacuum UV region ($<$ 90~nm) which is not accessible with conventional narrow-linewidth laser technology. 
\item [2)] Sensitivity to certain nuclear observables dramatically decreases with atomic number, for example the field shift as discussed in Sec.~\ref{sec:IS} and shown in Fig.~\ref{fig:fig2.2}. Therefore, extracting nuclear structure information from the HFS spectra requires high precision (an uncertainty of less than 1~MHz is required on isotope shifts)~\cite{Nor15}. 
\item [3)] Trapping and manipulating light atomic/ionic species is more challenging.
\item [4)] Light elements are reactive and exist mainly as molecular compounds (e.g. O$_{2}$, CO, CO$_{2}$, AlF$_{3}$), which are not yet suitable for HFS studies of exotic species. 
\end{itemize}

The new-generation RIB facility, FRIB at Michigan State University (MSU)~\cite{FRIB}, US, has recently become operational. This facility is expected to be able to produce short-lived light nuclei in sufficient yields, overcoming pre-existing production challenges. However, further developments of laser spectroscopy techniques and accurate atomic calculations are still required to reliably access the properties of light nuclear systems. 
A region of light nuclei, known as the \lq island of inversion' has long been the focus of experimental and theoretical investigation. In this region, collectivity was unexpectedly reported around the $N=20$ shell closure. Ground-state properties of neutron-rich sodium~($Z=11$) and magnesium~($Z=12$) isotopes studied by laser spectroscopy further verified their deformed intruder nature, resulting from multi-particle and multi-hole excitations across the $N=20$ shell gap, providing evidence of the disappearance of its magicity~\cite{Na-moment1978,Mg-moment2005,Mg-moment2007,Mg-moment2012}. 
Recent laser spectroscopy measurements of heavier aluminium~($Z=13$) isotopes revealed a sudden reduction in the charge radius of $^{32}$Al ($N=19$), supporting the possible magic character of $N=20$~\cite{Al-radii2021}. However, the general trend and this decrease in the charge radii of aluminium isotopes approaching $N=20$, in addition to that in neighbouring sodium and magnesium isotopes were not reproduced following recent shell-model and \textit{ab -initio} calculations~\cite{Al-radii2021,CC-Ne-Mg-radii,SM-Na-Mg-radii,CC-Na-radii}. This has motivated further experimental proposals and plans for laser spectroscopy studies of more neutron-rich isotopes of sodium, magnesium and aluminium at ISOLDE, CERN~\cite{MIRACLS-Mg,CRIS-Al}.

\subsection{\it The calcium region\label{sec:ca}}    
Isotopes situated in the vicinity of the magic calcium ($Z = 20$) chain constitute an important testing ground for developments in both experimental and theoretical nuclear physics. As was summarized in the preceding review article in this series in 2016~\cite{PPNP2016}, an extensive series of laser spectroscopy experiments has been undertaken in this region, in particular in the potassium and calcium isotope chains~\cite{K-moment2013,K-moment2014,K-radii2014-1,K-radii2014-2,K-radii2015,Ca-moment2015}. The spins and moments of odd-$A$ potassium isotopes from earlier measurements identified the inversion and subsequent re-inversion of the proton $\pi d_{3/2}$ and $\pi s_{1/2}$ orbitals as neutrons gradually fill the $\nu f_{7/2}$ and $\nu p_{3/2}$ orbitals~\cite{K-moment2013,K-moment2014}. In contrast, the spins and moments of odd-$A$ calcium isotopes are relatively single-particle in nature with their unpaired valence neutron occupying either the $\nu f_{7/2}$ and $\nu p_{3/2}$ orbitals~\cite{Ca-moment2015}. 

Since 2015, laser spectroscopy experiments have extended their reach to more exotic cases in this region towards both driplines in addition to entirely new isotopic chains (e.g. scandium, manganese). This has enabled a comprehensive series of studies of the evolution of nuclear properties across the neutron numbers $N=16, 20, 28, 32$ and 34~\cite{Ca-radii2016,Ca-radii2019,Ca-moment2019,K-radii2021,Sc-moment2022-40Sc,Sc-moments2022, Mn-radii2016,Fe-radii2016}.
Critical insights relating to different facets of nuclear structure have resulted from these studies, such as the \lq pure single-particle nature' of isotopes whose valence nucleons occupy the $f_{7/2}$ orbital~\cite{Ca-moment2015,Sc-moments2022}, challenging the proposed new \lq magic' number at $N=32$ driven by the tensor part of the proton-neutron monopole interaction~\cite{Ca-radii2016,K-radii2021}, the onset of deformation known as the `fifth island of inversion'~\cite{Mn-moment2016,Mn-radii2016}.
Details of these will be discussed in the following.
\begin{figure*}[h!]
\begin{center}
\includegraphics[width=0.99\textwidth]{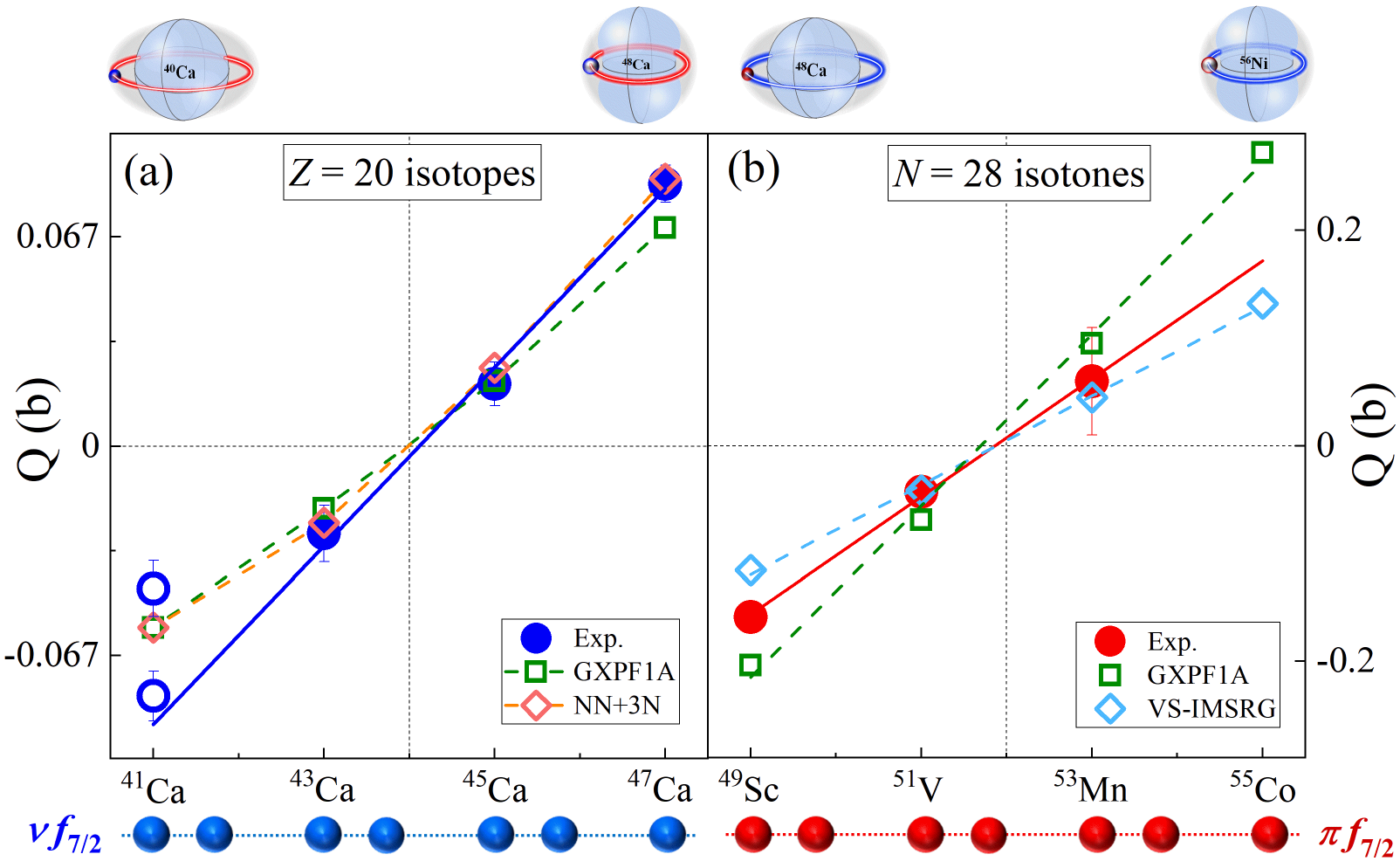}
\caption{\label{fig:fig5.3}\footnotesize{(a) Quadrupole moments of $Z=20$ isotopes as valence neutrons fill the $\nu f_{7/2}$ orbital. (b) Quadrupole moments of $N=28$ isotones as valence protons fill the $\pi f_{7/2}$ orbital. Experimental quadrupole moments are taken from Refs.~\cite{Ca-moment2015,Sc-moments2022}, which are compared to shell-model calculations using the effective interaction (GXPF1A), and calculations using nuclear forces derived from $\chi$EFT~(NN+3N, VS-IMSRG).}}
\end{center}
\end{figure*}

\subsubsection{Single-particle behaviour around doubly magic nuclei}  
As discussed in Sec.~\ref{sec:moment}, in a simple independent-particle shell-model picture, the nuclear ground-state spin and electromagnetic moments of an odd-$A$ nucleus are mainly determined by the unpaired nucleon~\cite{Mayer1949,Jensen1949}. Nuclear spins and electromagnetic moments in these cases can therefore act as sensitive probes of the ground-state wavefunction in near-magic isotopes. This was demonstrated in earlier studies measuring the moments of potassium and calcium isotopes~\cite{K-moment2013,K-moment2014,Ca-moment2015}. Recent extensions of laser spectroscopy studies to scandium isotopes have proven that the electromagnetic moments of the odd-$A$ isotopes are highly sensitive probes of single-particle behavior in addition to $M1$ and $E2$ nucleon-nucleon correlations~\cite{Sc-moment2022-40Sc,Sc-moments2022}. These measurements offered an excellent means in which the large-scale shell model, and recent advances in \textit{ab initio} many-body methods and microscopic interactions derived from $\chi$EFT, could be tested. This is partly due to the fact that the $0f_{7/2}$ orbital, located between the magic numbers $N,Z=20,28$, is unique in terms of its relative isolation in energy from the orbitals which neighbor it. Isotopes with valence protons and/or neutrons occupying the $0f_{7/2}$ orbital are therefore excellent probes to experimentally investigate both the single-particle nature and role of correlations in nuclei in this region of the nuclear chart.

Figure~\ref{fig:fig5.3} (a) and (b) presents the evolution of quadrupole moments of nuclei which are predominately determined by an unpaired neutron and an unpaired proton, respectively. The left-hand side of the figure shows experimental quadrupole moments of odd-$N$ calcium ($Z=20$) isotopes, which are expected to be described by the valence neutron in the $\nu f_{7/2}$ orbital. On the right-hand side, the quadrupole moments of the isotones at the neutron closed shell $N=28$ which are expected to be dominated by valence protons added to the $\pi f_{7/2}$ orbital, are shown~\cite{Ca-moment2015,Sc-moments2022}. In both cases, clear linear trends are observed from the available experimental $Q_{\rm{s}}$ of $Z=20$ isotopes and $N=28$ isotones. This experimentally confirms that the simple shell-model picture, as was introduced in Sec.~\ref{sec:Q} and in Eq.~\ref{eq28}, holds true in these semi-magic nuclei. In particular, the experimental values of $Q_{\rm{s}}$ cross zero at exactly the point in which the $\pi f_{7/2}$ orbital is half-filled. These systems represent a textbook example of nuclei behaving according to the independent-particle shell model~\cite{Ca-moment2015,Sc-moments2022}. 

\begin{figure*}[h!]
\begin{center}
\includegraphics[width=0.99\textwidth]{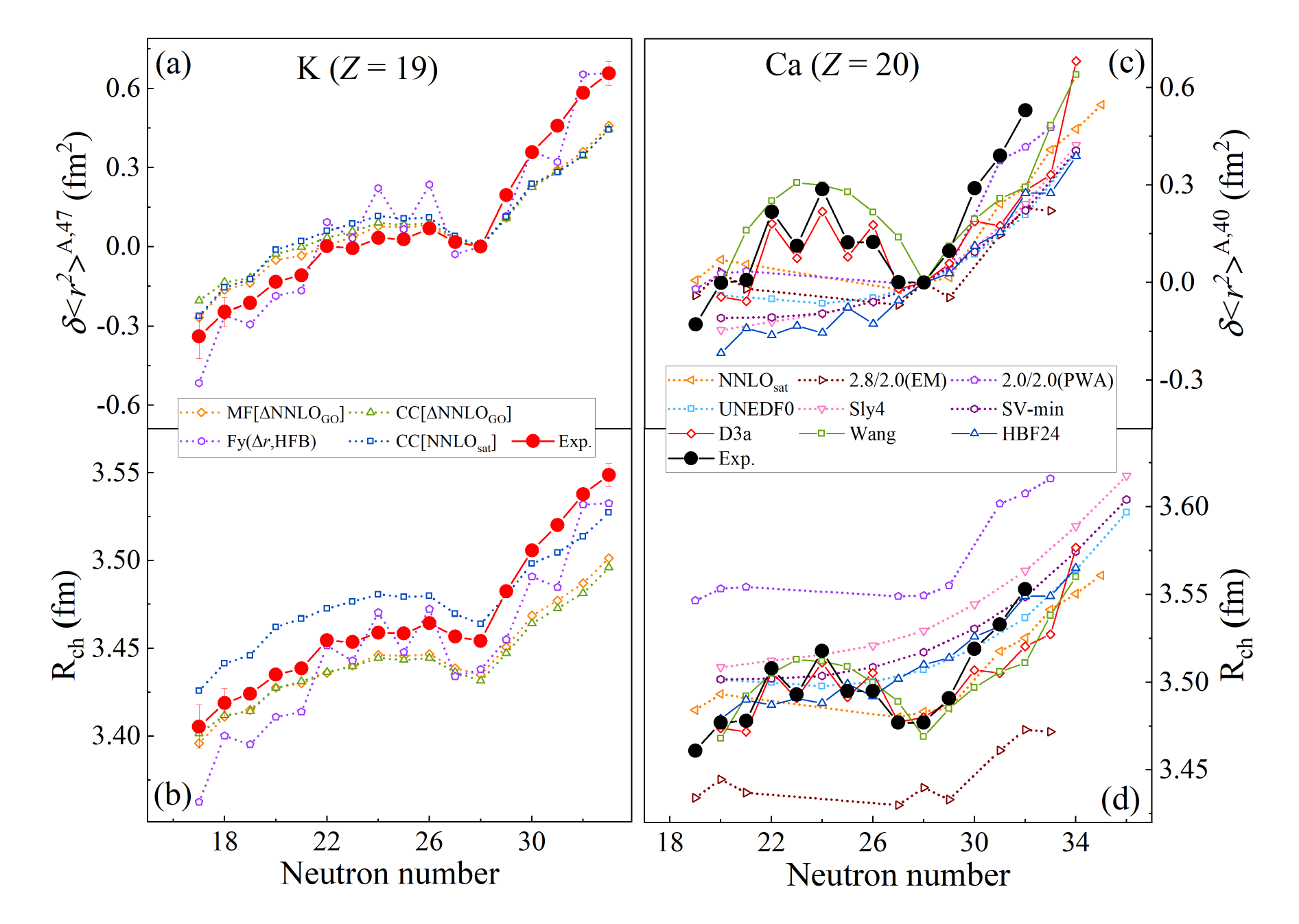}
\caption{\label{fig:fig5.4}\footnotesize{Evolution of the absolute nuclear charge radii, R$_{\rm{ch}}$, and changes in mean-square charge radii, $\delta \langle r^2 \rangle$ , of potassium and calcium isotopes. (a-b) Experimental values for potassium isotopes are compared to theoretical predictions from density function theory (Fy($\Delta r$,HFB)), mean-field calculations using $\Delta$NNLO$_{\rm{GO}}$ interaction (MF[$\Delta$NNLO$_{\rm{GO}}$]), and coupled cluster calculations using $\Delta$NNLO$_{\rm{GO}}$ and NNLO$_{\rm{sat}}$ interactions (CC[$\Delta$NNLO$_{\rm{GO}}$] and CC[NNLO$_{\rm{sat}}$]). (c-d) Experimental values for calcium isotopes are compared to theoretical predictions from density functional theory (UNEDF0, D3a, Sly4, SV-min, HBF24), coupled-cluster (NNLO$_{\rm{sat}}$, 2.8/2.0 (EM), 2.2/2.0 (PWA)), and global fits (Wang). Part of the data is taken from Ref.~\cite{Ca-radii2016,Ca-radii2019,K-radii2021}.}}
\end{center}
\end{figure*}

For isotopes with a simple nuclear configuration, one naturally expects that the shell model can accurately describe their electromagnetic moments. As shown in Fig.~\ref{fig:fig5.3}, a good overall agreement is found for the calcium isotopes. However, for the $N=28$ isotones, shell-model calculations using the effective GXPF1A interaction~\cite{GXPF1A} only achieve an adequate description of the general trend and overestimate the absolute single-proton quadrupole moment. After a systematic investigation, this was interpreted in Ref.~\cite{Sc-moments2022} to be attributed to a possibly underestimated $N=28$ shell gap in the GXPF1A interaction. VS-IMSRG~\cite{VS-IMSRG} calculations based on the chiral interaction $\Delta$NNLO$_{\rm GO}$~\cite{NNLOgo}, on the other hand, better capture the linear trend of quadrupole moments with respect to $Z$ of the $N=28$ isotones in addition to their absolute values. These results highlight the impressive progress made in advanced nuclear many body methods using microscopic interactions derived from $\chi$EFT.

\subsubsection{Structural changes at $N=32$ and $N=34$; Benchmarks for nuclear theory}
A great deal of attention in recent years has been devoted to study neutron-rich isotopes, in particular around the neutron numbers $N=32$ and $N=34$. Experimental results of binding energies and 2$_{1}^{+}$ excitation energies suggest the existence of additional closed shells at these neutron numbers~\cite{Wie13,Ste13,Rosen15,Michi18,Liu19,Xu19,Chen19,Leis21,Browne21}. However, laser spectroscopy results have revealed a large increase in the nuclear charge radii of neutron-rich calcium and potassium isotopes~\cite{K-radii2014-1,Ca-radii2016,K-radii2021}. These measurements challenge the interpretation of the proposed closed-shell nature of these $N=32$ nuclei far from stability. 

Experimental data and theoretical results for the changes in the mean-square charge radii of potassium ($Z=19$) and calcium ($Z=20$) isotopic chains are shown in Fig.~\ref{fig:fig5.4}. Describing the distinct parabolic behaviour observed in the radii of calcium isotopes, between $N=20$ and $N=28$, has remained a challenge for more than four decades~\cite{cau80}. Recent experiments in this region have extended our knowledge of the nuclear radii beyond the neutron number $N=28$~\cite{K-radii2014-1,Ca-radii2016,Fe-radii2016,K-radii2019,Tan20} and reveal a rapid increase of the nuclear size for these neutron-rich systems. The various features of nuclear charge radii observed in this region of the nuclear chart have motivated various theoretical developments for different many-body methods and interactions~\cite{sap16,Uta16,Bon16,mil19,Nak19,wu20,Wan19,Bho20,Per21,hor21,K-radii2021,Don22,Co22,kor22}. As shown in Fig.~\ref{fig:fig5.4}(b) and (d), accurately reproducing the absolute value of charge radii is a long-standing challenge for microscopic nuclear theory~\cite{Ca-radii2016,NNLOsat}. However, theoretical predictions can provide a relatively good description of the changes of the mean-square charge radii (Fig.~\ref{fig:fig5.4}(a) and (c)). Both density functional theory (DFT) and \textit{ab initio} calculations of absolute charge radii show a strong dependence on the choice of the functional or chiral nuclear force employed. The Fayans functional, labeled as D3a in Fig.~\ref{fig:fig5.4}(c) and (d), describes both absolute values and $\delta\langle r^2\rangle$ of the data well. In the \textit{ab initio} framework, the interactions that best describe the experimental data are NNLO$_{\rm{sat}}$~\cite{NNLOsat} and $\Delta$NNLO$_{\rm{GO}}$~\cite{NNLOgo}. The former is constrained by properties of selected nuclei up to $A=25$, while the latter uses nuclei with $A\leq4$ and also the properties of nuclear matter at saturation point when fitting its coupling constants. 

\subsubsection{Onset of collectivity towards $N=40$}  
Nuclear deformation has been suggested to appear towards $N=40$ in the open-shell isotopic chains between $Z=20$ and $Z=28$. These isotopes form a new island of inversion as evidenced by both experimental and theoretical investigations~\cite{IOICr,IOICr64,mass-Cr,PFSDG-U}. Laser spectroscopy experiments of open-shell manganese ($Z=25$) isotopes show increased collectivity in their magnetic and quadrupole moments as well as their charge radii on the neutron-rich side. The $N=28$ shell closure is still clearly observed in the charge radii of the less neutron-rich manganese and iron isotopes (Fig.~\ref{fig:fig3.4}(a))~\cite{Fe-radii2016,Mn-radii2016}. Describing the magnetic moments of $^{61-64}$Mn with shell-model calculations requires both an increasing degree of neutron excitations across the $N=40$ sub-shell gap and proton excitations across the $Z=28$ shell gap~\cite{Mn-moment2015,Mn-moment2-2015}. In order to reproduce the observed increase in the quadrupole moments of manganese isotopes from $N=36$ onwards, an additional increase in neutron excitations across $N=50$ is necessary~\cite{Mn-moment2016}. This suggestion of increased collectivity is further substantiated by experimental charge radii of $^{51,53-64}$Mn where a sudden increase in nuclear size is observed beyond $N=35$~\cite{Mn-radii2016}. To further investigate the structural evolution from the magic $N = 28$ region to the $N = 40$ island of inversion, plans to measure the ground-state properties of $^{50-63}$Cr have been proposed using collinear laser spectroscopy methods at IGISOL and ISOLDE-CERN~\cite{Cr-proposal}.

\subsection{\it The nickel region\label{sec:ni}} 
The concept of magic numbers with \lq traditional ones being $N,Z=2, 8, 20, 28, 50, 82$ and $N=126$ and the associated nuclear shell-model ~\cite{Jensen1949,Mayer1949} have long been cornerstones in our understanding of structure of atomic nuclei across the nuclear chart~\cite{Solin08-PPNP,Nowacki21-PPNP}.
However, in regions far from stability with extreme proton-to-neutron ratios, experimental evidence increasingly points to a weakening or even disappearance of the \lq traditional' magic numbers. Furthermore, some measurements support the existence of new shell closures driven by shell evolution due to proton-neutron interactions~\cite{Solin08-PPNP,Nowacki21-PPNP}. 

The region around the theoretically predicted doubly magic neutron-rich nucleus $^{78}$Ni, with $Z=28$ and $N=50$, has been investigated for decades by the nuclear physics community. The major open question in this region is \lq Is $^{78}$Ni a doubly magic nucleus?'. This, together with the astrophysical interest of $^{78}$Ni~acting as an r-process waiting-point nucleus~\cite{78Ni-astro,78Ni-astro-2}, have drawn significant attention from both theorists and experimentalists. Experiments aiming to shed light upon this system are hindered by production challenges. This isotope is difficult to produce at current RIB facilities in sufficient quantities to allow for detailed spectroscopic investigations, particularly for laser spectroscopy measurements of its ground-state properties. As a consequence, most experimental studies in this region have been performed for systems in the vicinity of $^{78}$Ni such as the less neutron-rich nickel isotopes~\cite{Ni-radii2022}, adjacent copper isotopes ($Z=29$)~\cite{Cu-moment2017} in addition to zinc~\cite{Zn-radii2016} and gallium isotopes~\cite{Ga-radii2017}. 

A \lq relatively' long history of laser spectroscopy studies in this region exists which spans more than a decade. 
This endeavour has involved a series of research programs at different experimental setups, including COLLAPS~\cite{Zn-moment2017,Ni-radii2022}, CRIS~\cite{Cu-radii2020,Ga-radii2017}, RILIS~\cite{Cu-moment2011-2} and BECOLA~\cite{Ni-radii2021}. In the previous review in this series~\cite{PPNP2016} and in Refs.~\cite{JPG2010,JPG2017}, results in this region measured with laser spectroscopy techniques were described including shell effects at $N=28$~\cite{Cu-moment2008,Cu-moment2009-1,Cu-moment2011-1}, the sub-shell effect of $N=40$~\cite{Ga-moment2010-1}, as well as the inversion of the proton $\pi 2p_{3/2}$ and $\pi 1f_{5/2}$ single-particle levels as neutrons fill the $\nu 1g_{9/2}$ orbital~\cite{Cu-moment2009-2,Ga-moment2010-1}.
Since then, laser spectroscopy investigations have mostly centered around the even-$Z$ isotopic chains (i.e. nickel ($Z=28$)~\cite{Ni-radii2022}, zinc ($Z=30$)~\cite{Zn-radii2019} and germanium ($Z=32$)~\cite{Ge-moment2020}) but also include neutron-rich copper ($Z=29$) towards $N=50$~\cite{Cu-radii2020,Cu-radii2016}. Nuclear-structure results from these efforts will be detailed in the following.

\begin{figure*}[t!]
\begin{center}
\includegraphics[width=0.99\textwidth]{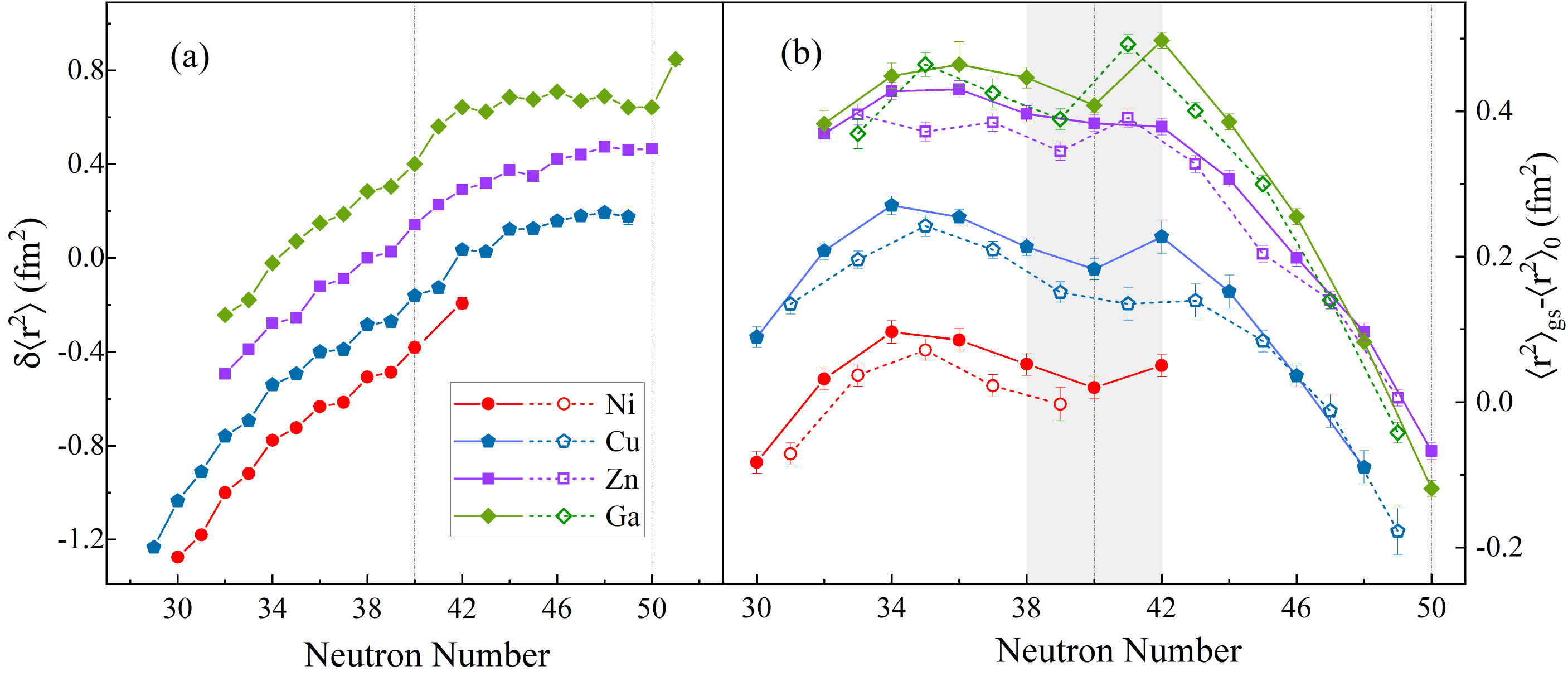}
\caption{\label{fig:fig5.5}\footnotesize{(a) Changes in mean-square charge radii of the ground states of nickel, copper, zinc and gallium isotopes, which are arbitrarily offset for better visualization. (b) Residual charge radii $\langle r^2 \rangle-\langle r^2 \rangle_{0}$ of ground states of the nickel, copper, zinc and gallium isotopes with the spherical volume contribution ($\langle r^2\rangle_{0}$) calculated from the droplet model subtracted. Data are taken from Refs.~\cite{Ni-radii2022,Cu-radii2016,Cu-radii2020,Zn-radii2019,Ga-radii2012,Ga-radii2017}.}}
\end{center}
\end{figure*}

\begin{figure*}[t!]
\begin{center}
\includegraphics[width=0.99\textwidth]{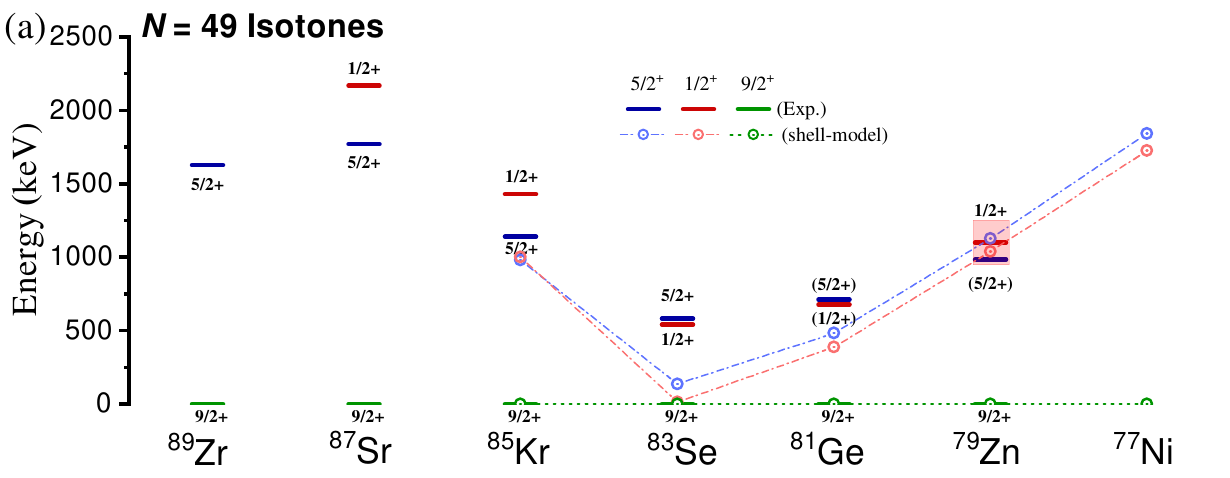}\\
\vspace{4mm}
\includegraphics[width=0.95\textwidth]{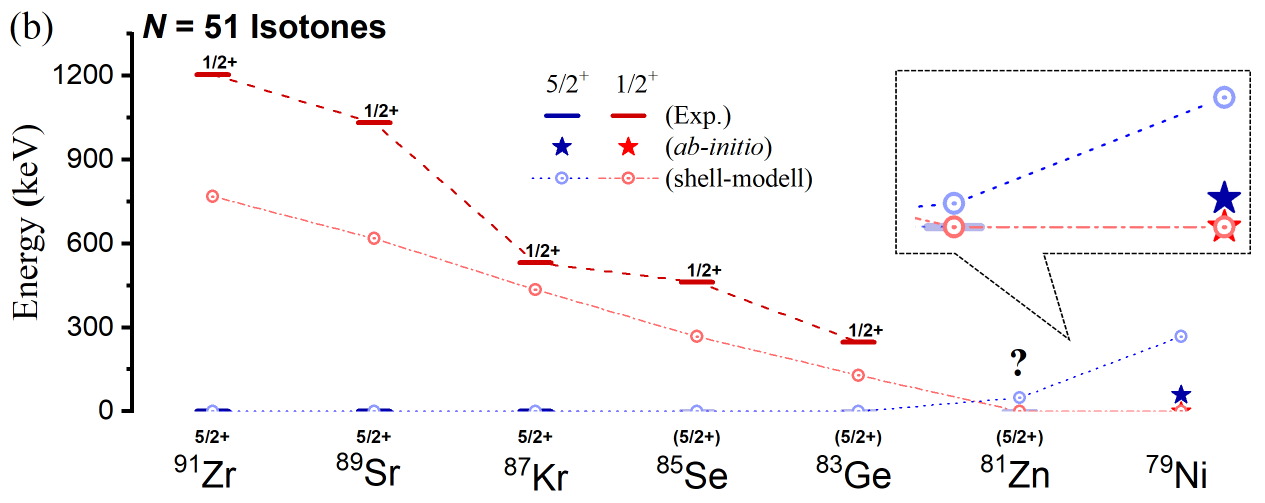}
\caption{\label{fig:fig5.6}\footnotesize{Energy level systematics for the low-lying states of odd-mass $N=49$ and $N=51$ isotones compared to the theoretical calculations. Data are taken from Refs.~\cite{Zn-radii2016,Zn-moment2017,Gaute-78Ni,Zn-proposal} and from the NNDC.}}
\end{center}
\end{figure*}

\subsubsection{$N = 40$ sub-shell effect}  
The $N=40$ sub-shell effect was clearly observed in the $E(2^{+}_{1})$ energy of $^{68}$Ni compared to that of its adjacent even-$N$ neighbors. This effect however is not seen in the neighboring even-$Z$ isotopic chains, such as zinc ($Z=30$) and iron ($Z=26$). In an attempt to answer whether the $N=40$ sub-shell is a local phenomenon that only exists in the magic nickel chain, measurements of the ground-state properties, in particular charge radii, of isotopes in the region~\cite{Ni-radii2022,Cu-radii2016,Cu-radii2020,Zn-radii2019,Ga-radii2012} have been carried out. After around a decade of experiments using the laser spectroscopy setups (COLLAPS, CRIS and RILIS) at ISOLDE-CERN, a comprehensive picture of the charge radii for the four isotopic chains was achieved, and shown in Fig.~\ref{fig:fig5.5} (a). To aid identifying a possible subtle sub-shell effect, the charge radii $\langle r^2\rangle-\langle r^2\rangle_{0}$ are subtracted by the spherical volume contribution ($\langle r^2\rangle_{0}$) from the droplet model and shown in Fig.~\ref{fig:fig5.5} (b). In Ref.~\cite{Cu-radii2016}, the obvious local minimum of the \mbox{$\langle r^{2}\rangle-\langle r^{2}\rangle_0$} observed at $N = 40$ for copper isotopes is explained as due to the sub-shell effect. This effect was also observed in the nickel isotopic chain based on recently reported experimental charge radii~\cite{Ni-radii2022}. However, a local minimum in \mbox{$\langle r^{2}\rangle - \langle r^{2}\rangle_0$} is barely visible in zinc. This is interpreted in Ref.~\cite{Zn-radii2019} as the disappearance of the $N$ = 40 sub-shell as just 2 protons are added above $Z = 28$. This rapid reduction in the sub-shell effect is also observed in other experimental observables such as the magnetic and quadrupole moments~\cite{Zn-moment2017,Cu-moment2017,Cu-moment2009-2,Ga-moment2010-1}, nuclear masses~\cite{mass-ISOLDE, mass-IGISOL}, $E(2^+$) excitation energies, and $B(E2)$ transition rates~\cite{AOI2010,Per2006,Lou2013}. The consensus reached by these measurements supports that the $N=40$ sub-shell is localized to a small region around $^{68}$Ni. It is important to note that the larger \lq dip' observed at $N = 40$ in the gallium isotopic chain is considered to be an effect of inverted odd-even staggering and deformation, as discussed in Refs.~\cite{Cu-radii2016,Ga-radii2012,Zn-radii2019}. 

\begin{figure*}[t!]
\begin{center}
\includegraphics[width=0.60\textwidth]{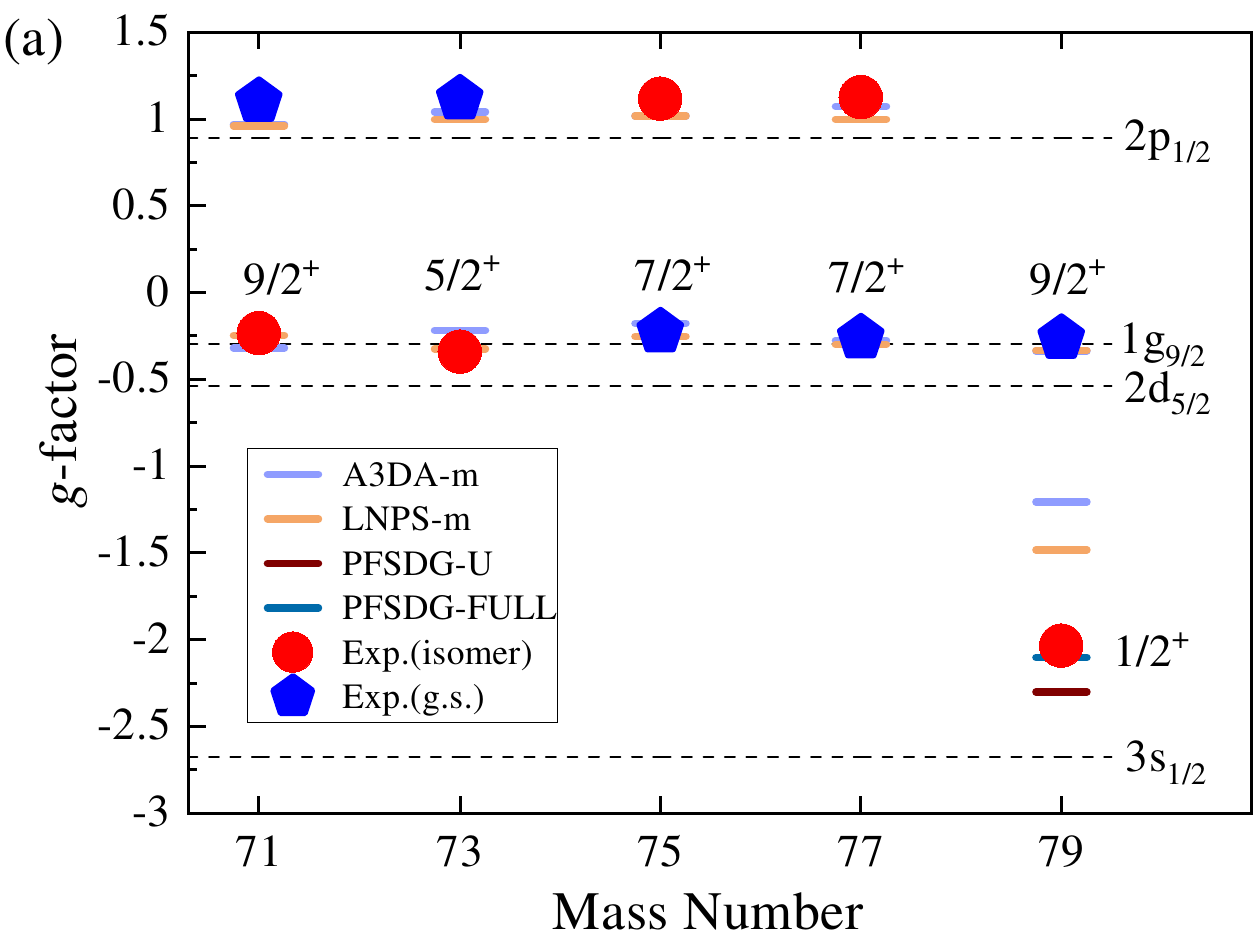}\\
\includegraphics[width=0.60\textwidth]{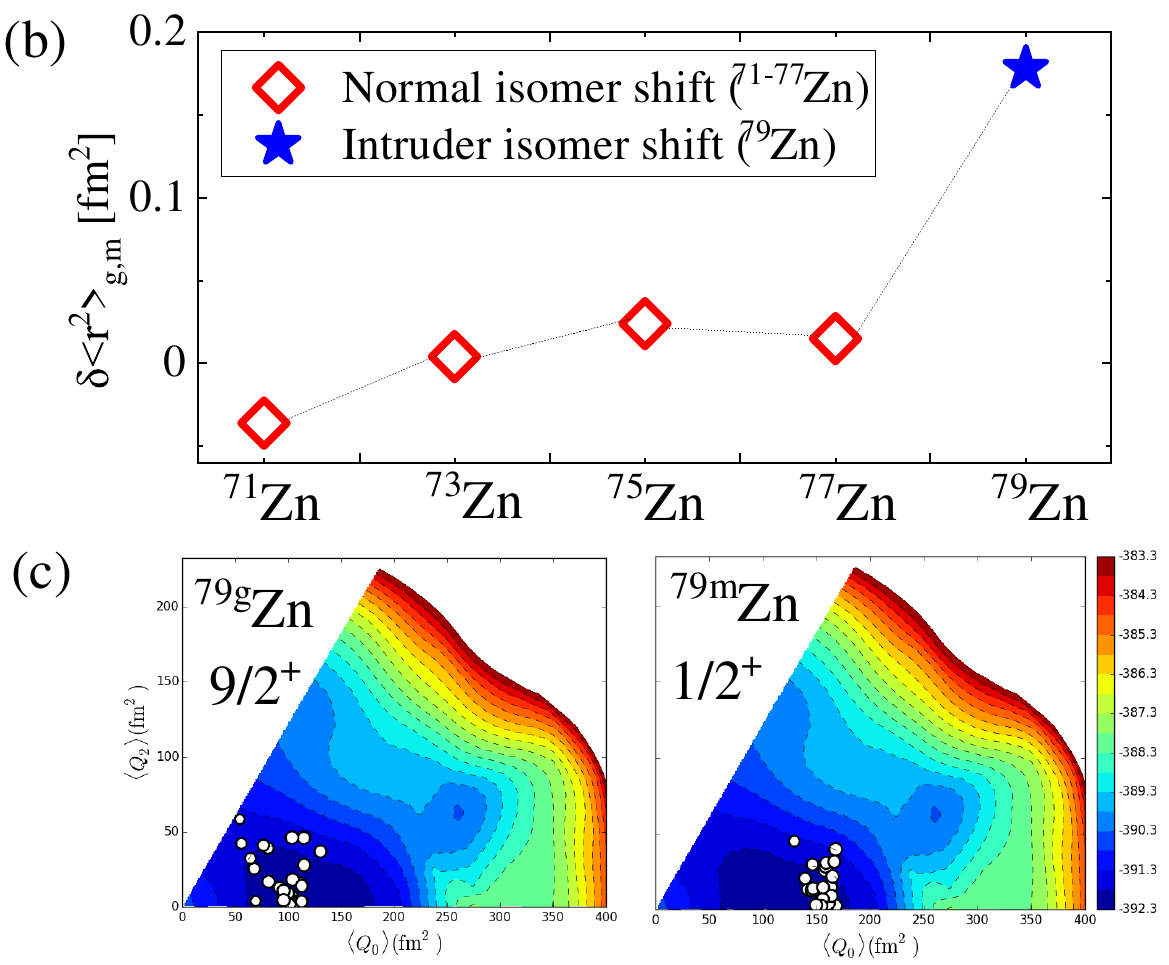}
\caption{\label{fig:fig5.7}\footnotesize{(a) Experimental $g$-factors of ground and isomeric states of $^{71-79}$Zn, compared with effective single-particle $g$-factors of neutron $2p_{1/2}$, $1g_{9/2}$, $2d_{5/2}$ and $2s_{1/2}$ orbitals, as well as with those from large-scale shell-model calculations employing different effective interactions~\cite{A3DA-m,IOICr64,PFSDG-U,pfsdg-full}. Data are taken from Refs.~\cite{Zn-radii2016,Zn-moment2017,Zn-radii2019}. (b) Isomer shift $\langle r^2 \rangle_{\rm m}-\langle r^2 \rangle_{\rm g}$ of $^{71-79}$Zn. Data are taken from Ref.~\cite{Zn-radii2019}. (c) Potential energy surface (T-plots) for the ground and isomeric states of $^{79}$Zn calculated by the Monte Carlo Shell Model (MSCM) based on the effective interaction PFSDG-FULL. Figures are provided by T.~Otsuka and Y.~Tsunoda~\cite{Zn-radii2019,pfsdg-full}}.}
\end{center}
\end{figure*}

\subsubsection{Shape coexistence around $^{78}\rm{Ni}$}  
Shape coexistence has long been a contemporary topic in nuclear physics and remains an active area of research to this day. It has been discussed in a series of dedicated reviews~\cite{SC-1983,RMP2011,SC-2016,SC-2022}. Shape coexistence is a nuclear structure phenomenon where states of different shapes coexist in the same isotope at low energy~\cite{SC-2022}. This is considered to result from the interplay between the stabilizing effect of closed shells and the residual interactions between protons and neutrons outside of them~\cite{SC-1983,SC-2016}. As has been summarized in the review by Heyde and Wood~\cite{RMP2011}, this phenomenon has been experimentally observed across much of the nuclear chart. In Ref.~\cite{SC-1983}, it is stated the combination of magnetic and quadrupole moments in addition to the isomer shift, which can be simultaneously accessed by laser spectroscopy, can provide a compelling and intuitive means to determine the presence of shape coexistence in a given nucleus. 
Around $N=50$, although intruder states were reported in some $N=49$ isotones (as is shown in Fig.~\ref{fig:fig5.6} (a)), such as the low-lying 1/2$^{+}$ intruder state in $^{83}$Se and $^{81}$Ge, no experimental evidence of shape coexistence was been reported in this region until recently. As stated in a recent review on the topic of shape coexistence~\cite{SC-2022}, the measurement of a large isomer shift for the 1/2$^{+}$ isomer in $^{79}$Zn~\cite{Zn-radii2016} is the first and most direct evidence of shape coexistence in the vicinity of $^{78}$Ni. The experiment, performed at COLLAPS at ISOLDE-CERN, identified a new long-lived isomeric state in $^{79}$Zn just one neutron and two protons away from $^{78}$Ni. The properties of this new isomer, including its magnetic moment and large isomer shift were reported in Refs.~\cite{Zn-radii2016,Zn-moment2017}. As can be seen in Fig.~\ref{fig:fig5.7}~(a), the negative sign of the $g$-factor of this 1/2 isomeric state in $^{79}$Zn confirms its positive parity. It is only possible to describe its $g$-factor using the shell model employing interactions with an extended model space that includes the neutron $sdg$ orbitals above $N=50$~\cite{Zn-moment2017, Zn-radii2019}, such as PFSDG-U~\cite{Zn-moment2017} and PFSDG-FULL~\cite{Zn-radii2019,pfsdg-full}. This supports the assignment of a $2h-1p$ intruder configuration of this isomer which mainly results from neutron cross-shell ($N=50$) excitations to the $3s_{1/2}$ orbital. The observed large isomer shift shown in Fig.~\ref{fig:fig5.7}~(b) suggests a deformation parameter of $\beta_{2} >0.2$. While for the normal 9/2$^+$ ground state, the deformation parameter, calculated from the measured quadrupole moment, gives $\beta_{2}=$0.15(2) which is consistent with that estimated from the experimental values of $B(E2\uparrow)$ in neighboring $^{78,80}$Zn. Therefore, it is concluded that a normal near-spherical ground state and a deformed long-lived intruder state coexist in $^{79}$Zn. This conclusion is further supported in T-plots, shown in Fig.~\ref{fig:fig5.7}~(c), calculated using the Monte Carlo Shell Model (MCSM) using the PFSDG-FULL interaction~\cite{Zn-radii2019, pfsdg-full}. The exact excitation energy of this isomeric state, suggested to be 1.10(15) MeV from a $^{78}$Zn($d,p$)$^{79}$Zn transfer reaction~\cite{Orlandi2015-79Zn} (Fig.~\ref{fig:fig5.6}~(a)), is not certain. Shell-model calculations~\cite{Zn-moment2017} and decay experiments at RIKEN-RIBF~\cite{Delattre-PhD} support a lower excitation energy of this isomer than the $5/2^{+}$ state and is in line with a simple estimation from its lifetime (several hundred milliseconds)~\cite{Zn-radii2016}.

As seen from the low-lying energy systematics of the $N=49$ isotones (Fig.~\ref{fig:fig5.6}~(a)), a long-lived isomer with a tentatively-assigned spin-parity of 1/2$^{+}$ was already reported in $^{81}$Ge, which is situated lower in energy than the normal excited 5/2$^{+}$ state. Recently, a $^{80}$Ge($d,p$)$^{81}$Ge transfer reaction experiment validated the 1/2$^{+}$ spin-parity assignment and measured its spectroscopic factor~\cite{81Ge-new}, confirming its $2h-1p$ intruder configuration from neutron excitations across $N=50$ to the $3s_{1/2}$ orbital. $^{81}$Ge therefore constitutes an ideal next candidate in which shape coexistence around $^{78}$Ni can be studied using laser spectroscopy. Plans to realize this are underway at COLLAPS at ISOLDE-CERN~\cite{Ge-proposal}. 

A future study of the properties of the low-lying state in $^{77}$Ni, a more exotic nucleus in the $N=49$ isotone series, in which a 1/2$^{+}$ level is predicted to be the first excited state by shell-model calculations~\cite{PFSDG-U,Zn-moment2017}, will deepen our understanding of the $2h-1p$ intruder nature and the phenomenon of shape coexistence near the doubly magic $^{78}$Ni. It is also worth mentioning that shape coexistence in the even-even isotopes $^{78}$Ni~\cite{PFSDG-U, Gaute-78Ni} and $^{80}$Ge~\cite{Gottarto2016-80Ge} have also been investigated by both theory and experiment. However, further studies are required before any firm conclusions may be drawn.

\subsubsection{Nuclear structure at $N = 50$}  
While the observation of shape coexistence in $^{79}$Zn might call the magicity of $^{78}$Ni into question, the systematic study of nuclear moments~\cite{Zn-moment2017,Cu-moment2017} and charge radii~\cite{Zn-radii2019,Cu-radii2016} in zinc and copper isotopes broadly support its magic nature.
Corroborating findings include the relatively pure single-particle configuration concluded from the moments of the ground state of $^{79}$Zn~\cite{Zn-moment2017}, reduced proton occupations above $Z=28$ and neutron occupations above $N=50$ in $^{78}$Cu~\cite{Cu-moment2017}, reduced \lq residual charge radii' when approaching $N=50$ as shown in Fig.~\ref{fig:fig5.5}(b) and the reduced odd-even staggering around $N=50$ in the charge radii of the zinc and copper isotopes~\cite{Zn-radii2019,Cu-radii2020}. All of these observations point to the enhanced stability of $Z=28$ and $N=50$, supporting the conclusion that $^{78}$Ni is doubly magic. Experiments measuring other observables such as the binding energies up to $^{79}$Cu~\cite{Welker2017-Cu} as well as $\gamma$-spectroscopy of  $^{79}$Cu~\cite{Olivier2017-Cu} are consistent with this conclusion. Recently, direct experiments on low-lying states in$^{78}$Ni were performed at RIBF-RIKEN via in-beam $\gamma$-spectroscopy~\cite{Taniuchi2019-Ni}. This result, which established the excitation energy of the 2$^{+}_{1}$ state, provided the first direct experimental evidence for the doubly magic nature of $^{78}$Ni. 

This in-beam experiment~\cite{Taniuchi2019-Ni} also intriguingly confirms the existence of a deformed second 2$^+$ state at low energy, supporting the prediction of shape coexistence in $^{78}$Ni by large-scale shell-model calculations~\cite{PFSDG-U}. The theoretical prediction presented in Ref.~\cite{Taniuchi2019-Ni} suggests a breakdown of the proton $Z = 28$ shell closure above $N=50$, and that deformed ground states are favored in the more neutron-rich nickel isotopes~\cite{Taniuchi2019-Ni}. Figure~\ref{fig:fig5.6}(b) summarizes the systematics of the low-lying 5/2$^+$ and 1/2$^+$ states of $N=51$ isotones from $Z=40$ down to $Z=28$ showing a dramatic decrease in energy of the 1/2$^+$ state. Approaching $Z=28$, an inversion of the 1/2$^+$ and 5/2$^+$ states is expected and theoretically predicted by both shell-model and \textit{ab initio} calculations~\cite{Gaute-78Ni}, although experimental data is scarce (Fig.~\ref{fig:fig5.6}(b)). This strongly motivates laser spectroscopy experiments to be performed on more neutron-rich isotopes above $N=50$ to determine their spins, moments and radii of their ground states and possible long-lived isomeric states. Challenges remain in directly determining the properties of neutron-rich nickel isotopes experimentally, as they are not yet producible in sufficient quantities at current RIB facilities. Nevertheless, experiments to study the ground-state structure of $^{81,82}$Zn, the closest even-$Z$ isotopes to $^{79,80}$Ni, have been proposed using CRIS at ISOLDE-CERN~\cite{Zn-proposal}.

\subsubsection{Charge radii and nuclear matter}
In addition to the connection between mirror nuclei and parameters contained within the equation of state discussed in Sec.~\ref{sec:matter} and Fig.~\ref{fig:fig3.6}, charge radii measurements can also provide constraints to the neutron skin and nuclear dipole polarizability. Recently, the charge radius of $^{68}$Ni, measured at COLLAPS at ISOLDE-CERN, provided an important benchmark for \textit{ab initio} coupled-cluster calculations allowing the neutron skin of $^{68}$Ni constrained to be within the range of $0.18-0.20$~fm~\cite{Ni-radii2020}. Furthermore, measurements of the charge radii of $^{63-78}$Cu~\cite{Cu-radii2020} and $^{58-60,70}$Ni~\cite{Ni-radii2022} have also offered stringent tests for theoretical calculations based on nuclear density functional theory and multiple \textit{ab-initio} many-body methods. This in turn offers valuable input into understanding the saturation density of nuclear matter~\cite{Cu-radii2020}.

\subsection{\it The tin region\label{sec:sn}}    
Isotopes in the neighborhood of the doubly magic $^{100}$Sn and $^{132}$Sn nuclei have been the subject of numerous recent experimental and theoretical studies~\cite{hin12,fae13,bad13,all15,cor15,Morris2018,togashi2018}. 
In these, laser spectroscopy experiments have provided critical experimental data to shape our understanding of this key region of the nuclear chart. These measurements have allowed the extraction of nuclear electromagnetic properties in the isotopic chains of palladium($Z=46$)~\cite{Pd-radii2022}, silver($Z=47$)~\cite{Ag-radii2021}, cadmium ($Z=48$)~\cite{Cd-radii2016}, indium ($Z=49$)~\cite{In-moment2022}, tin ($Z=50$)~\cite{Sn-moment2020,Sn-radii2020}, and antimony ($Z=51$)~\cite{Sb-moment2021} in the vicinity of $^{100}$Sn and/or $^{132}$Sn. These studies have proven essential in our understanding of the evolution of nuclear structure properties in this frontier region of the nuclear chart motivating developments in nuclear theory. A few recent highlights of laser spectroscopy results obtained since the last review in this series~\cite{PPNP2016} are detailed in the following. 
\begin{figure*}[t!]
\begin{center}
\includegraphics[width=0.99\textwidth]{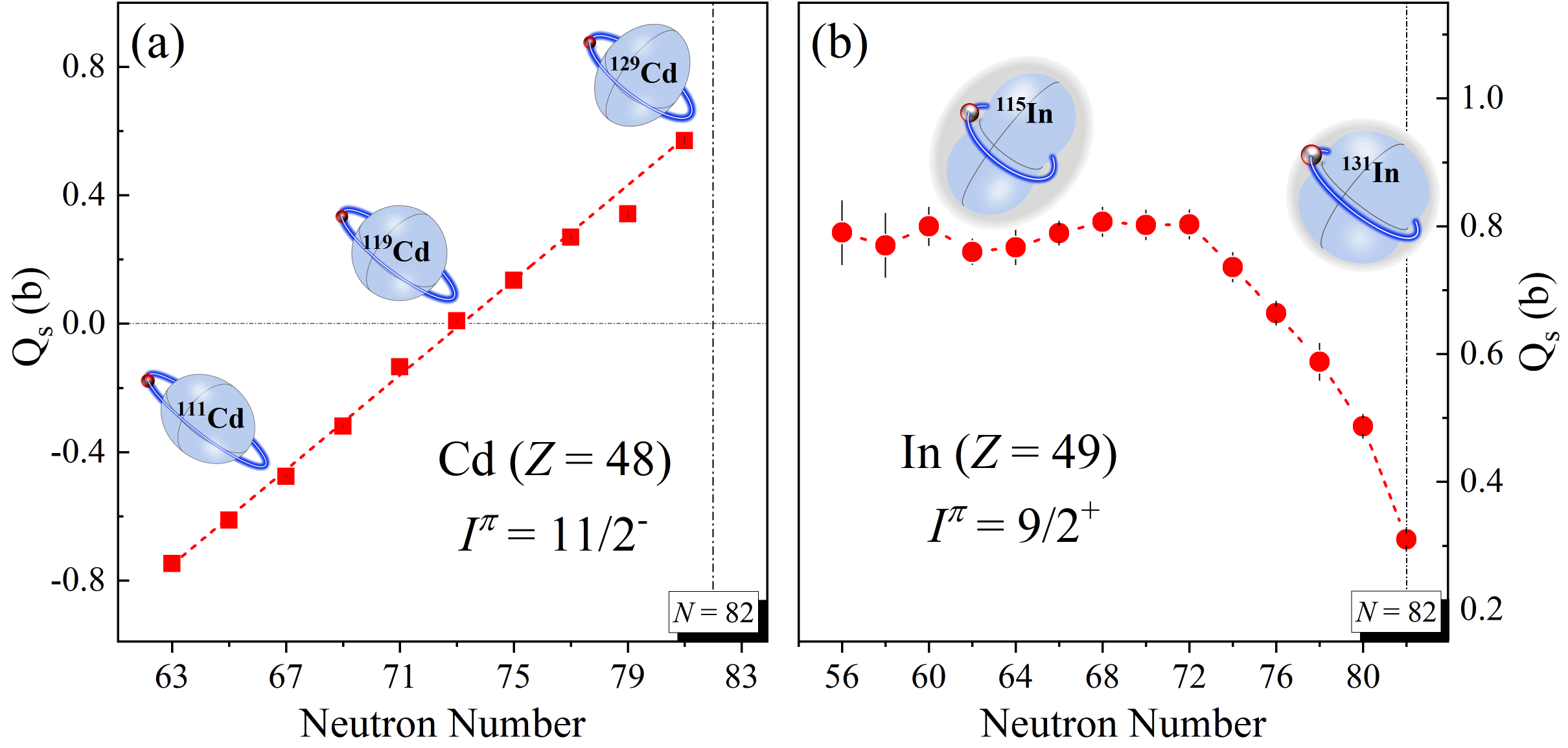}
\caption{\label{fig:fig5.8}\footnotesize{Experimental quadrupole moments of the cadmium ($Z=48$) and indium ($Z=49$) isotopes as a function of the neutron number. A linear relation observed for cadmium isotopes is in agreement with the expectations from a single neutron occupying a shell-model orbital. For indium isotopes a decrease of the magnitude of the quadruple moments towards the magic number $N=82$ suggests a rapid reduction of collective contributions. Data are taken from Refs.~\cite{Cd-moment2013,In-moment2022}.} }
\end{center}
\end{figure*}

\subsubsection{Evolution of electromagnetic properties in the proximity of $^{100}\rm{Sn}$}
The evolution of properties in proximity of $^{100}$Sn remains as an open question in nuclear structure. Contradictory experimental evidence has been reported about the level ordering of shell-model orbitals in the neutron-deficient tin isotopes~\cite{sew07,dar10}, Furthermore, several open questions remain in connection with the robustness of the $N = Z = 50$ shell closures~\cite{hin12,ban05,vam07}. While the observation of the largest Gamow-Teller decay strength found to date in $^{100}$Sn supports a shell-model interpretation with strong $N = Z = 50$ shell closures~\cite{hin12}, the existence of low-lying states and relatively large $B(E2\uparrow)$ values (known down to $^{104}$Sn~\cite{vam07}) points to a weakened $^{100}$Sn core. 

Recent charge radii measurements of neutron-deficient silver isotopes at IGISOL discovered a pronounced discontinuity in their charge radii, with a local minimum at $N = 50$~\cite{Ag-radii2021}. The magnitude of this discontinuity was however not well-reproduced by nuclear theory. Further studies on nearby palladium isotopes reveal complementary details of how pairing and deformation impacts upon the reproduction of local variations in nuclear charge radii~\cite{Pd-radii2022}.  

Precision laser spectroscopy experiments of neutron-deficient $^{103-109}$Sn~\cite{Sn-proposal,gus20} and $^{101-109}$In~\cite{In-proposal} isotopes have already been performed using the CRIS experiment at ISOLDE, allowing systematic studies of nuclear spins, electromagnetic moments, and charge radii of their ground and long-lived isomeric states. These observables will provide a comprehensive picture of the evolution of nuclear properties in the proximity of $N=Z=50$. 

\subsubsection{Simple structure of complex nuclei}
The nuclear properties of certain isotopes of elements around $Z=50$ have become textbook examples of single-particle behaviour in atomic nuclei~\cite{Cd-moment2013,Cd-radii2016,Cd-moment2018,Heyde2004}. Figure~\ref{fig:fig5.8} shows two prominent examples of the evolution of nuclear quadrupole moments as a function of the neutron number for odd-even isotopes of cadmium ($Z=48$) and indium ($Z=49$). The linear trend observed in cadmium isotopes has been interpreted to originate mainly from a single valence neutron in the $h_{11/2}$ orbital. As discussed in Sec.~\ref{sec:Q}, in the single-particle picture, a neutron polarizes the core towards oblate deformation, thus the quadrupole moments exhibit a linear trend as a function of the neutron number. A change in the sign of the quadrupole moment, corresponding to prolate deformation, is then interpreted in this picture as resulting from core polarization induced by a single neutron hole.   

Beyond this single-particle picture, the quadrupole moments of indium isotopes are expected to be strongly influenced by the interplay between the unpaired proton-hole in the $g_{9/2}$ orbit, and the collective behaviour of the remaining nucleons. Recent observations~\cite{In-moment2022} have revealed an abrupt change in the magnetic moments and a rapid reduction of the magnitude of the quadrupole moments when approaching $N=82$, indicating a significant reduction of collectivity at $N=82$. 

In contrast to the linear trend of quadrupole moments observed in the neutron-rich cadmium isotopes, an unexpected distinct quadratic trend was recently observed in those of the ground- and isomeric-states of tin isotopes. This behaviour is not yet understood~\cite{Cd-radii2016,Sn-radii2020}. Differences in the charge radii of cadmium isotopes between the lowest $I=1/2^{+}$ and $I=3/2^{+}$ states to the $I=11/2^{-}$ isomers, were found to follow a parabolic dependence as a function of the neutron number~\cite{Cd-radii2016}. A similar trend was also observed in tin~\cite{Sn-radii2020}.

\subsubsection{Structural changes at $N=82$}
Recent charge radii measurements of tin isotopes showed a clear kink at $N=82$~\cite{Sn-radii2019}, which is a feature commonly observed at shell closures~\cite{gar20a}. 
The results have provided important benchmarks to test our understanding of the origin of the discontinuities in nuclear charge radii observed at nuclear closed-shells. DFT calculations, in particular those employing the Fayans functional~\cite{Cd-radii2018,Sn-radii2019}, have simultaneously provided a relatively good description of the general trends observed in both tin and cadmium isotopes~\cite{Cd-radii2018,Sn-radii2019,Sn-radii2020,Cd-radii2022}. 

Laser spectroscopy experiments also provided measurements of the nuclear electromagnetic moments of tin ($Z=50$) and antimony ($Z=51$) isotopes around the neutron closed-shell $N=82$~\cite{Sn-moment2020,Sb-moment2021}. The magnetic moments of $^{133}$Sn and $^{133}$Sb act as probes of the single-particle nature of a single neutron and a single proton around the doubly magic nucleus $^{133}$Sn, respectively. Systematic studies of the magnetic moments of indium isotopes ($Z=49$) revealed an abrupt and sudden change at $N=82$~\cite{In-moment2022}. Theoretical and experimental evidence in the region therefore suggests that the magnetic moment of $^{131}$In is dominated by a single-proton hole, while other configurations contribute significantly to the structure of less neutron-rich indium isotopes. This contradicts previous interpretations of the trend in the lighter indium isotopes being a textbook example of single-particle behaviour~\cite{Heyde2004}. Recent DFT calculations have shown that the inclusion of time-symmetry-breaking mean fields are essential for a correct description of nuclear magnetic properties~\cite{In-moment2022,sas22}.

Further laser spectroscopy studies in this region, e.g. the neutron-rich tellurium~\cite{Te-proposal} and silver~\cite{Ag-proposal} isotopes will provide further input into our understanding of nuclear properties around $^{132}$Sn.

\subsection{\it The lead region\label{sec:pb}}   
The region surrounding the doubly magic isotope $^{208}$Pb has been the subject of sustained experimental and theoretical interest since the discovery of the large odd-even staggering in neutron-deficient mercury isotopes~\cite{Hg-radii1972}. 
Despite the first measurements of this now iconic example of nuclear shape coexistence being reported five decades ago, the region remains an active area of research today.
Laser spectroscopy experiments, including both in-source RIS and CLS techniques, form a large part of this endeavour.

\subsubsection{Shape-staggering in neutron-deficient mercury and bismuth isotopes}
\begin{figure*}[t!]
\begin{center}
\includegraphics[width=0.99\textwidth]{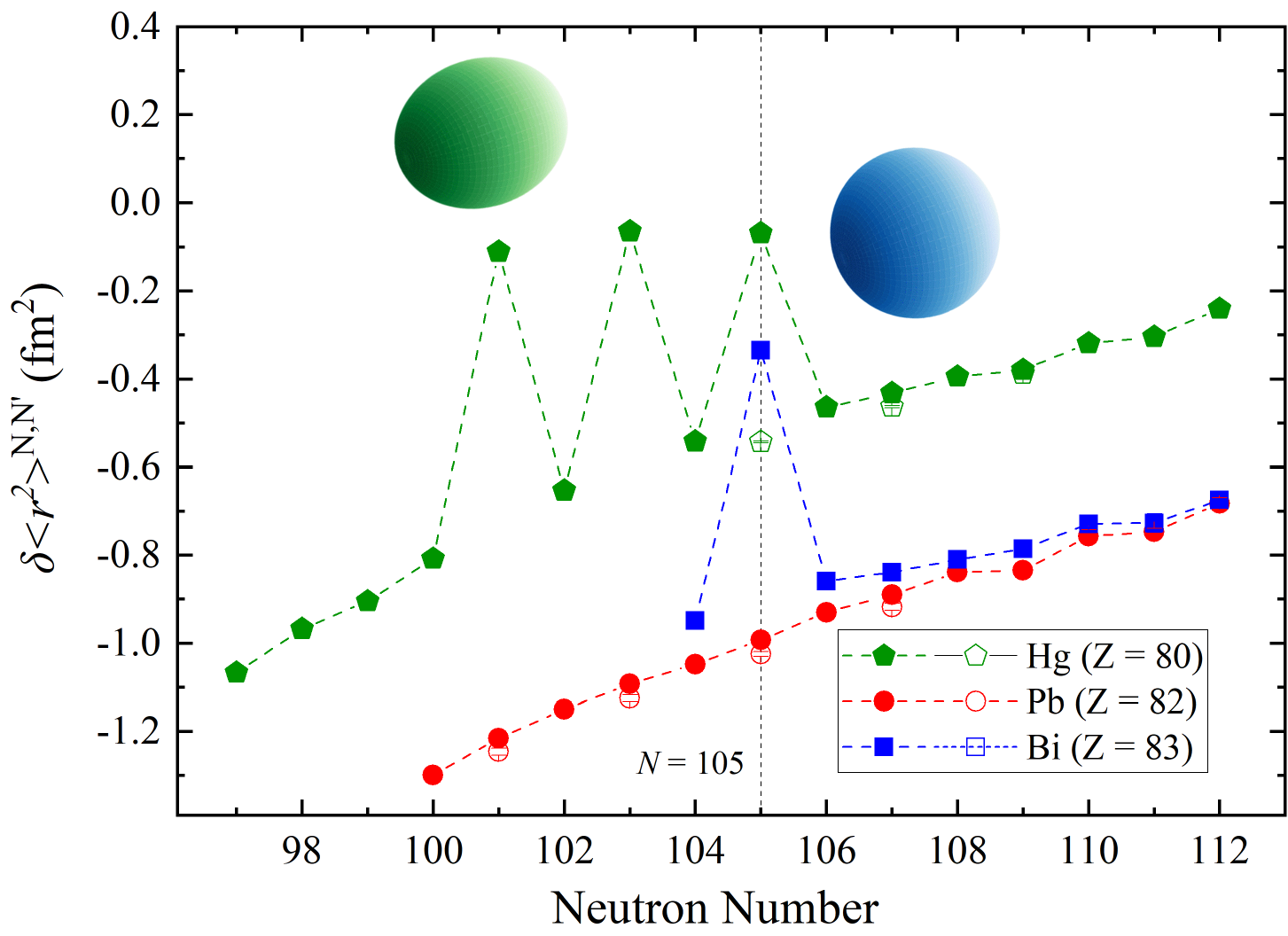}
\caption{\label{fig:fig5.9}\footnotesize{Changes in the mean-square charge radii of mercury, bismuth and lead isotopes. The filled and open markers indicate the data for the ground and isomeric states, respectively. Data are taken from Refs.~\cite{Hg-radii2018,Bi-radii2021,Pb-radii2007}. }}
\end{center}
\end{figure*}

Several decades after the first optical measurements of neutron-deficient mercury isotopes ($Z=80$)~\cite{Hg-radii1972}, the chain was revisited at ISOLDE-CERN.
The properties of four new isotopes, $^{177-180}$Hg, were successfully measured with in-source RIS \cite{Hg-radii2018,Hg-radii2019} allowing isotopes down to $N=97$ to be accessed. 
The new data reveal no additional large shape staggering beyond what was previously measured, with the new isotopes exhibiting a more normal odd-even staggering, as shown in Fig.~\ref{fig:fig5.9} with green pentagons.
Large-scale MCSM calculations were employed and highlighted the role of multiple proton and neutron excitations to the $\pi 1h_{9/2}$ and $\nu 1i_{13/2}$ orbitals in driving this phenomenon, with the latter resulting in large quadrupole deformation in the $N=104$ mid-shell region.

The onset of another example of large shape staggering was recently discovered in the very neutron-deficient bismuth isotopes ($Z=83$)~\cite{Bi-radii2021}.
Relative to its neighbours $^{187,189}$Bi, the ground state of $^{188}$Bi ($I=(10)^{-}$) possesses a significantly larger charge radius resulting in another remarkable example of this phenomenon, which is also presented in Fig.~\ref{fig:fig5.9} with blue squares. The onset of this newly observed example of large shape staggering in the bismuth isotopic chain occurs at the exact same neutron number ($N=105$) as in the mercury isotopes and also possesses a similar magnitude. A stark contrast between these isotopes and the intermediate semi-magic lead chain (red circles in Fig.~\ref{fig:fig5.9}) can be seen where the neutron-deficient lead isotopes remain largely spherical in nature~\cite{Pb-radii2007}. 

\subsubsection{Intruder states and shape coexistence in even-$N$, odd-$Z$ nuclei}
Additional examples of shape coexistence in this region observed through large isomer shifts have been measured in neutron-deficient even-$N$ isotopes of the odd-$Z$ gold~($Z=79$), thallium~($Z=81$), bismuth~($Z=83$) and astatine~($Z=85$) isotopic chains. In thallium, excitation of a proton from the $3s_{1/2}$ to $1h_{9/2}$ orbital across the $Z=82$ shell gap becomes energetically favorable with decreasing $N$ towards the mid-shell region. The resulting deformed intruder states become low-lying 9/2$^{-}$ isomers that are concluded to coexist with 1/2$^{+}$ ground states.

\begin{figure*}[t!]
\begin{center}
\includegraphics[width=0.99\textwidth]{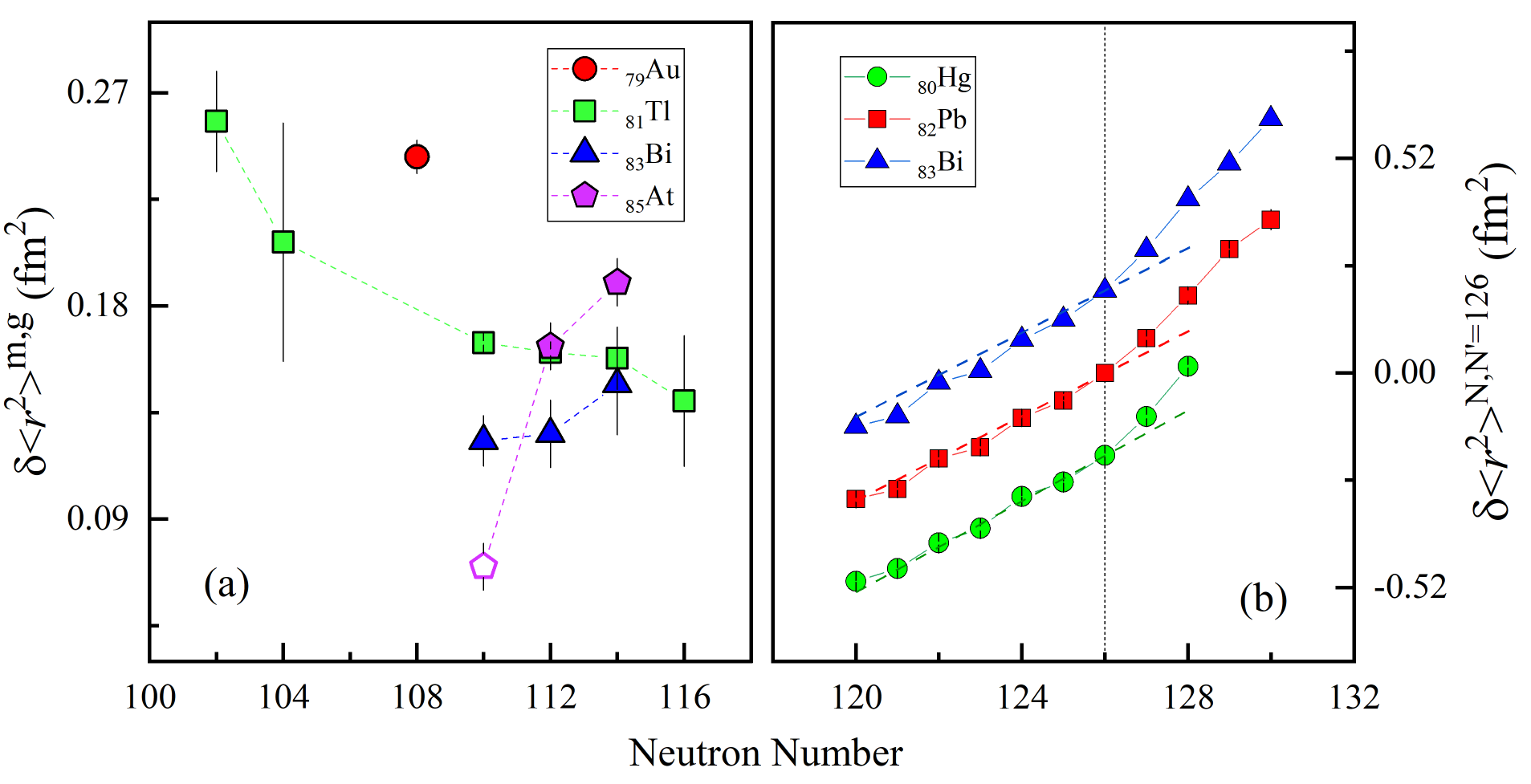}
\caption{\label{fig:fig5.10}\footnotesize{ (a) Isomer shifts of intruder states in the odd-$Z$ thallium, bismuth and astatine isotopes. The open markers indicate nuclei where the intruder state becomes the ground state. (b) Changes in mean-square charge radii of mercury, bismuth and lead isotopes, showing a clear `kink' at the $N=126$ shell closure~\cite{Hg-radii2021_PRL,Bi-radii2021,Pb-radii2007}.}}
\end{center}
\end{figure*}

Large isomer radii were measured previously in even-$N$ thallium ($Z=81$) with $N= 104-116$, shown as green squares in Fig.~\ref{fig:fig5.10}(a), where the isomeric $9/2^{-}$ states exhibit a larger charge radius due to a significant degree of deformation with respect to their almost spherical $1/2^{+}$ ground states (no ground-state data exist for $N=106,108$). 
A recent study extending to the more neutron-deficient thallium isotopes was able to measure an additional $9/2^{-}$ isomer at $N=102$~\cite{Tl-radii2017}. No isomeric states were measured in $^{179,181}$Tl ($N=98,100$) whose ground-state spins were unambiguously determined to be $1/2^{+}$. 
The magnitude of measured isomer shifts exhibits a relatively weak dependence upon $N$ at higher neutron numbers but becomes significantly larger at $N=102,104$ (Fig.~\ref{fig:fig5.10}(a)).

In the gold isotope chain, shape coexistence was recently reported in $^{187}$Au ($N=108$) through observation of a large isomer shift between the $1/2^{+}$ and $9/2^{-}$ states~\cite{Au-radii2020-1} (the red circle in Fig.~\ref{fig:fig5.10}(a)).
Located at $N=108$, this isotope directly precedes an abrupt onset of ground-state deformation which occurs at lower neutron number to form an elevated plateau in their charge radii~\cite{Au-radii1987}.

Above $Z=82$, the spin/parity of the ground and isomeric states in odd-$A$ bismuth and astatine are reversed such that the isomeric intruder state typically possesses a spin/parity of $I=1/2^{+}$, and for the most part exhibits a larger charge radius than its respective $9/2^{-}$ ground state. 
Measured isomer shifts for bismuth (blue triangles) and astatine (purple pentagons) are shown in Fig.~\ref{fig:fig5.10}(a). In bismuth, a large but relatively consistent isomer shift is seen in $^{193,195,197}$Bi ($N=110,112,114$)~\cite{Bi-radii2016}. However, in astatine, a large reduction in the magnitude of isomer shift is observed in $^{195}$At at $N=110$, shown as an empty purple pentagon in Fig.~\ref{fig:fig5.10}(a). 
In this isotope, the intruder-configuration $1/2^{+}$ state becomes the ground state. In addition, a strong onset of deformation in the non-intruder state which is now isomeric and tentatively assigned a spin of $7/2^{-}$) results in it possessing a larger charge radius.
The behaviour of these isomer shifts exhibits a strong dependence upon the number of valence protons (protons holes) with respect to the magic $Z=82$ lead chain.

\subsubsection{Changes in mean-square charge radii beyond $N=126$}
The characteristic change in the slope observed in the trend of nuclear charge radii crossing neutron shell closures was recently measured in mercury isotopes~\cite{Hg-radii2021_PRL,Hg-radii2021-PRC}.
These data reveal the first measurements of this so called `kink' at $N=126$ for an isotope chain below $Z=82$.
Figure~\ref{fig:fig5.10} (b) presents the changes in mean-square charge radii of mercury~(green circles), lead~(red squares) and bismuth~(blue triangles) around $N=126$. The slope of the kink crossing $N=126$ in mercury is similar in magnitude to that of what was reported in the proton-magic lead~\cite{Pb-radii2009} and bismuth chains~\cite{Bi-radii2018-2}.

The new measurements in mercury and previous ones in lead were compared to theoretical calculations from both relativistic Hartree-Bogoliubov (RHB) and non-relativistic Hartree-Fock-Bogoliubov (NR-HFB) approaches, where it is shown that both the kink at $N=126$ and the OES are better reproduced with the former~\cite{Hg-radii2021_PRL,Hg-radii2021-PRC}.
In the framework of RHB, the kink is shown to originate from occupation of the $\nu 1h_{11/2}$ orbital above $N=126$. Contrary to recent interpretations from Fayans DFT describing the kinks at $N=126$ and $N=82$ in lead and tin isotopes, respectively ~\cite{Sn-radii2019}, a significant contribution from pairing is not needed to simultaneously reproduce the observed kink at $N=126$ and the OES surrounding it.

\subsubsection{Octupole deformation from inverted odd-even staggering of nuclear charge radii}
One major avenue of research in the lead region involves investigating nuclei which are octupole deformed. 
Early decay studies inferred the existence of these `pear-shaped’ nuclei through identification of low-lying opposite parity doublets in their excitation spectra.
More recently, experiments were able to directly measure non-zero $E3$ transition matrix elements using Coulomb excitation of post-accelerated beams, which revealed static octupole deformation in the ground states of certain neutron-rich radium isotopes~\cite{Ra-pearshape,Ra-pearshape2}.
The reflection-asymmetric charge distributions of these deformed systems are predicted to largely amplify their symmetry-violating properties, making these nuclei of particular interest for experiments searching for physics beyond the Standard Model~\cite{Ra-pearshape1,Ra-pearshape2,Ra-pearshape}. 

Laser spectroscopy experiments investigating the changes in mean-square charge radii have long observed that the odd-even staggering of this observable is inverted in regions associated with high octupole deformation.
In the odd-$Z$ actinium isotopes \footnote{Although actinium is usually classified as being part of the actinide series, measurements of its nuclear properties are included in this section to aid the discussion of octupole deformation in lighter non-actinide elements.}, laser spectroscopy performed at the ISAC-TRIUMF, was used to search for signatures of octupole deformation in $^{225-229}$Ac~\cite{Ac-radii2019}.
The isotope $^{226}$Ac was shown to exhibit an inverted odd-even staggering in its charge radius. 
In addition, comparison of the magnetic moments of $^{225,227,229}$Ac with values calculated using the Nilsson model support the assumption that $^{225,227}$Ac are octupole deformed with $^{229}$Ac being so to a lesser degree.

A large inversion of odd-even staggering was also reported in the neutron-rich astatine isotope $^{218}$At where it was shown to possess a larger radius than the average of its even-$N$ neighbours $^{217,219}$At~\cite{At-radii2019}, suggesting possible octupole collectivity.
In radium, a recent study of its neutron-rich isotopes employing high- and low-resolution spectroscopy provided first measurement of the quadrupole moment of $^{231}$Ra and charge radius of $^{233}$Ra~\cite{Ra-radii2018}.
These data show that the normal odd-even staggering continues to persist in the most neutron-rich radium isotopes indicating an end to the region of octupole deformation. These systems instead become more quadrupole deformed with increasing $N$ as previously suggested~\cite{Ra-mass2014}.

\subsubsection{Precision tests of QED}
Bound electrons in highly charged ions (HCI) experience extremely strong fields up to a million times larger than those experienced by those in light atoms.
HCIs therefore constitute a powerful environment in which to test predictions from QED~\cite{HCI2019}.

Recent measurements of the hydrogen-like ($^{209}$Bi$^{82+}$) and lithium-like $^{209}$Bi$^{80+}$ HCI in bismuth at the Heavy Ion Storage Ring (ESR) at GSI determined the specific difference between their ground-state hyperfine splittings~\cite{Bi-HCI2017}.
This quantity, believed to be relatively insensitive to nuclear structure effects, was compared to predictions from QED where a 7-$\sigma$ discrepancy was observed constituting the so-called `hyperfine puzzle'.

Complementing this study, collinear laser spectroscopy of atomic $^{208}$Bi combined with hyperfine anomaly calculations in addition to the aforementioned measurements of $^{209}$Bi$^{80+,82+}$, yielded a high-precision value of $\mu(^{208}$Bi$)$~\cite{Bi-radii2018}.
This measurement was then used to predict the hyperfine splittings in $^{208}$Bi$^{80+,82+}$.
A future experiment is proposed in order to experimentally confirm the cancellation of nuclear structure effects in the specific difference which will enable such contributions to be excluded as the cause of the hyperfine puzzle.

\subsubsection{Fundamental atomic properties of heavy elements}
Laser-based techniques at RIBs have also yielded fundamental atomic properties of heavy elements which do not occur in large quantities in nature.

Previously, in-source RIS at ISOLDE was used to precisely measure the ionization potential of astatine~\cite{At-IP2013}.
More recently, laser photodetachment of astatine anions using GANDALPH~\cite{GANDALPH2020} at the same facility was used to determine the element's electron affinity~\cite{At-EA2020}.
In this study, astatine anions were produced directly through negative surface ionization, which can be efficient for elements possessing a high electron affinity.
An alternative approach for negative ion production utilizes collisions with an alkali vapor to achieve double charge exchange on incoming positive ions~\cite{Po-proposal}. 
Whereas negative ions are only directly deliverable for a small class of elements at ISOLDE, double charge exchange could enable radioactive anions of a wider range of elements to be studied.

In the neighboring polonium chain, two experimental campaigns at ISAC-TRIUMF and ISOLDE-CERN measured its first ionization potential.
At ISAC, over 100 new even-parity Rybderg states belonging to 5 series were identified~\cite{Po-IP2} whereas 110 odd-parity Rydberg states were measured at ISOLDE~\cite{Po-IP2019-1}. 
The two experiments were able to independently determine the ionization potential of polonium with over a two-orders-of-magnitude improvement in precision compared to existing values.

\subsection{\it Towards heavier elements in the actinides~\label{sec:heavy}}  
Tremendous progress has been made in the past decade in performing laser spectroscopy of the heaviest elements.
A dedicated review on this topic was recently published in Ref.~\cite{PPNP2021}. Here, a brief overview of the recent main highlights will be given.

\subsubsection{Nobelium ($Z = 102$)}
One of the most substantial milestones in this region is the first laser spectroscopy study of nobelium~\cite{No-atom2016,No-radii2018,No-IP2018}. With an atomic number of 102, these isotopes represent the heaviest species to be studied with laser spectroscopy techniques so far.

Using the RADRIS technique described previously in~ Sec.~\ref{sec:RADRIS}, the strong $^{1}S_{0} \rightarrow ^{1}P_{1}$ transition was identified in $^{254}$No, produced at rate of around 15 particles per second at GSI~\cite{No-atom2016}.
Spectroscopy was performed on the isotopes $^{252-254}$No yielding their changes in mean-square charge radii in addition to the magnetic dipole and electric quadrupole moments of $^{253}$No~\cite{No-radii2018}.
DFT calculations of these isotopes suggest a significant reduction in charge density at the centres of these nuclei.
In addition, the ionization potential of nobelium was measured through extrapolating the convergence limit of high-lying Rydberg states~\cite{No-IP2018}. Series originating from two distinct states ($^{1}$P$_{1}$ and $^{3}$D$_{3}$) were observed where the $^{3}$D$_{3}$ state was populated through collisional de-excitation of excited atoms in the $^{1}$P$_{1}$ state. 

Beyond nuclear structure, an important role of this work is its function as a stringent benchmark for atomic theory. Methods predicting properties of heavy elements face significant challenges as their atomic structure is strongly influenced by relativistic and electron-correlation effects. Experimental measurements of actinide elements are critical to validate techniques whose predictions will form an important part of extending laser spectroscopy studies to even heavier systems.

\subsubsection{$^{229m}${\rm{Th}} ($Z=90$)}
The low-lying isomer in $^{229}$Th has garnered significant attention owing to its exceptionally low excitation energy~\cite{Th-2021review}. The most precise determination of this energy to date is 8.28(17)~eV~\cite{229mTh-energy}, an energy which may be soon accessible by precision laser technology.
This unique characteristic makes it the most promising candidate for a nuclear clock which could exceed the stability of state-of-the-art atomic clocks based upon forbidden optical transitions in ions in the future~\cite{Th-2021review}.

Recently, laser spectroscopy of $^{229}$Th was performed on trapped Th$^{2+}$ ions at two institutes~\cite{229mTh-radii2018,229mTh-moment2018}. In one setup, $^{229}$Th$^{2+}$ ions were trapped following laser ablation of a $^{229}$Th-containing target~\cite{229mTh-radii2018}. At the other, the ions were loaded through catching recoiling $\alpha$-particles resulting from the decay of $^{233}$U. The isomeric state is populated with a branching ratio of 2$\%$ in this process~\cite{229mTh-moment2018}. In the apparatus of the latter, additional HFS components were observed which could be unambiguously assigned to belong to the isomeric state. This allowed the magnetic dipole and electric quadrupole moments of the isomer to be determined in addition to its mean-square charge radius.
These measurements are important for investigating feasibility of a solid-state clock wherein the nuclear quadrupole moment of the isomer interacts with electric field gradients within the crystal structure in which it is implanted.
In addition, they allow the HFS of the isomer to be estimated in any transition across all charge states of thorium, if measurements exist in $^{229g}$Th and any other isotope.
This, for example, will allow the HFS of the isomer to be derived in transitions which form part of the closed laser-cooling cycles of Th$^{3+}$~\cite{Th3+lasercool}.

\subsubsection{Other heavy elements}
Beyond the cases described above, laser spectroscopy experiments have been undertaken in additional elements in the actinide region.
A combined CLS and high-resolution RIS study in plutonium yielded the changes in mean-square charge-radii of the long-lived $^{238-242,244}$Pu isotopes~\cite{Pu-radii2017}.
The resulting radii were shown consistent with results from muonic x-ray experiments.

An off-line RIS study on einsteinium ($Z=100$) isotopes performed at RISIKO was able to determine the magnetic moment of $^{255}$Es and the quadrupole moments of $^{254,255}$Es for the first time~\cite{Es-moment2022}. Isotope shifts were also measured, however no changes in mean-square charge radii were deduced owing to the lack of available field- and mass-shift factors in this chain.

Recent studies in protactinium ($Z=91$) at RISIKO measured a multitude of newly observed transitions that involve over 1500 states, demonstrating the highly complex atomic structure of actinide species~\cite{Pa-levels2018}.

\subsection{\it Molecules containing unstable nuclei\label{sec:Molecular}} 
Developments in precisely controlling and interrogating molecules have opened new doors in recent years in the study of fundamental interactions~\cite{acme18,cairncross:2017,alt18}. Molecules containing radioactive atoms, in particular, are expected to provide a new window into our study of atomic nuclei, offering unique opportunities for systematic measurements of symmetry-violating nuclear properties~\cite{flambaum2020electric,Isa10,Gar21,Yu21,Isa21}. In certain molecules, parity- and time-reversal violation effects can be massively enhanced when compared to atomic systems~\cite{Saf18,hut20}. As these symmetry-violating effects scale with the atomic number, nuclear spin and nuclear deformation (both quadrupole and octupole), molecules containing heavy, deformed radioactive nuclei have been predicted to provide a superior sensitivity in several theoretical studies~\cite{Isa10,flambaum2020electric,Mit21,zul22}. Experimental measurements of such radioactive systems are however scarce, and quantum chemistry calculations often constitute the only source of information available. Recent pioneering experiments have demonstrated the ability to perform laser spectroscopy measurements of molecules containing short-lived isotopes of radium, such as radium monofluoride (RaF)~\cite{Gar20,Udre21,Gar21}. The structure of RaF molecules was shown to be favorable for laser cooling~\cite{Isa10,Gar20}, and at the same time highly sensitive to the short-range electron-nucleon interaction~\cite{Udre21}. These studies therefore confirmed RaF as a compelling system in which to explore opportunities for precision nuclear structure and fundamental symmetry tests. Complementary studies have demonstrated the ability to create polyatomic molecules containing laser-cooled radium isotopes in ion traps~\cite{Fan21}.
\begin{figure*}[h!]
\begin{center}
\includegraphics[width=0.99\textwidth]{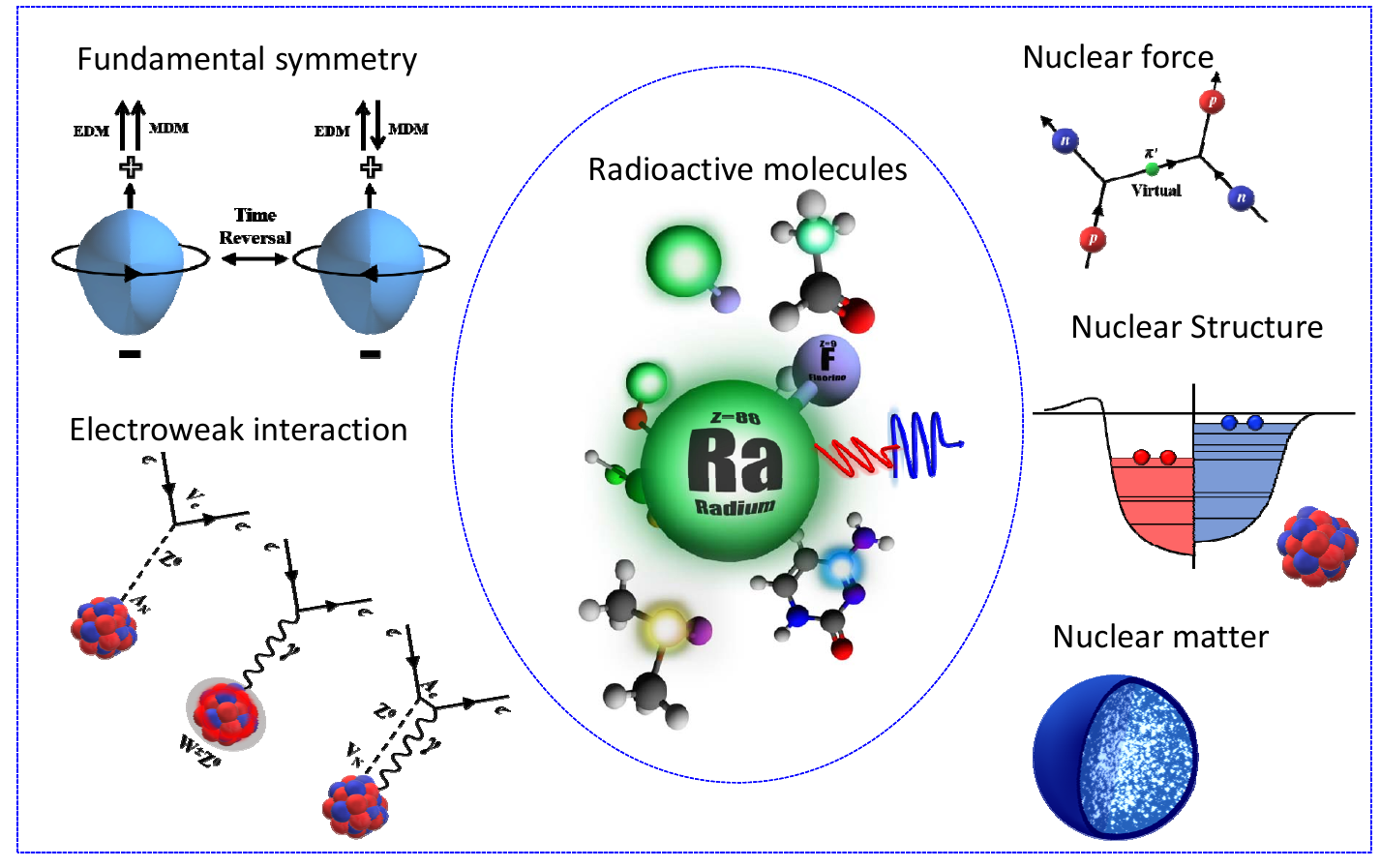}
\caption{\label{fig:fig5.11}\footnotesize{Radioactive molecules offers several, and in some cases unique, opportunities in the study of fundamental symmetries, nuclear structure, and astrophysics. See text for more details.}}
\end{center}
\end{figure*}

A sketch of the different physics that can be studied with radioactive molecules is shown in Fig.~\ref{fig:fig5.11}. Complementary to atoms, molecules can be formed to exhibit a hyperfine structure in their ground states, hence, offering superior, and in some cases unique, sensitivity to measure nuclear electromagnetic moments~\cite{teodoro:2015}. A huge benefit is the enhanced sensitivity to hadronic parity violation, which can be more than 11 orders of magnitude higher than that in atoms. In addition, diatomic molecules can be highly polarizable, reaching internal effective electric fields that can enhance the sensitivity to time-reversal violating properties by more than 3 orders of magnitude when compared to atoms~\cite{Saf18,acme18}. This can be combined with the nuclear enhancement of time-violating nuclear properties, e.g Schiff moments, found in octupole-deformed nuclei~\cite{Isa10,flambaum2020electric,zul22,ala22}.

In addition to fundamental physics research, molecules composed of radioactive nuclei are expected to generate new opportunities in astrophysics.
Radioactive nuclei act as sensitive probes of stellar nucleosynthesis processes as their lifetimes are short with respect to typical stellar and galactic timescales~\cite{RadMolAst2021,astrophysics-RIB}.
Any radioactive nuclei that are present must therefore originate from current generation stars which allows observations of them to stringently test models of galactic chemical evolution.
Precisely characterizing the rotational structure of molecules containing radioactive isotopes of astrophysical interest will enable their unambiguous identification in space and allow their presence to be pinpointed to individual stellar objects~\cite{26AlF-2018}. 
This is currently not possible as the majority of observations of them to date involve detecting characteristic $\gamma$- or x-rays resulting from their decay~\cite{SPI26Al}.
Such observations are hindered by their poor spatial resolution due to the difficulty in manipulating high-energy photons and are strictly limited to species which decay in this manner.
Therefore, radioactive molecules are expected to enable new radioisotopes to be detected around single stellar objects across our galaxy.
Their radioactive half-life can also act as a means to probe the timescales involved with their journey from their parent star as they propagate towards the interstellar medium where they can form new-star forming regions.

\section{Perspectives and Challenges\label{outlook}}
Although considerable progress has been made studying unstable nuclei with laser spectroscopy techniques across much of the nuclear chart, only about one third of the total isotopes producible at current-generation RIB facilities have been addressed and some key regions of the nuclear chart remain unexplored. 
As summarized in Fig.~\ref{fig:fig1.2}, several gaps remain in our knowledge, even close to the valley of stability. This is mainly due to challenges involving the production of isotopes of these elements and/or their structure hindering studies.

To overcome these challenges and expand the ability to perform laser spectroscopy experiments on more isotopes, elements, and even to yield additional nuclear observables, further improvements and new developments of both RIB facilities and experimental techniques are being pursued, as will be detailed in the following.

\subsection{\it RIB production and experiments at existing and future facilities}
Until now, the majority of laser spectroscopy experiments of unstable isotopes have been performed at ISOL-type RIB facilities (see Fig.~\ref{fig:fig1.1}), such as ISOLDE-CERN. These facilities directly produce low-energy RIBs with favorable properties for laser spectroscopy experiments. As has been discussed in Sec.~\ref{sec:ISOL}, the efficient production and extraction of RIBs at thick-target ISOL facilities depends strongly upon the chemical properties of the produced element. This can be challenging for certain elements such as light and refractory elements. Efforts to produce more elements at ISOL facilities are ongoing by utilizing different target materials and/or ion source combinations. Improving laser ionization schemes for RIB production is a significant part of this effort.

Alternative techniques to produce RIBs have been developed to provide complementary access to radioactive isotopes of other elements, e.g. the Ion Guide Isotope Separator On-Line (IGISOL) technique~\cite{IGISOL2014}. This approach offers much-improved access to refractory elements.
The PF technique, as was discussed in Sec.~\ref{sec:PF}, can in principle produce RIBs of all elements that are lighter than the heavy projectile. However, isotopes produced in this fashion have a high kinetic energy and possess a large energy spread. By using a gas cell, the RIB can be fully stopped and cooled to produce low-energy beams for laser spectroscopy experiments~\cite{Gas-cell}. This has been realized at NSCL/FRIB in the US, enabling unique access to neutron-deficient isotopes of calcium and nickel~\cite{Ca-radii2019, Ni-radii2021}. 

Gas-cell methods offer further opportunities for laser spectroscopy experiments~\cite{Gas-cell,Gas-cell2}, and several related techniques are currently under development and/or planned at existing and upcoming RIB facilities. Some examples of laser spectroscopy experiments using gas-cell techniques are S3-LEB at SPIRAL2-GANIL~\cite{S3-LEB}, PALIS at RIKEN-RIBF~\cite{PALIS}. New examples using collinear laser spectroscopy following a gas stopper include BECOLA and RISE at NSCL/FRIB~\cite{BECOLA}, LaSpec at FAIR~\cite{Laspec}, and CLS at SLOWRI of RIKEN-RIBF~\cite{RIKEN-CLS}. 

The development of new-generation facilities, such as the recently operational FRIB in the US~\cite{FRIB}, as well as those still under construction like SPIRAL2-GANIL in France~\cite{SPIRAL2-GANIL}, FAIR in Germany~\cite{FAIR}, HIAF in China~\cite{HIAF} and RAON in South Korea~\cite{RISP-RAON}, will allow unprecedented access to short-lived isotopes.
Upgrades of existing facilities are also planned, largely aiming to increase the usable intensity and/or energy of driver beams to enhance exotic isotope production, such as the EPIC project at ISOLDE-CERN~\cite{EPIC}. Future opportunities to explore ever-more exotic nuclei are foreseen with the next-generation facilities EURISOL in Europe~\cite{EURISOL} and BISOL in China~\cite{BISOL}.

\subsection{\it Improving experimental sensitivity and resolution}
Efforts are being undertaken to improve both the overall sensitivity and spectral resolution of laser spectroscopy techniques.
A higher experimental sensitivity enables the study of more exotic isotopes, which are produced with small yields and often delivered alongside large amounts of isobaric contamination. On the other hand, a higher spectral resolution will allow access to more subtle details of the nuclear distribution. In addition to the magnetic dipole, and electric quadrupole moments that are routinely measured today, higher-order moments could provide additional information for nuclear structure studies~\cite{chi92,Sc-moment2020-45Sc,rei22}. 

\subsubsection{Experimental sensitivity}
In-source laser spectroscopy experiments have reached record sensitivities employing sensitive detection schemes utilizing MR-TOF-MS devices, Penning traps and decay detection~\cite{Hg-radii2018,Ag-radii2021}. Efforts to incorporate these detection approaches to the CRIS technique are planned which will further boost the sensitivity of the method.

As discussed in Sec~\ref{sec:iontrap}, ion traps, such as the widely used gas-filled radiofrequency Paul trap~\cite{Nb-radii2009,Mn-moment2016}, have long been used to aid laser spectroscopy experiments. While in-trap laser spectroscopy experiments are well-established techniques for precision studies of stable isotopes, they have only been realized in a handful of radioactive species. A program aiming to perform laser spectroscopy measurements of radioactive ions circulating a MR-TOF-MS device is being developed, named MIRACLS, at ISOLDE-CERN~\cite{sel21,MIRACLS}. The sensitivity of this approach is expected to be significantly improved when compared to conventional fluorescence-detected CLS in elements which possess a suitable spectroscopic transition and structure in the their ionic form. This setup is expected to be used in the coming years for the study of neutron-rich $^{33,34}$Mg and neutron-deficient $^{20}$Mg at ISOLDE-CERN~\cite{MIRACLS-Mg}. 

\subsubsection{Spectral resolution}
The achievable spectral resolution of collinear laser spectroscopy experiments used for the study of radioactive isotopes at on-line facilities is generally limited to a few tens of MHz. This limitation mainly results from combination of factors related to the laser technology employed, the natural linewidth of the transitions measured in addition to the experimental interaction time, and the energy spread of the species interrogated. 
In order to meaningfully reduce the spectral resolution to a few hundreds or even few tens of kHz, in-trap and/or RF/MW double resonance laser spectroscopy methods can be employed. This increased resolution will enable the access to higher-order moments of the nuclear distribution. Double resonance spectroscopy methods have been intensively used for atomic and molecular physics research~\cite{atomicbeam,molecularbeam,chi92}. Application of these methods to radioactive isotopes have been limited so far and there are ongoing efforts to realize them on both trapped and in-flight unstable nuclei~\cite{OROCHI}.

Progress in improving the resolution of in-source RIS techniques promise to not only yield higher-precision nuclear observables but also broadly extend their applicability.
For hot-cavity in-source RIS, the ability to circumvent Doppler broadening using PI-LIST will allow medium-mass elements to be investigated at thick-target ISOL facilities~\cite{Tc-radii2020,Pm-radii2020}.
Furthermore, PI-LIST will enable an isomer-selective mode of RILIS operation to be a more universal feature across the nuclear chart.

The first realization of the in-gas jet method confirmed its potential for performing efficient, high-resolution~(few hundred MHz) spectroscopy of species produced at gas-cell facilities~\cite{Ac-moment2017}. 
Setups employing this method at S3-LEB of SPIRAL2-GANIL~\cite{S3-LEB} and GSI~\cite{GasJetGSI} are being commissioned.

A new technique proposed, named Laser Resonance Chromatography (LRC), plans to make use of the difference in the mobility of ions occupying different electronic states as they pass through a gas-filled drift tube~\cite{LRC2020,PPNP2021}. 
Optically pumping these ions in the supersonic jet following the stopping cell could enable a general approach for performing high-precision spectroscopy of the heaviest elements which avoids the need to neutralize and re-ionize these species.

\subsection{\it Precision measurements of radioactive molecules}
The study of radioactive molecules requires an exceptional combination of spectral resolution and sensitivity. The additional vibrational and rotational degrees of freedom can result more than $10^4$ molecular states being populated at room temperature. To overcome this challenge, multiple strategies are being pursued by different research groups~\cite{Gar20,Fan21}. The production of cold molecules by using cryogenic atom/ion traps and laser cooling of certain molecules are two such routes towards achieving high-precision studies of these complex systems. While some molecules of interest have been identified at current RIB facilities, developing fully controllable methods to produce specific families of radioactive molecules is an ongoing effort~\cite{Ballof-Phd,LOI-molecular}.

\section{Conclusion\label{summary}}

Laser spectroscopy is well-established as a unique and powerful tool in our global understanding of the structure of atomic nuclei and the interactions that govern them. The nuclear properties of short-lived nuclei (spins, electromagnetic moments, and charge radii) that can be obtained from these experiments have proven essential to guide the development of nuclear theory. 

In recent years, significant progress has been made in the development of laser spectroscopy techniques, enabling the study of exotic nuclei at the extremes of existence. Combined improvements in sensitivity and resolution have motivated the development of theoretical models of atomic and nuclear structure. These achievements have expanded our knowledge of weakly bound nuclei towards the \textit{terra incoginta} of the nuclear chart and deepened our understanding of complex nuclear phenomena and the forces which drive them. The main purpose of this article was to introduce the versatile range of laser spectroscopy methods emphasizing recent technological breakthroughs in addition to their application in studying exotic nuclei.

Some of the technical achievements that were recently realized include: the first on-line application of laser ionization spectroscopy in a supersonic gas jet~\cite{Ac-moment2017,Ac-radii2019}, the combination of in-source spectroscopy with sensitive ion traps~\cite{Hg-radii2018,Ag-radii2021}, the first study of isotopes of nobelium ($Z=102$), the heaviest element investigated by laser techniques to date~\cite{No-atom2016,No-radii2018}, improvements of sensitivity in collinear laser spectroscopy~\cite{K-radii2015,ROC-COLLAPS,Ca-radii2019}, and simultaneously achieving a high resolution and high sensitivity with the CRIS technique~\cite{Cu-radii2020,K-radii2021}. These developments continuously yield new results and allow us to further advance our understanding of nuclear structure.

Some noteworthy highlights in the study of atomic nuclei in different mass regions of the nuclear chart include: the measurement of nuclear charge radii in the calcium region ~\cite{Ca-radii2016,K-radii2021,Ca-radii2019}, the identification and measurement of a long-lived intruder isomer~\cite{Zn-radii2016} and a rich series of nuclear properties in isotopes around $^{78}$Ni~\cite{Cu-radii2020,Cu-moment2017,Ni-radii2022}, extensive studies of the nuclear properties around the doubly magic nuclei $^{100,132}$Sn~\cite{Cd-radii2018,Sn-radii2020,Ag-radii2021,In-moment2022,Pd-radii2022}, a microscopic interpretation of underlying mechanism responsible for the famous shape staggering in mercury~\cite{Hg-radii2018}, and the extension of experiments to measure the nuclear properties of actinide isotopes up to nobelium~\cite{No-atom2016}. These experimental findings have in turn stimulated significant progress in improving nuclear interactions and the many-body methods that employ them, in order to better understand the properties of exotic nuclei~\cite{Cu-radii2020,Sn-radii2019}. The first laser spectroscopy experiment that was able to measure the structure of molecules containing short-lived isotopes~\cite{Gar20,Udre21}, has paved the way for additional opportunities in using these systems in nuclear physics, fundamental symmetries and astrophysics studies. 

The achievements summarized in this review would not have been possible without continuous efforts to improve all aspects of this multidisciplinary endeavour ranging from the advancements in laser spectroscopy techniques and approaches to the production of more exotic isotopes at current and new-generation RIB facilities around the world. 
Efforts to further surpass the current limits of experimental sensitivity and precision promise to enable the study of isotopes towards the limits of existence at next-generation facilities. Combining these measurements alongside state-of-the-art nuclear theory will enable the emergence of new nuclear phenomena and their connection with the fundamental forces of nature to be examined in unprecedented detail.

\section*{Acknowledgments}
This work is supported by the National Key R\&D Program of China (Contract No. 2018YFA0404403), the National Natural Science Foundation of China (No:11875073,12027809,U1967201,11961141003), the U.S. Department of Energy, Office of Science, Office of Nuclear Physics under award numbers DE-SC0021176 and DE-SC0021179. Support from the John W. Jarve (1978) Seed Fund for Science Innovation is also acknowledged.

\bibliographystyle{elsarticle-num}
\bibliography{laser}

\end{document}